\documentclass[a4paper,11pt]{article}
\usepackage{jcappub}
\usepackage{amsmath, amssymb, graphics, epsfig}
\usepackage{subfig}
\usepackage{caption}
\usepackage{hyperref}\usepackage{booktabs}
\usepackage{amsmath}
\usepackage{mathtools}
\usepackage{color}
\usepackage{verbatim}
\usepackage{float}
\usepackage{amsfonts}
\usepackage{amssymb}
\usepackage{bm}
\usepackage{ulem}\usepackage{graphicx}
\usepackage[usenames,dvipsnames,svgnames,table]{xcolor}

\def\beq{\begin{equation}}
\def\eeq{\end{equation}}

\def\bea{\begin{eqnarray}}
\def\eea{\end{eqnarray}}

\def\C{{\cal C}}
\def\0{{\boldsymbol 0}}

\def\A{{\cal A}}
\def\B{{\cal B}}
\def\T{{\cal T}}
\def\C{{\cal C}}

\def\lsim{\mathrel{\rlap{\lower3pt\hbox{\hskip0pt$\sim$}}
   \raise1pt\hbox{$<$}}}         
\def\gsim{\mathrel{\rlap{\lower4pt\hbox{\hskip1pt$\sim$}}
   \raise1pt\hbox{$>$}}}         

 \newcommand{\sfootnote}[1]{}
\definecolor{bluc}{cmyk}{1,1,0,0.1}
\definecolor{rossoCP3}{cmyk}{0,.88,.77,.40}
\definecolor{rosso}{cmyk}{0,1,1,0.4}
\definecolor{rossos}{cmyk}{0,1,1,0.55}
\definecolor{rossoc}{cmyk}{0,1,1,0.2}
\definecolor{verdes}{cmyk}{0.92,0,0.59,0.4}

\hypersetup{colorlinks, bookmarksopen, bookmarksnumbered,
citecolor=verdes, linkcolor=bluc, pdfstartview=FitH, urlcolor=rossos}

\newcommand{\mio}[1]{}

\definecolor{Gray}{gray}{0.95}

\renewcommand\&{&}

\def\circa#1{\,\raise.3ex\hbox{$#1$\kern-.75em\lower1ex\hbox{$\sim$}}\,}

\newcommand{\be}{\begin{equation}}
\newcommand{\ee}{\end{equation}}
\newfam\rsfsfam
\def\mathscr#1{{\fam\rsfsfam\relax#1}}

\def\circa#1{\,\raise.3ex\hbox{$#1$\kern-.75em\lower1ex\hbox{$\sim$}}\,}

\title{Gravitational waves and geometrical optics in scalar-tensor theories}
\author[a,b,c]{Alice Garoffolo,}
\emailAdd{garoffolo@lorentz.leidenuniv.nl}
\author[d]{Gianmassimo Tasinato,}
\emailAdd{g.tasinato2208gmail.com}
\author[e,f]{Carmelita Carbone,}
\emailAdd{carmelita.carbone@inaf.it}
\author[c,g,h]{Daniele Bertacca,}
\emailAdd{daniele.bertacca@pd.infn.it}
\author[c,g,h,i]{and Sabino Matarrese}
\emailAdd{sabino.matarrese@pd.infn.it}
\affiliation[a]{Institute Lorentz, Leiden University, PO Box 9506, Leiden 2300 RA, The Netherlands}
\affiliation[b]{Dipartimento di Fisica ``Aldo Pontremoli'', Universit\`{a} degli Studi di Milano, via Celoria 16, I-20133 Milano, Italy}
\affiliation[c]{Dipartimento di Fisica e Astronomia G. Galilei, Universit\`a degli Studi di Padova, I-35131 Padova, Italy.}
\affiliation[d]{{Department of Physics, Swansea University, Swansea, SA2 8PP, UK}}
\affiliation[e]{INAF - Istituto di Astrofisica Spaziale e Fisica cosmica di Milano, via Alfonso Corti 12, 20133, Milano, Italy}
\affiliation[f]{INFN, Sezione di Milano, via Celoria 16, I-20133 Milano, Italy}
\affiliation[g]{INFN, Sezione di Padova, via F. Marzolo 8, I-35131, I-35131 Padova, Italy.}
\affiliation[h]{INAF - Osservatorio Astronomico di Padova, vicolo dell'Osservatorio 5, I-35122 Padova, Italy.}
\affiliation[i]{Gran Sasso Science Institute, viale F. Crispi 7, I-67100 L'Aquila, Italy.}

\abstract{ The detection of gravitational waves (GWs) propagating through cosmic structures can provide invaluable information
on the geometry and content of our Universe as well as on the fundamental theory of gravity. In order to test possible departures from General
Relativity, it is essential to analyze, in a modified gravity setting,   how GWs propagate through a perturbed cosmological space-time. Working
within the framework of geometrical optics, we develop tools to address this topic for a broad class of scalar-tensor theories,  including scenarios
with non-minimal, derivative couplings between scalar and tensor modes.
{By focussing on a set-up where scalar modes propagate with the same speed as tensor
degrees of freedom, }
we determine the corresponding  evolution equations  for
 the GW amplitude and polarization tensor. The former satisfies a generalized evolution equation that includes possible effects due to a
 variation of the effective Planck scale; the latter can fail to be parallel transported along  the GW geodesics unless certain conditions are satisfied.
We apply our general formulas to specific scalar-tensor theories with unit tensor speed, and then focus on GW propagation on a perturbed space-time.
We determine corrections to standard formulas for the GW luminosity distance and for the evolution of the polarization tensor, which
depend both on modified gravity and on the effects of cosmological perturbations. Our results can constitute a starting point to disentangle among degeneracies
from different sectors that can influence GW propagation through cosmological space-times.
}

\begin{document}

\maketitle

\section{Introduction}

The $\Lambda$CDM model of cosmology (see e.g. \cite{Bahcall:1999xn} for
an introduction) offers a well defined, simple framework
 for analyzing current cosmological data, and it   agrees with
 observations. In this set up, a cosmological constant $\Lambda$ is responsible
 for the present-day acceleration of our universe. Nevertheless, the extreme smallness
 of the value of $\Lambda$ with respect to all other energy scales one
 encounters in particle physics \cite{Weinberg:1988cp}, as well as anomalies in
 observational results (the most recent being the so-called $H_0$ tension, see
 e.g. the assessment in \cite{Freedman:2017yms}),
  motivate  the  exploration of  alternative scenarios, possibly based on modifications
  of General Relativity (GR). In modified
  gravity, additional degrees of freedom are introduced, whose dynamics drive present-day
  cosmological acceleration (see e.g. \cite{Frieman:2008sn,Clifton:2011jh,Joyce:2014kja} for reviews).
  The minimal  possibility for modifying gravity is
   to consider scalar-tensor theories, where a single scalar field is added
   to the spin-2 degrees of freedom characterising GR. Thanks
  to interactions with itself and with spin-2 tensor modes (or matter fields),
   scalar-tensor theories can exhibit {\it self-accelerating solutions}, where
   the scalar profile sources accelerating cosmologies with no need of cosmological
   constant (see e.g. \cite{Joyce:2014kja} for a  review). Moreover, scalar
   fifth forces can be suppressed thanks to powerful
   {\it screening mechanisms}, which are able to hide the
   effects of a light scalar  nearby sources (see e.g. \cite{Burrage:2017qrf,Babichev:2013usa} for reviews on chameleon
   and
   Vainshtein screening mechanisms).

  \smallskip

  The rapidly developing field of gravitational wave cosmology offers
  new avenues for testing modified theories of gravity (see e.g. the specific analysis of \cite{Yunes:2016jcc,Pardo:2018ipy,Lagos:2019kds,Abbott:2018lct,Nishizawa:2019rra,DiValentino:2018jbh}).
  The single event GW170817 \cite{TheLIGOScientific:2017qsa,Goldstein:2017mmi,Savchenko:2017ffs,Monitor:2017mdv}
and its EM counterpart \cite{GBM:2017lvd}
impose strong constraints on modified theories of gravity predicting
a speed of gravitational waves different than light \cite{Creminelli:2017sry,Sakstein:2017xjx,Ezquiaga:2017ekz,Baker:2017hug} (see also
the analysis in  \cite{Lombriser:2015sxa,Bettoni:2016mij}).  In general, the physics of standard  sirens offer great promises for cosmology
 \cite{Schutz:1986gp,Holz:2005df,Dalal:2006qt,MacLeod:2007jd,Nissanke:2009kt,Cutler:2009qv,Camera:2013xfa, D_Agostino_2019}, and
 studies of cosmological parameter estimation have been carried on for
2nd generation GW experiments (see e.g. \cite{DelPozzo:2011yh}),  for ET (see \cite{Sathyaprakash:2009xt,Zhao:2010sz,Taylor:2012db}), for
 LISA and space-based detectors \cite{Nishizawa:2011eq,Tamanini:2016zlh,Caprini:2016qxs,Belgacem:2019pkk}.
 In the future,
GW
observables related with GW frequency, chirp mass, and luminosity distance of  coalescing events
%
 will be able to further constrain
 (or discover) modified gravity effects with GW observations, see e.g.
 \cite{Deffayet:2007kf,Saltas:2014dha,Nishizawa:2017nef,Belgacem:2017ihm,Belgacem:2018lbp,Ezquiaga:2018btd,Belgacem:2019pkk} for works
 discussing the potential of GW cosmology
  in probing modified gravity.

  \smallskip

  The purpose of this work is to develop {the first steps towards}
  a systematic and consistent approach for studying the propagation of GWs around arbitrary space-times, based on
  the separation by {\it high} and {\it low} frequency modes
  pioneered by Isaacson \cite{Isaacson:1967zz,Isaacson:1968zza} (see also \cite{Misner:1974qy,Anile:1976gq} for early works on the subject). This formalism
  is at the basis of a geometrical optics approach to GW propagation; we extend it for accounting the possibility of having rapidly moving scalar fluctuations, with
  non minimal couplings to the metric.
  Within this framework,
  in Sections \ref{sec-gen} and \ref{sec_geo_optics}  we
    derive general formulas for
    describing the propagation of GWs, whose structure does not depend on the specific choice
  of the scalar-tensor theory under consideration, nor on the particular background over which the high-frequency
  GW is travelling.  We only need to make (physically well founded) assumptions on the structure of the EMT controlling the propagation
  of high-frequency fluctuations\footnote{We focus on set-up with luminal speed for the spin-2 modes, as suggested by the GW170817 observations.},
  and then apply our arguments within a geometrical optics approximation.
   Besides well-known modified gravity  consequences for the evolution of the amplitude of tensor fluctuations --  related with non-conservation of the
   effective Planck mass
      along
   the null GW geodesics -- we also point out  potentially interesting effects associated with the failure of parallel transport the polarization  tensors along the GW
   geodesics. Although these latter effects can be vanishingly small for GWs propagating through smooth, homogeneous space-times, they can instead contribute
    to the evolution of the polarization tensors through the
    inhomogeneous Universe. GW evolution equations within GR, which  include effects of non-parallel transport for the polarization
    tensors have been studied  in a framework {\it beyond} geometrical optics, see e.g.
      \cite{Misner:1974qy,Anile:1976gq,Harte:2018wni,Dolan:2018ydp,Lagos:2019kds}, associating it with (higher
      order) GW lensing effects (see e.g.  \cite{Takahashi:2005sxa,Oancea:2019pgm,Hou:2019wdg}). Interestingly, we find that these effects can be present also within a geometrical optics approximation in theories of modified gravity. Over the past decades, various classes
      scalar-tensor theories have been applied to cosmology: Brans-Dicke \cite{Brans:1961}; K-essence \cite{ArmendarizPicon:2000ah}; Galileons \cite{Nicolis:2008in}, Horndeski \cite{Horndeski:1974wa,Deffayet:2011gz,Kobayashi:2011nu}, beyond-Horndeski \cite{Zumalacarregui:2013pma,Gleyzes:2014dya}, and DHOST \cite{Langlois:2015cwa,Langlois:2018dxi,Crisostomi:2016czh,Achour:2016rkg,BenAchour:2016fzp,deRham:2016wji,Langlois:2017mxy}. These are increasingly  complex theories with specific
      derivative couplings between scalar and tensor degrees of freedom.
       Our general
       {approach can be in principle applied to any of these theories (and
      possibly beyond, if one admits Lorentz-violating scenarios)  although for this work we focus only
      on scenarios in which the scalar mode propagates at the same speed of
      tensor degrees of freedom. In Section \ref{sec-ST} we apply our approach to a specific case
      of the  kinetic-gravity braiding set-up developed in  \cite{Deffayet:2010qz}, analyzing the differences
  with respect to GR for what concerns the evolution of GWs.   }
  Finally, in Section \ref{c} we apply these formulas to GW propagation through a perturbed
  space-time, to understand the role of modified gravity effects for propagation in an inhomogeneous universe.

  \smallskip

  In fact, we wish to close our Introduction with more detailed arguments for motivating our study.
The physics of GWs travelling through cosmological distances can provide valuable information on  gravitational interactions
 and on the properties of space-time between the  emission and
 detection of GWs.
 In the literature, the effect of cosmological perturbations on the propagation of  GWs has been often neglected. On the other hand, we find it timely to investigate these subtle effects
 to understand possible contaminations and degeneracies, so  to obtain a more reliable estimates of cosmography. Obviously a similar analysis, for the correlated fluctuations in luminosity distance of  the electromagnetic spectrum, has already and widely been discussed in literature, e.g. see \cite{Sasaki:1987ad,Pyne:2003bn,Holz:2004xx,Bonvin:2005ps,Hui:2005nm}.
Recently there have been several initial attempts to investigate the Integrated-Sachs Wolfe (ISW) effect on GWs from supermassive black hole mergers and in particular its impact on the system's parameter estimation \cite{Laguna:2009re}, the ISW of a primordial stochastic background
\cite{Alba:2015cms, Contaldi:2016koz, Bartolo:2019oiq} and for astrophysical stochastic background in
 \cite{Cusin:2017fwz, Cusin:2018rsq, Cusin:2017mjm, Jenkins:2018lvb, Jenkins:2018uac, Jenkins:2018kxc, Cusin:2019jpv, Cusin:2019jhg, Jenkins:2019uzp, Jenkins:2019nks, Bertacca:2019fnt, Canas-Herrera:2019npr}. In these works the authors consider the presence of inhomogeneities in the matter distribution and allow to probe GW's sources on cosmological, galactic and sub-galactic scales, peculiar velocity \cite{Seto:2001qf,Kocsis:2005vv,Bonvin:2016qxr} and lensing effects~\cite{Holz:2005df, Hirata:2010ba, Dai:2016igl, Dai:2017huk,Mukherjee:2019wfw,Mukherjee:2019wcg,Nakamura:1997sw,Takahashi:2003ix}.
 \cite{Bieri:2017vni} examines the GW memory, a permanent displacement of the gravitational wave detector after the wave has passed, in a highly inhomogeneous universe.
 It is worth noticing that environmental effects can also influence estimates of the luminosity distance~\cite{Barausse:2014tra}.  Precisely, these sources are affected by coherent peculiar velocity of the merging at low redshift and weak gravitational lensing by intervening inhomogeneities in the cosmic mass distribution. Both these systematic errors could introduce a misidentification of the host's redshift.
 Consequently, changes of typically a few percent (but occasionally much larger) in the flux are introduced which do not significantly affect the redshift, providing a source of noise in the luminosity distance - redshift relation \cite{Sathyaprakash:2009xt, Hirata:2010ba}.
 Very recently, in \cite{Bertacca:2017vod}, the authors discuss the effect of cosmological perturbations and inhomogeneities on estimates of the luminosity distance of black hole (BH) or neutron star (NS) binary mergers through gravitational waves.
They  applied the ``Cosmic Rulers'' formalism~\cite{Schmidt:2012ne} and considered {\it the observer frame} as reference system and they derived a different expression {\it wrt} \cite{Laguna:2009re}, which is correct for the effect of large-scale structures on GW waveforms, accounting for lensing, Sachs-Wolfe, integrated Sachs-Wolfe, time delay and volume distortion effects, and evaluate their importance for future GW experiments.
\cite{Bertacca:2017vod} showed that  the amplitude of the corrections is important and cannot be negligible for future interferometers as the ET, DECIGO and  Big Bang Observer (BBO). 
 From \cite{Bertacca:2017vod}, the additional luminosity distance uncertainty, arising because of the inclusion of perturbations, has a peak at low-$z$ due to velocity contributions that surpasses the predicted measurement errors for all the experiments considered here. However, peculiar velocity effects rapidly decrease as $z$ increases.
On the other hand lensing, which is an integrated effect, increases with the redshift of the source $z$, and its amplitude is of a factor $\sim 10 $ smaller than ET forecast precision ($\sim2$ for DECIGO).  For  BBO,  the correction to the luminosity distance  is consistently twice the predicted errors, making it a very relevant correction, that one will need to take into account.
Making use of the weak-lensing magnification effect on a GW from a compact binary object, in \cite{Camera:2013xfa} is showed that it is possible to discriminate the concordance $\Lambda$CDM cosmological model and up-to-date competing alternatives as dynamical dark energy models (DE) or modified gravity theories (MG) parametrized with the usual two free  phenomenological functions that modify  the perturbed Einstein  equations relating the matter density contrast to the lensing and the Newtonian potential (e.g. see \cite{Zhao:2010dz}).
Finally, for the effect of lensing magnification, in \cite{Hirata:2010ba}, the authors pointed out that for the uncertainty in the distance to an ensemble of GW's sources is not completely correct to use the standard deviation of the lensing magnification distribution divided by the square root of the number of sources. They showed  that by exploiting the non-Gaussian nature of the lensing magnification distribution, it is possible to improve this distance determination, typically by a factor of $2$ to $3$.

\section{General formalism}\label{sec-gen}

In this Section we develop a general formalism
  aimed to   describe the propagation
  of high-frequency
  gravitational waves (GWs), in  the framework of scalar-tensor theories.

  \smallskip

   While the evolution of spin-2 and spin-0 degrees of freedom decouple around
   conformally flat space-times (as for example homogeneous FRW universes), more care
   is needed when describing GW propagation around perturbed backgrounds -- above
   all in scalar-theories with  kinetic mixing between different fields, motivated
   by recent approaches to the dark energy problem.
   To address this topic in general terms, in this Section
   we propose an
   approach  based on an expansion of the evolution
   equations in an high-frequency
   parameter $\epsilon$.  Such approach  makes use of a convenient unitary gauge to deal
   with scalar fluctuations. It {\it does not} require us to specify the exact structure of the
   scalar-tensor theory we consider  but only at which  order in the small parameter $\epsilon$
    it contributes to the energy momentum
   tensor controlling the high-frequency fluctuations. The general
   set of evolution equations we obtain in this Section shall be then
   applied in Section \ref{sec_geo_optics} to a specific geometrical optics Ansatz.

\subsection{Linearized equations of motion: high and low frequency modes}\label{sec_sep_gauge}
Geometrical optics for gravitational waves (GWs) is a well developed subject, which started with the systematic analysis of Isaacson \cite{Isaacson:1967zz,Isaacson:1968zza}, and that is now part of well-known  textbooks discussing GW physics (see e.g. \cite{Maggiore:1900zz}).
It corresponds to the limit where GWs are characterized by a short wavelength, while the background space-time, over which the wave propagates,  varies over
 much larger scales. Such distinction between  two characteristic length scales allows one to define a hierarchy  and a `small' parameter used  for a derivative expansion. Most works consider geometrical optics for scalar-tensor systems in which the scalar is minimally coupled with gravity (see e.g. the book \cite{Will:2018bme}
  and references therein). On the other hand, for analyzing more modern
 scalar-tensor theories aimed to explain dark energy, the standard formalism
  is generally not sufficient, since it does not satisfactorily catch  possible effects associated with
 derivative self-interactions and with non-minimal couplings of scalar to tensor degrees of freedom. The latter scenarios are well motivated by theories of dark energy exhibiting  self-accelerating solutions  (see e.g. \cite{Joyce:2014kja} for a review).
   Our purpose in this Section is to provide an extension able to eventually include also these models (and possibly
  additional ones),  trying to  be as  general and flexible as possible.

  \smallskip

In the system we consider, the metric $g_{\mu \nu}$ satisfies the gravitational field equations
\be\label{EinsteinEq}
 R_{\mu\nu}\,=\,\frac{8 \pi\,G_N}{c^4}\,\left( T_{\mu\nu}-\frac12  g_{\mu\nu}  T \right)
\ee
where the left hand side (LHS) contains only geometrical quantities depending on the metric (the Ricci tensor), while the right hand side (RHS) contributions from the extra scalar
coupled with the spin-2 excitations,  and possibly additional  matter fields. We concentrate all the effects of `modified gravity'  in the RHS of the equation.

We assume that the tensor $g_{\mu\nu}$ can be separated into a slowly varying background  $\bar{g}_{\mu\nu}$, and high frequency small fluctuations $h_{\mu\nu}$, which represent the GWs propagating over the smooth space-time:
\bea\label{MetricSplit}
g_{\mu\nu}&=&\bar{g}_{\mu\nu}+h_{\mu\nu}\,, \qquad \mbox{with} \qquad |h_{\mu\nu}| \ll |\bar{g}_{\mu\nu}|\,.
\eea
The background metric is characterized by the  length scale $L_B$, while the wavelength of the wave is $\lambda$, such that the derivatives of these quantities scale as
\be\label{OrdersMagnitude}
\partial \bar{g} \sim \frac{1}{L_B}\,, \qquad \partial h \sim \frac{1}{\lambda} \qquad \mbox{and} \qquad \lambda \ll L_B \,.
\ee

{ In this context, the background metric $\bar{g}_{\mu\nu}$ is  general as long as it satisfies Eq.~\eqref{MetricSplit} and Eq.~\eqref{OrdersMagnitude}. In particular, on cosmological scales it  does not necessarily describe an isotropic and homogeneous Universe. We shall make use of this fact in Section \ref{c}.}

Our system contains also a scalar field $\varphi$, whose value
can be   separated in a smooth contribution $\bar  \varphi$, plus a rapidly varying fluctuation
\be\label{ScalarSplit}
 \varphi\,=\,\bar \varphi +\varphi_{r}\,,
\ee
analogously to the metric decomposition  in Eq.~\eqref{MetricSplit}. 
We assume that  the typical wavelength of the scalar fluctuation $\varphi_r$ of order $\lambda$ and the variation length scale of $\bar{\varphi}$ of order $L_B$. As discussed
 in the Introduction, for the class of
 scalar-tensor theories we are interested in, the background scalar field acquires a non-null vacuum expectation value ({\it vev})    $\bar \varphi$, associated with the physics governing the  late-universe cosmological acceleration.  The smooth {\it vev} $\bar \varphi$ normally depends on coordinates, and its gradient allows us to define a vector
 \be \label{defvm}
 \bar v^\mu\,=\,\bar \nabla^\mu\bar \varphi\,,
 \ee
slowly varying over the  background. For cosmological applications
 such vector is typically  time-like, $\bar v^\mu\,\bar v_\mu\,<\,0$.

\smallskip
We plug the decomposition of the metric Eq.~\eqref{MetricSplit} in the field equations of Eq.~\eqref{EinsteinEq}.
  We  organize the expansion of the Ricci tensor as
\begin{equation}\label{RicciSplit}
    R_{\mu\nu} = \bar{R}_{\mu\nu} + R^{(1)}_{\mu\nu}+ R^{(2)}_{\mu\nu} + {\cal O} (h^3)\,,
\end{equation}
where $R^{(1)}_{\mu\nu}$ and $R^{(2)}_{\mu\nu}$ denote respectively linear and second order terms in $h_{\mu\nu}$ and are of the order $\sim h / \lambda^2$ and $\sim h^2 / \lambda^2$ respectively, while $\bar{R}_{\mu\nu}$ only depends on $\bar{g}_{\mu\nu}$.
The RHS of Eq.~\eqref{EinsteinEq} may depend on the metric and on the scalar and matter fields. We can apply the same reasoning as for the Ricci and expand it at second order in the metric fluctuations. In this way Eq.~\eqref{EinsteinEq} becomes
\begin{equation}\label{EisteinSplit}
    \bar{R}_{\mu\nu} + R^{(1)}_{\mu\nu}+ R^{(2)}_{\mu\nu} =  \frac{8 \pi\,G_N}{c^4}\,\left[\left(\Bar{T}_{\mu\nu}-\frac{1}{2}\Bar{g}_{\mu\nu}\bar{T} \right) + \left(T_{\mu\nu}-\frac{1}{2}g_{\mu\nu}T\right)^{(1)} + \left(T_{\mu\nu}-\frac{1}{2}g_{\mu\nu}T\right)^{(2)} \right]\,.
\end{equation}

\subsection{On the averaging procedure}

It is possible to separate between high and low-frequency modes by averaging Eq.~\eqref{EisteinSplit} over length scales large compared to $\lambda$ and small compared to $L_B$, i.e. over volumes of size $\bar{l}$ with
$$\lambda \ll \bar{l} \ll L_B\,.$$ This procedure has been known in literature as ADM-averaging scheme and is well described in \cite{Isaacson:1967zz}. The effect of such averages is to extract the slowly varying part of a quantity as its contribution is almost constant throughout the volume of integration, while the rapidly oscillating one averages out to zero. We can find the equation of motion of the high-frequency part by subtracting from Eq.~\eqref{EisteinSplit} the slow part extracted through the averaging.
In particular, in Eq.~\eqref{EisteinSplit} we have
\begin{itemize}
    \item[i)] Barred quantities that survive the averaging, e.g. $\langle \bar{R}_{\mu\nu} \rangle_{\bar{l}} = \bar{R}_{\mu\nu}$.
    \item[ii)] Linear quantities in $h_{\mu\nu}$ that do not survive the averaging, e.g. $\langle R^{(1)}_{\mu\nu} \rangle_{\bar{l}} = 0$.
    \item [iii)]Quadratic quantities in $h_{\mu\nu}$ that may survive the averaging. For instance, in a quadratic contribution as $h_{\mu\nu}h_{\rho\sigma}$, a mode with a high frequency wave-vector $\textbf{k}_1$ from $h_{\mu\nu}$ can combine with a mode with a high wave-vector $\textbf{k}_2 \approx - \textbf{k}_1$ from $h_{\rho\sigma}$ to give a low frequency wave-vector mode.
\end{itemize}
Using these prescriptions, after taking the average of Eq.~\eqref{EisteinSplit}, we can split Eq.~\eqref{EisteinSplit} into
\begin{align}
        & \Bar{R}_{\mu\nu} = \left(\Bar{T}_{\mu\nu}-\frac{1}{2}\Bar{g}_{\mu\nu}\bar{T}\right) - \left\langle R^{(2)}_{\mu\nu} \right\rangle_{\bar{l}} +\left\langle \,\left(T_{\mu\nu}-\frac{1}{2}g_{\mu\nu}T\right)^{(2)} \right\rangle_{\bar{l}}\,, \label{Background} \\
        & R^{(1)}_{\mu\nu}  = \left(T_{\mu\nu}-\frac{1}{2}g_{\mu\nu}T\right)^{(1)} - \left[ R^{(2)}_{\mu\nu}\right]^{\begin{tiny}high\end{tiny}} +  \bigg(T^{(2)}_{\mu\nu} - \frac{1}{2}(g_{\mu\nu}T)^{(2)}\bigg)^{\begin{tiny}high\end{tiny}}\,, \label{Pert}
\end{align}
where the over-script \textit{high} means  $\left[R^{(n)}_{\mu\nu}\right]^{\begin{tiny}high\end{tiny}} = R^{(n)}_{\mu\nu} - \langle R^{(n)}_{\mu\nu} \rangle_{\bar{l}}$.

\subsection{The derivative expansion}
In  light of these arguments,
  we can define  separate expansions for the quantities entering in the field equations:
 \begin{itemize}
\item[i)]
 An expansion in small fluctuations $h_{\mu\nu}$, and (independently)
 \item[ii)] an expansion in {\it derivatives} of the high-frequency modes, controlled by the  small parameter
\be
\epsilon\equiv \lambda/L_B
\ee
\end{itemize}
The mutual relationship between the two parameters is set by Eq.~\eqref{Background}.
If the background curvature is dominated by the matter contribution, Eq.~\eqref{Background} sets
\begin{equation}
    \frac{1}{L_B^2} = \frac{h^2}{\lambda^2} + \mbox{(matter contribution)} +  \mbox{(dark energy contribution)} \gg \frac{h^2}{\lambda^2}\,,
\end{equation}
meaning that the hierarchy between the amplitude of GW and ratio of GW/background characteristics wavelengths is  \cite{Maggiore:1900zz}
\be\label{hierar}
1\gg\frac{\lambda}{L_B}\gg h\, \qquad \mbox{thus} \qquad \epsilon \gg h\,.
\ee
This is not always the case:  for instance, there may be situations in which the background curvature is determined by the gravitational waves content. In this case the relationship between the two parameters would be $\epsilon = h$ \cite{Maggiore:1900zz}. However, one of our  goals is to study the propagation of GWs through the cosmic inhomogeneities where the background space-time is completely determined by the large-scale structures present in the Universe. In this case the hierarchy in Eq.~\eqref{hierar} holds.
The $\epsilon$-expansion is meant to single out  high frequency contributions to the equations: each derivative of high-frequency fluctuations  ($h_{\mu\nu}$, $\varphi_{r}$) collects  a factor $1/\epsilon$, which  allows one to clearly separate independent contributions to  the evolution equations, which are   controlled by powers of $\epsilon$.

The hierarchy Eq.~\eqref{hierar} allows us to discard the quadratic terms in $h_{\mu\nu}$ of Eq.~\eqref{Background} and Eq.~\eqref{Pert}, and expand what is left in powers of $\epsilon$ because $\epsilon \gg h\epsilon \gg h \epsilon^2 \gg \dots \gg h^2$. We have  the background metric field equation
\begin{align}
        & \Bar{R}_{\mu\nu} \,=\,\,\frac{8 \pi\,G_N}{c^4}\, \left(\Bar{T}_{\mu\nu}-\frac{1}{2}\Bar{g}_{\mu\nu}\bar{T}\right)\,,
\end{align}
which is the fully non-linear Einstein equation  solved by the slowly-varying metric $\bar g_{\mu\nu}$, turning off the high-frequency fluctuations. The linearized  equations
 for the perturbations are
\bea
\label{eqeps2}
&& \left[R_{\mu\nu}^{(1)}\right]_{1/\epsilon^{2}}\,=\,0 \,,
\\
\label{eqeps1.0}
&&  \left[R_{\mu\nu}^{(1)}\right]_{1/\epsilon}\,=\,\frac{8 \pi\,G_N}{c^4}\,\left[\left(T_{\mu\nu}-\frac12 g_{\mu\nu} T \right)^{(1)}\right]_{1/\epsilon}\,,
\\
\label{eqeps0}
&& \left[R_{\mu\nu}^{(1)}\right]_{\epsilon^0}\,= \,
\frac{8 \pi\,G_N}{c^4}\,\left[\left(T_{\mu\nu}-\frac12 g_{\mu\nu} T \right)^{(1)}\right]_{\epsilon^0}\,,
\eea
the subscript $\epsilon^p$ means that the equality holds at that order in the expansion in power of $\epsilon$.

 The limit of geometrical optics consists in focusing on  the previous equations at leading and next-to-leading order in an $\epsilon$ expansion: that is, on Eq.~\eqref{eqeps2} and Eq.~\eqref{eqeps1.0}.
For sake of clearness, we write the RHS of the equations above explicitly
\be\label{RHS1}
\left(T_{\mu\nu}-\frac12 g_{\mu\nu} T \right)^{(1)} = T^{(1)}_{\mu\nu} - \frac12 \bar{g}_{\mu\nu} T^{(1)} - \frac12 h_{\mu\nu} \bar{T} + \frac12 \bar{g}_{\mu\nu} h^{\alpha\beta}\bar{T}_{\alpha\beta}
\ee
where we defined
\begin{itemize}
  \item[-] $\bar{T}_{\mu\nu}$ is the background stress-energy tensor built with $\bar{g}_{\mu\nu}$ and $\bar{T} = \bar{g}^{\mu\nu}\bar{T}_{\mu\nu}$\,,
    \item[-] $T^{(1)}_{\mu\nu}$ is the linear in $h_{\mu\nu}$ stress-energy tensor and $T^{(1)} = \bar{g}^{\mu\nu} T^{(1)}_{\mu\nu}$.
\end{itemize}
The last two terms of Eq.~\eqref{RHS1} do not contain derivatives of the metric perturbation $h_{\mu\nu}$, so they will not contribute to Eq.~\eqref{eqeps1.0} but only to Eq.~\eqref{eqeps0}, namely
\bea
&&  \left[T^{(1)}_{\mu\nu} - \frac12 \bar{g}_{\mu\nu} T^{(1)} - \frac12 h_{\mu\nu} \bar{T} + \frac12 \bar{g}_{\mu\nu} h^{\alpha\beta}\bar{T}_{\alpha\beta}\right]_{1/\epsilon^{1}} = \left[\,T^{(1)}_{\mu\nu} - \frac12 \bar{g}_{\mu\nu} T^{(1)}\,\right]_{1/\epsilon^{1}}\,,
\eea
therefore we can rewrite Eq.~\eqref{eqeps1.0} as
\bea
\label{eqeps1}
&&  \left[R_{\mu\nu}^{(1)}\right]_{1/\epsilon^{1}}\,=\,\frac{8 \pi\,G_N}{c^4}\,\left[\,T^{(1)}_{\mu\nu} - \frac12 \bar{g}_{\mu\nu} T^{(1)}\,\right]_{1/\epsilon^{1}}\,.
\eea

\subsection{On the structure of the energy-momentum tensor $T_{\mu \nu}$ }

Usually, when describing propagation  of GWs over cosmological distances,
it is assumed that the energy-momentum tensor (EMT)  $T_{\mu \nu}$ experienced by  GWs is {\it smooth}, and does not affect the high-frequency GW evolution in the limit of geometrical optics (see e.g. \cite{Maggiore:1900zz}). Under such hypothesis, both the
RHS of Eq.~\eqref{eqeps2} and of Eq.~\eqref{eqeps1} vanish, and the EMT does
not contribute to the GW evolution  in the geometrical optics limit. But this hypothesis can be too restrictive
in scalar-tensor theories, where derivative scalar self-interactions can lead to derivative contributions to the
  EMT involving high-frequency modes. As an example,  the EMT associated with a
  cubic Galileon model
 described by the  Lagrangian density
 \be
 {\cal L}\,=\,-\frac12\,\left(\partial_\mu \varphi \partial^\mu \varphi\right)\,\Box \varphi\,
 \ee
contains second covariant derivatives of the scalar, $\nabla_\mu \nabla_\nu \varphi$ that -- through the Christoffel symbols contained
in the $\nabla_\mu \nabla_\nu$ operator  --
lead to first derivatives of the high-frequency metric fluctuations. (We will
study in more detail  this and other models in Section \ref{sec-ST}.)

In this case, as well as in
   other modified
gravity scenarios, one then expects contributions to the EMT  at next-to-leading
order in an $\epsilon$-expansion, affecting a geometrical optics description. For the rest of this Section and in Section \ref{sec_geo_optics}, therefore, we assume to have a non-vanishing contribution to the RHS of Eq.~\eqref{eqeps1},
 associated with single derivatives of the high-frequency modes. Without further specifying the theory
one considers, we can then  derive general consequences of this hypothesis for what concerns  GW  propagation in the framework
of geometrical optics\footnote{For simplicity we shall only  allow for contributions to the EMT with a single derivative acting on high frequency fields, i.e. at order $1/\epsilon$. As it will become clear in section \ref{sec_geo_optics}, the presence of two or higher derivatives in such fields could affect our hypothesis that gravitons propagate on null geodesics, and at the speed of light.}.

\smallskip
Eq.~\eqref{eqeps0} at order ${\cal O}(\epsilon^0)$, as well as other equations characterized by  higher powers of $\epsilon$,  will not be considered, since they do not contribute in the limit of geometrical optics (where $\epsilon \ll1$).
 When expanded at order ${\cal O}(\epsilon^0)$, Eq.~\eqref{eqeps0}   acquires extra contributions after performing a coordinate transformation, and  it is not
easily related to a physical observable. This fact is discussed in \cite{Isaacson:1967zz} for
 the case of GR. In next Section,  we show that the same argument holds also in the more general scenario
 we study.

\subsection{Gauge choice and evolution equations}\label{sec-gau}
We now discuss how to choose a convenient gauge choice for high frequency fluctuations, and
the resulting equations  of motion for the propagating modes.
 We shall adopt a unitary gauge that
sets to zero high frequency fluctuations associated with the scalar $\varphi_r$.  This choice turns out to be
convenient for physically motivating our  Ansatz, and  the  applications we develop  in Section  \ref{sec_geo_optics}.
Note that, within this gauge that sets to zero
the rapidly-varying scalar fluctuations,  the dark energy scalar field can still be  characterized by  slowly varying
 dynamical
 perturbations.

\subsubsection{A convenient gauge for theories with a single extra high-frequency mode}
We make the hypothesis that the scalar-tensor system one considers transforms in the standard  way under
diffeomorphism transformations.  As discussed in \cite{Isaacson:1967zz,Maggiore:1900zz},
gauge transformations can be organized in inverse powers of an expansion in  $\epsilon$. Calling $\xi^\mu$ the high-frequency part of the the coordinate transformation $x^\mu\to x^\mu+\xi^\mu$, the
high frequency GW metric fluctuations and high frequency scalar field fluctuation transform in curved space as
\bea
&&h_{\mu\nu}\to h_{\mu\nu}-\left(\bar \nabla_\mu \xi_\nu+\bar \nabla_\nu \xi_\mu \right)\,, \label{trh} \\
&&\varphi_{r}\to \varphi_{r}+\bar v^\rho\,\xi_\rho\,,\label{scaldifr}
\eea
where the bars on $\bar \nabla_\mu$ mean covariant derivatives with respect to the background curved metric $\bar g_{\mu\nu}$ and $\bar v^\mu\,=\,\bar \nabla^\mu \bar \varphi$ as defined in Eq.~\eqref{defvm}.  From now on, space-time indexes are always raised and lowered with the background metric $\bar g_{\mu\nu}$.

As the decompositions  Eq.~\eqref{MetricSplit} and Eq.~\eqref{ScalarSplit} have to hold also in the new coordinate system, we require that the vector field $\xi_\mu$ satisfy the following inequalities
\be
|\xi_\nu|\lesssim\, h\,L_B \qquad \mbox{and}\qquad | \partial_\mu \xi_\nu| \lesssim\, h  \qquad \mbox{and}\qquad |\xi_\nu|\lesssim\, \varphi_r \,L_B \label{ordersMagVectors}
\ee
in this way the background is left unchanged when performing a gauge transformation. These requirements together with Eq.~\eqref{scaldifr} imply that
\be
 \varphi_r \sim h\, \epsilon\,, \label{SWamplitude}
\ee
so, when considering an expansion in the quantity
 $\epsilon$,
 the amplitude of the rapidly varying scalar  fluctuation is one order of magnitude smaller than the amplitude of the metric fluctuations.

The high-frequency contributions to the linearized Ricci tensor and of the EMT are gauge invariant up to  next-to leading order in the $1/\epsilon$ expansion \cite{Isaacson:1967zz,Maggiore:1900zz}.
Indeed, under a gauge transformation, they  transform as (we focus on the Ricci tensor as an example)
\begin{equation}\label{GaugeTranformation}
     R^{(1)}_{\mu\nu} \rightarrow  R'^{(1)}_{\mu\nu} =R^{(1)}_{\mu\nu} - \mathcal{L}_{\xi} \Bar{R}_{\mu\nu}\,,
\end{equation}
where $\Bar{R}_{\mu\nu}$ refers to the quantity evaluated with the background metric, and $\mathcal{L}_{\xi}$ is the Lie derivative along the vector field $\xi$.
In general the Lie derivatives will be different from zero, in fact $\mathcal{L}_{\xi} \Bar{R}_{\mu\nu} \sim \epsilon^0$, therefore $ R^{(1)}_{\mu\nu} \rightarrow R'^{(1)}_{\mu\nu}$ only at $1/\epsilon^2$ and $1/\epsilon$ orders.
This means that Eq.~\eqref{eqeps2} and Eq.~\eqref{eqeps1} are gauge invariant, while Eq.~\eqref{eqeps0} is not.
The reason behind this fact is that, on a scale of distance of order $\lambda$, the space-time appears locally flat and the curvature is locally gauge invariant. As long as $\lambda \ll L_B$, perturbations do not have any long-wavelength component and this local behavior carries over to curved background,  to give a global gauge invariance \cite{Isaacson:1967zz} which is a result of our   high-frequency assumptions.
Since we will never use quantities at ${\cal O}(\epsilon^0)$,  we adopt the following prescription in our computation: we keep only terms at $1/\epsilon^2$ and $1/\epsilon$ orders, and neglect those at $\epsilon^0$ onward.

It is  convenient to work with the trace-reversed metric perturbation defined as
\be
\hat h_{\mu\nu}\,=\,h_{\mu\nu}-\frac12\,\bar{g}_{\mu\nu}\,h\,,
\ee
with $h_\mu^{\,\,\mu}\,=\,h$.

We are interested in gauge-fixing  $\hat h_{\mu\nu}$ so that it is transverse, and set to zero the high-frequency scalar field oscillation $\varphi_r= 0$. So we  focus on the transformation laws (the primes  indicate quantities after acting with a gauge transformation)
\bea
&& \varphi'_{r}=\varphi_{r}+\bar v^\rho\,\xi_\rho\,,\\
&&\bar \nabla^\nu\,\hat h_{\mu\nu}'\,=\,\bar \nabla^\nu\,\hat h_{\mu\nu}+
\bar \nabla^\rho \bar\nabla_\rho\,\xi_\mu-\bar R_{\mu\nu}\,\xi^\nu\,,
\eea
and choose $\xi^\mu$ such that
\be \label{harmonicgauge}
\bar \nabla^\rho \bar\nabla_\rho\,\xi_\mu = - \bar \nabla^\nu\,\hat h_{\mu\nu}\,,
\ee
where we neglect the contribution proportional to  $\bar R_{\mu\nu}$  in the transformation law of $\bar{\nabla}^\mu \hat{h}_{\mu\nu}$, since  it is lower at order ${\cal O}(\epsilon^0)$ in an $\epsilon$-expansion. We then find
\bea
&&\bar \nabla^\nu\,\hat h_{\mu\nu}'\,=0 \,, \qquad \mbox{and}\qquad  \varphi'_{r}=\varphi_{r}+\bar v^\rho\,\xi_\rho\,.
\eea
We can make the further transformation $x^\mu \rightarrow x^\mu + \chi^\mu$ with $\Bar{\Box} \chi^\mu = 0$ that does not spoil the harmonic gauge,
\bea
&& \varphi''_{r}=\varphi_{r}+\bar v^\rho\,(\xi_\rho + \chi_\rho)\,,\qquad \mbox{and}\qquad  \bar \nabla^\nu\,\hat h''_{\mu\nu}\,=0\,,
\eea
which allows us to set $\varphi''_r =0$ by taking
\be  \label{cond1}
v^\mu \,\chi_\mu\,=\,-\left( \varphi_{r}+\bar v^\rho\,\xi_\rho  \right)
\ee
so that we are finally left with
\bea \label{harmonic+unitary}
&& \varphi''_{r}=0\,,\qquad \mbox{and}\qquad  \bar \nabla^\nu\,\hat h''_{\mu\nu}\,=0\,,
\eea
as desired.

We point out that there is another residual gauge freedom given by $x^\mu \rightarrow x^\mu + \sigma^\mu$ with $\Bar{\Box} \sigma^\mu = \bar{v}^\mu \sigma_\mu = 0$ that leaves Eq.~\eqref{harmonic+unitary} unchanged.
We  opt to make use of this remaining freedom to fix the quantity
\be
c_\mu \equiv \bar{v}^\nu \hat h_{\mu\nu}
\ee
that will play an important role in our discussion, and
 whose transformation law follows from the one of $\hat{h}_{\mu\nu}$.  Indeed, under $x^\mu \rightarrow x^\mu + (\xi^\mu + \chi^\mu + \sigma^\mu)$ the vector $c_\mu$  transforms as
\bea
c'''_\mu &=& c_\mu - \bar{v}^\nu \left(\bar \nabla_\nu (\xi_\mu+ \chi_\mu + \sigma_\mu) + \bar \nabla_\mu (\xi_\nu + \chi_\nu + \sigma_\nu) -\bar{g}_{\mu\nu}\,\bar \nabla^\rho (\xi_\rho+ \chi_\rho + \sigma_\rho)\right) \nonumber \\
&=& c''_\mu - \bar{v}^\nu \left(\bar \nabla_\nu \sigma_\mu + \bar \nabla_\mu  \sigma_\nu -\bar{g}_{\mu\nu}\,\bar \nabla^\rho \sigma_\rho\right) \label{gaugecmu}
\eea
with $c''_\mu = \bar{v}^\nu\,\hat h''_{\mu\nu}$. From the transversality of $\hat{h}'_{\mu\nu}$ it follows that $c'_\mu$ is divergenceless at leading and next-to-leading order, in fact
\be
 \bar{\nabla}^\mu c'_\mu = \bar{\nabla}^\mu ( \bar{v}^\nu\,\hat h'_{\mu\nu} ) =  \bar{v}^\nu\, \bar{\nabla}^\mu  \,\hat h'_{\mu\nu} = 0\,,
\ee
and the same holds for $c''_\mu$ and $c'''_\mu$. This can be checked also by taking the divergence of Eq.~\eqref{gaugecmu}. However, from Eq.~\eqref{harmonicgauge} we see that $\bar{\nabla}^\mu c_\mu = - \bar{v}^\mu \bar{\Box}\xi_\mu$. Hence, from Eq.~\eqref{gaugecmu} we learn that using the residual gauge condition associated with the vector $\sigma_\mu$, we can make our desired choice for the vector $c_\mu$, fixing the last remaining gauge freedom. { Notice that whatever our initial choice for $c_\mu$  is, this vector shall have to satisfy conditions of compatibility with the metric equations of motion, implying that in general
its value will not be covariantly preserved over the GW null geodesics.
We shall concretely discuss in Section \ref{sec_geo_optics} how the $c_\mu$-evolution is controlled by the equations of motion for the high-frequency metric fields.}
In the case of null waves, both scalar and tensor, one can choose to use the last gauge freedom associated to $\sigma_\mu$ to fix the scalar longitudinal and vector modes present in the wave content. In Appendix \ref{app_gauge} we show how it is possible to do so.

\subsubsection{Compatibility among our gauge choices}
We have learned that it is in principle possible to select at the same time a unitary gauge for the rapid
scalar fluctuations, and a transverse gauge for the rapid metric fluctuations, as in Eq.~\eqref{harmonic+unitary}. We
now examine whether there are requirements to impose on the system for ensuring the compatibility of
these two gauge conditions.  As we
have seen,
the possibility of choosing  the gauge Eq.~\eqref{harmonic+unitary} is guaranteed if there exist a vector $\chi^\mu$ such that
\bea
&& \bar{\Box} \chi_\mu = 0 \,,\\
&& \bar v^\rho\,\chi_\rho = - (\varphi_{r} + \bar v^\rho\,\xi_\rho)\,. \label{vrhochirho}
\eea
In order to understand
  weather this is allowed, we take $\bar{\Box}$ on both the left and right hand sides of Eq.~\eqref{vrhochirho}, and after some manipulations, we arrive to
\bea
\bar{\Box} \varphi_{r} - \bar{\nabla}^\mu c_\mu &=& -(\chi_\rho + \xi_\rho) \bar{\Box}\bar v^\rho\,-\,2 \bar{\nabla}^\mu \bar v^\rho \, \bar{\nabla}_\mu ( \chi_\rho + \xi_\rho)\,. \label{consistencyEquation}
\eea
The orders of magnitude of the vector fields on the RHS of the previous equation is restricted by Eq.~\eqref{ordersMagVectors} so that
\bea
&& -(\chi_\rho + \xi_\rho) \bar{\Box}\bar v^\rho \lesssim h L_B^{-2} \\
&& -\,2 \bar{\nabla}^\mu \bar v^\rho \, \bar{\nabla}_\mu ( \chi_\rho + \xi_\rho) \lesssim h L_B^{-2}
\eea
so the RHS of Eq.~\eqref{consistencyEquation}  does not have contributions at leading ($\epsilon^{-2}$) and next to leading ($\epsilon^{-1}$) and therefore the compatibility condition is given by
\be
[\bar{\Box} \varphi_{r} - \bar{\nabla}^\mu c_\mu]_{1/ \epsilon^{2}, 1 / \epsilon^1} = 0\,. \label{CompatibilityCondition}
\ee

The compatibility condition Eq.~\eqref{CompatibilityCondition} can be also stated in a more physically transparent way in terms
of a linearized equation for the scalar fluctuation
\be
[(\Box \varphi)^{(1)}]_{1/ \epsilon^{2}, 1 / \epsilon^1} = 0\,, \label{CompatibilityCondition1}
\ee
because, up to sub-leading contributions that we neglect, we have
\bea
 (\Box \varphi)^{(1)}= \bar{g}^{\mu\nu}(\varphi_{\mu\nu})^{(1)} &=& \bar{g}^{\mu\nu}(\bar{\nabla}_\mu \bar{\nabla}_\nu \varphi_r - \frac{\bar{\varphi}^\sigma}{2} (\bar{\nabla}_\mu h_{\sigma \nu}+\bar{\nabla}_\nu h_{\sigma \mu}-\bar{\nabla}_\sigma h_{\mu \nu})) = \nonumber\\
&=& \bar{\Box}\varphi_r - v^\sigma \bar{\nabla}^\mu \hat{h}_{\sigma \mu} =\bar{\Box}\varphi_r -  \bar{\nabla}^\mu c_\mu \,.
\eea

\smallskip

We conclude that it is possible to choose simultaneously the gauge choices in Eq.~\eqref{harmonic+unitary} {\bf only if}
Eq.~\eqref{CompatibilityCondition} -- or equivalently  Eq.~\eqref{CompatibilityCondition1} --  is satisfied. This condition amounts to impose that
 rapidly moving linearized scalar fluctuations  propagate freely -- and at light speed -- around the cosmological background under
 consideration.  This fact typically imposes constraints on the scalar-tensor theories  we can consider with our formalism,
 and on their background
 solutions.
  In Appendix \ref{app-scal-vel} we discuss in the concrete  setting of a general scalar-tensor system  such constraint conditions.
  For the rest of our work, we then assume that the scalar-tensor
  systems and background configurations around which our rapidly moving fluctuations propagate are such that  Eq.~\eqref{CompatibilityCondition1}  is satisfied.

\subsubsection{Summary of the gauge conditions and of the relevant equations of motion}\label{summary_gauge}

To briefly summarize, our system satisfies a unitary gauge for the high-frequency perturbations of the scalar field
\be\label{gu1a}
\varphi_{r}\,=\,0
\ee
and a transverse (but not traceless) gauge for the fast moving part of the metric fluctuations:
\be \label{trgc}
\bar \nabla^\mu\,\hat h_{\mu \nu}\,=\,0
\ee
As a final condition, that exhausts our gauge freedom,  we  impose 
\be\label{gd1a}
\bar v^\mu\, \hat h_{\mu\nu}\,=\,c_\nu \qquad \mbox{with} \qquad \bar{\nabla}^\mu c_\mu = 0 \,,
\ee
where the initial value of $c_\mu$ can be chosen arbitrarily and its evolution along GW geodesics will be controlled by the condition of compatibility with the equations of motion for the high-frequency metric fields.

The number of propagating degrees of freedom (dofs) can be counted as follows. We start from 11 dofs: 1 in the scalar, 10 in the symmetric tensor $h_{\mu\nu}$. We have 1 gauge condition Eq.~\eqref{gu1a}, 4 gauge conditions Eq.~\eqref{trgc}, 3 independent gauge conditions Eq.~\eqref{gd1a} ($c_\nu$ is divergenceless). In total, we generically have $11-1-4-3\,=\,3$ independent propagating dofs, as expected in a scalar-tensor theory of gravity.

\smallskip

We shall use this gauge fixing and rewrite the relevant equations of motion Eqs.~\eqref{eqeps2}--\eqref{eqeps1}, they become\footnote{ The linearized Ricci tensor, neglecting its contribution at $\epsilon^0$ order, in this gauge reads
\be
R_{\mu\nu}^{(1)}\,\,
=-\frac12\left(
\,\,\bar \Box \,\hat h_{\mu\nu}-\frac12\,\bar{g}_{\mu \nu}\,\bar \Box \hat h\right)\,.
\ee
}

\bea
&& \left[\,\bar \Box \,\hat h_{\mu\nu}\,\right]_{1/\epsilon^{2}}\,=\,0 \,,\label{eqeps2b}\\
&&  \left[\,\bar \Box \,\hat h_{\mu\nu}\,\right]_{1/\epsilon^{1}}\,=\,-\frac{16 \pi\,G_N}{c^4}\,\left[\,T^{(1)}_{\mu\nu}\,\right]_{1/\epsilon^{1}}\,.\label{eqeps1b} 
\eea

\bigskip

These are the evolution equations for the high frequency modes in our setting. Given
that the only high frequency fields are the metric fluctuations  $\hat h_{\mu\nu}$, the
 EMT appearing in Eq.~\eqref{eqeps1b} is built in terms of single derivatives of these quantities, that select
 the $1/\epsilon$ contributions to the equations. (The concrete structure  of $T^{(1)}_{\mu\nu}$ depends
 of course on the scalar-tensor theory one considers.)

Since we work in an unitary gauge for the high frequency scalar fluctuations, the metric fluctuations $\hat h_{\mu\nu}$  contain the extra  degree of freedom which we expect to generically propagate in our system. Consequently, it {\it does not} correspond to a pure spin-2, transverse-traceless fluctuations of GR.

\smallskip
The compatibility of equations of motion and gauge conditions leads to an additional relation.
We apply a covariant derivative $\bar \nabla^\mu$ on both sides of Eq.~\eqref{eqeps1b}, and we select the $1/\epsilon^2$ contributions. In the LHS we can interchange the order of covariant derivatives (the error being of order $1/\epsilon$ hence negligible) and find $\bar \Box \,\bar \nabla^\mu \,\hat h_{\mu\nu}\,=\,0$ for the transverse gauge condition Eq.~\eqref{trgc}. In the RHS, we obtain a condition corresponding to the conservation of the energy momentum tensor:
\be \label{consEMT}
\left[\,\bar \nabla^\mu\,T^{(1)}_{\mu \nu}\,\right]_{1/\epsilon^2} = 0
\ee

As typical in scalar-tensor systems, we expect such  condition to be equivalent to the conditions
provided by the high-frequency contributions to the scalar field equation in unitary gauge.
Therefore, the three dynamical equations, at separate orders in $\epsilon$ expansion, are
\bea
\left[\, \bar \nabla^\rho \bar \nabla_\rho\,\hat h_{\mu\nu}\,\right]_{1/\epsilon^2}\,&=&\,0\,,\label{coep2}
\\
\left[\, \bar \nabla^\rho \bar \nabla_\rho\,\hat h_{\mu\nu} \,\right]_{1/\epsilon}\,&=&\,-\frac{16 \pi\,G_N}{c^4}\, \left[\,T^{(1)}_{\mu\nu}\,\right]_{1/\epsilon}\label{coep1}
\\
\left[\,\bar \nabla^\mu\,T^{(1)}_{\mu\nu} \,\right]_{{1}/{\epsilon^2}}\,&=&\,0\,.\label{coepc}
\eea
We now analyse all these equations in the framework of geometrical optics. Our considerations
will be sufficiently general to be applied to any scalar-tensor system with a high-frequency contribution
to its EMT, as described above.


\section{Geometrical optics with a single  extra degree of freedom}\label{sec_geo_optics}


The analysis of Section \ref{sec-gen} sets the stage to consistently
treat the evolution of high frequency fluctuations in scalar-tensor settings,
without  specific hypothesis on the scalar-tensor theory and the background one considers.
 The effects of `modified gravity' in the  high-frequency
field dynamics are enclosed in contributions to the EMT, built in terms of high-frequency fluctuations\footnote{In fact, we made
the choice to  include all the effects of modified gravity on the RHS of Einstein equations, while on the LHS we keep the traditional
Einstein tensor.}. The scope of this Section is to
apply this formalism
to a framework of geometrical optics. We adopt a specific Ansatz that allows us to follow the evolution of the GW amplitude, phase, and GW polarization tensor, as a function of certain combinations of the EMT.
 We determine modified evolution equations for these quantities with respect to GR, and discuss the potential physical
consequences.

We  find that, for scalar-tensor theories of the kind we are interested in, GWs follow null geodesics,
but the
 evolution equation for the GW amplitude  is modified
with respect to GR, leading to the phenomenon of non-conservation of the effective Planck mass. This
is a well known property found in various works that study  GW propagation in FRW space-times:
our results generalize those findings to more general settings.

Interestingly, we also find that the polarization tensor  is generally {\it not parallel transported}
along the GW null geodesics.  Possible implications of this fact for the evolution
of GWs will then be discussed analyzing specific examples in the next Sections.

\subsection{Ansatz and evolution equations}

We start defining the geometrical optic Ansatz we adopt.
The metric fluctuations can be  expressed in terms of an amplitude and a phase as \cite{Isaacson:1967zz} 
\bea\label{ansgeo}
\hat h_{\mu\nu}&\equiv&\left[{\cal A}_{\mu\nu}+\epsilon \cdots 
+\dots\right]
\,e^{i \theta/\epsilon}\,,
\eea
where all the parts with dots are higher order in the $\epsilon$ expansion described in Section \ref{sec-gen}, and will be neglected in the
geometrical optics approximations we adopted here. The tensor ${\cal A}_{\mu\nu}$ contains both spin-2 and spin-0 contributions. Since they are part of the same metric tensor $\hat h_{\mu\nu}$ in the unitary gauge we adopt,
 we make the hypothesis that they have the same phase, hence they follow the same geodesics.
 This hypothesis is backed up by the fact that {in our approach we   restrict ourselves only to systems
 in which    scalar waves and gravitational waves travel at the same speed.}
   (In Section \ref{sec_sep20} we also investigate at what extent,
 within this hypothesis,  the evolution of tensor and scalar modes can be studied independently.)

The tensorial quantity  ${\cal A}_{\mu \nu}$ is decomposed into an amplitude  ${\cal A}$ and a tensor polarization ${\bf e}_{\mu\nu}$ with unit norm
 \bea \label{fdecA}
 {\cal A}_{\mu \nu}&\equiv& {\cal A}\,{{\bf e}}_{\mu\nu}\,,
 \eea
 with
\begin{align}\label{PropAE}
&{\cal A}= \sqrt{ {\cal A}_{\mu \nu}\, {\cal A}^{\mu \nu}}\,, \qquad {\bf e}_{\mu\nu}= \frac{ {\cal A}_{\mu \nu}}{{\cal A}}\,, \qquad {\bf e}_{\mu\nu}\,{{\bf e}^{\mu\nu}}\,=\,1\,.
\end{align}

In the unitary gauge adopted, the high frequency contribution to the EMT is built with derivatives of the metric fluctuations, therefore we can assume it
  has the same phase as the metric fluctuation (we will give an explicit example in Section \ref{sec-ST}).  We then adopt the following Ansatz for the  EMT at order $1/\epsilon$ (the overall coefficient is chosen such to simplify the following  equations):
\be
 \label{defHS}
\left[\,T^{(1)}_{\mu \nu}\,\right]_{1/\epsilon}\,\equiv\, \frac{i}{\epsilon}\,\frac{c^4}{16\,\pi\,G_N}\,t_{\mu \nu}\,e^{i\, \theta/\epsilon}\,.
\ee

\smallskip

As explained in the previous Sections, the contributions that
scale as $1/\epsilon$ collect the terms proportional to  first derivatives of the high-frequency fields.
 The tensor $t_{\mu \nu}$, consequently, is proportional to the coefficients
 of these terms. Importantly, notice that we collect in the coefficient of the RHS of Eq.~\eqref{defHS} the quantity $G_{N}\,=\,\bar M_{\rm Pl}^{-2}$ with $\bar M_{\rm Pl}$ a
 reference, constant mass scale. The effective Planck scale experienced by the GWs can on the other hand vary along the GW trajectory - a phenomenon found in various
 examples of scalar-tensor theories. We make the hypothesis that the tensor  $t_{\mu\nu}$ includes contributions containing this possibility, if realized in the
  scalar-tensor theories under consideration.
   \smallskip

 So far, we learned that the unitary gauge condition
we adopted allow us to make the reasonable assumption that all the quantities we are dealing with
in geometrical optics share the same phase. We will focus on this simplifying assumption throughout all this work.
We also write
\be
\bar v^\mu\,\hat h_{\mu\nu}\,=\,c_\nu\,=\,\left[{\cal C}_\nu + \epsilon...\right]\,e^{i \theta/\epsilon}\,.
\ee
Finally, the wave-vector of the propagating GW is defined as\footnote{The minus sign in the definition of $k_\mu$ is because we use the comoving distance $\chi$ as affine parameter. The opposite convention is used in \cite{Laguna:2009re} because they parametrize the GW's geodesics with $\lambda = - \chi$.}
\bea
k_\mu&\equiv&-\partial_\mu\,\theta\,,
\eea
and identifies the GW geodesics as we now  show.
 The gauge conditions Eqs.~\eqref{trgc}--\eqref{gd1a}, at leading and next-to-leading order, 
 are given by
\bea
k^\mu {\bf e}_{\mu \nu}&=&0\,, \label{polttra}
\\
\bar v^\mu\,\A_{\mu \nu}&=&{\cal C}_\nu\,, \label{gaugeCb}
\\
k^\mu {\cal C}_\mu&=&0\,,  \label{citra}
\eea
so that the polarization tensor ${\bf e}_{\mu \nu}$ and the   vector ${\cal C}_\mu$ are transverse to the GW propagation.
We can now plug the Anstaz~\eqref{ansgeo} and~\eqref{defHS} in the equations of motion Eqs.~\eqref{coep2}, \eqref{coep1} and \eqref{coepc}.

From Eq.~\eqref{coep2} we obtain
\be
k_\mu k_\nu\,\bar g^{\mu\nu}=0 \label{nullco}\,
\ee
which means that {\it the GWs wave-vector is a null vector of the slowly varying background $\bar{g}_{\mu\nu}$}.  Moreover, using the definition of $k_\mu$, one can show   that it also satisfies
\be
k^\mu \bar \nabla_\mu k_\nu\,=0 \label{partrk}\,.
\ee
Thus, {\it the GWs wave-vector is a geodesic vector}.
These two results allow us to conclude that, even in the more general case where $T^{(1)}_{\mu\nu}$ contains contributions at $1 /\epsilon$, GWs travel along null geodesics of the background manifold as in the case of GR with a smooth matter content.
The hypothesis that $T^{(1)}_{\mu\nu}$ only contribute starting at order $1/\epsilon$, and not $1/\epsilon^2$, is crucial to ensure this property. Our assumption is motivated by the fact that the multi-messenger
event  GW170817 established that, in excellent approximation, GWs
travel at the speed of light.

From Eq.~\eqref{coepc} we obtain
\be\label{ttransv}
k^\mu t_{\mu\nu} =0 \,,
\ee
which states that the $1/\epsilon$ order of $T^{(1)}_{\mu\nu}$ is transverse. This property is a direct consequence of the fact that $t_{\mu\nu}$ is built by the high-frequency metric fluctuations,  which are transverse.

From Eq.~\eqref{coep1} we obtain
\be \label{eomG1A}
2 k_\rho\,\bar \nabla^\rho\left( {\cal A}_{\mu\nu}\right) +{\cal A}_{\mu\nu} \,\,\left(\bar \nabla^\rho k_\rho\right)\,=\, t_{\mu\nu}\,,
\ee
which can be separated
into an equation for the amplitude ${\cal A}$
\be
\boxed{
\bar \nabla_{\rho}\left( k^\rho \,{\cal A}^2\right)\,=\,{\cal A} \,\,t_{\mu\nu}\,{\bf e}^{\mu\nu}
}
\label{res1A}
\ee
\noindent{and} equation for the polarization ${\bf  e_{\mu \nu}}$:
\be
\boxed{
k_\rho\,\bar \nabla^\rho\,{{\bf e}}_{\mu \nu}\,=\,\frac{1}{2\,{\cal A}}\,\left[
t_{\mu\nu}-{{\bf e}}_{\mu\nu} \left( t_{\rho\sigma} {{\bf e}}^{\rho \sigma} \right)
\right]\label{res2A}
}
\ee
\noindent{by}  multiplying Eq.~\eqref{eomG1A} by ${{\bf e}}^{\mu\nu}$ and using Eq.~\eqref{PropAE}.
Eqs.~\eqref{res1A}--\eqref{res2A} control the evolution of the metric's fluctuations in the limit of geometrical optics. The  difference with respect to standard geometrical optics in GR lies in their potentially non-vanishing right-hand-sides, controlled by the high-frequency contribution to the EMT $t_{\mu \nu}$, as defined in Eq.~\eqref{defHS}.

 We propose the following interpretation of these results:
   \begin{itemize}
 \item
 Eq.~\eqref{res1A} controls the evolution of the GW amplitude of the high-frequency GWs as they travel
 along a null geodesics.
 A non-vanishing     RHS is associated  with the non-conservation of the current ${\cal A}^2\,k^\mu$.

 In scalar-tensor
 theories where  the graviton number is conserved, the physical         interpretation of this fact is related with the non-conservation
 of effective Planck mass,  as experienced by travelling GWs. Indeed, we expect its RHS  to be proportional to the rate of change
 of the effective Planck mass, a phenomenon already discussed in related settings, see e.g. \cite{Belgacem:2018lbp,Belgacem:2019pkk}.
 Eq.~\eqref{res1A} provides
  a `covariant' version of this result in the gauge we are adopting.
 This fact has interesting phenomenological consequences being related with observables associated with
     the GW luminosity distance,
  see e.g. \cite{Deffayet:2007kf,Saltas:2014dha,Nishizawa:2017nef,Belgacem:2017ihm,Belgacem:2018lbp,Ezquiaga:2018btd,Belgacem:2019pkk}.    In the next Sections, we shall make this connection
 more explicit when analyzing specific models.

  There can also be theories where the graviton number is not conserved (as theories with extra dimensions, see e.g. \cite{Pardo:2018ipy,Visinelli:2017bny,Abbott:2018lct,Calcagni:2019kzo,Calcagni:2019ngc} for
  recent studies in the framework of GW cosmology) which
   might be described in terms of effective scalar-tensor systems; in this case, a non-vanishing RHS of Eq.~\eqref{res1A} controls the
   amount of graviton number non-conservation.
     \item
  Eq.~\eqref{res2A} controls the evolution of the polarization tensor. If its RHS is non-vanishing, it
   implies  that   this quantity  is {\it not}   parallel  transported along  GW null geodesics.
  This is a novel effect, which again depends on the structure of the high-frequency  EMT $t_{\mu\nu}$. Similar phenomena have been noticed in standard gravity working   beyond a geometrical optics approximation in GR -- see e.g. \cite{Misner:1974qy,Dolan:2018ydp,Dolan:2018nzc,Anile:1976gq,Harte:2018wni,Cusin:2019rmt} -- and might
  be related  with lensing of GWs. Interestingly,  we find that these effects can be present also {\it at leading order} in geometrical optics for certain scalar-tensor theories. We shall return
  in the next Sections on the physical implications of this result, when specializing to particular models.
 \end{itemize}
Solving the  independent Eqs.~\eqref{res1A}--\eqref{res2A} allows one to determine the most general solution for ${\cal A}$ and for the transverse, normalized tensor
${\bf e}_{\mu\nu}$.
Moreover,  contracting Eq.~\eqref{eomG1A} with $\bar{v}^\mu$, and using the gauge conditions  Eq.~\eqref{gaugeCb} and Eq.~\eqref{citra}, we find the following condition on the transverse vector
\be \label{eqfCmu}
\boxed{
2 k_\rho \bar \nabla^\rho {\cal C}_\nu\ = \A_{\mu\nu} \left[ \,2 \,k^\rho\, \bar \nabla_\rho \bar v^\mu\, - \bar v^\mu \, \bar \nabla_\rho k^\rho \right] + \,\bar v^\mu\, t_{\mu\nu} }
\ee


\noindent{This} equation informs us that the vector ${\cal C}^\mu$ is in general {\it not} parallel transported
along GW null geodesics (unless its RHS vanish).
The RHS of Eq.~\eqref{eqfCmu} is made of two contributions:  the first only depends on the background gradient $\bar v^\mu$ of the scalar field; the second depends on $t_{\mu\nu}$, hence its structure depends on the specific modified gravity model.

\subsection{On the decomposition of the energy-momentum tensor $t_{\mu\nu}$}\label{subsec_dec_t}

Even without relying on any specific scalar-tensor  theory, we now have sufficient ingredients to pin down the general structure of the EMT for the systems we are interested in.
In fact, it has to be: symmetric, transverse (in the sense that it has to satisfy Eq.~\eqref{ttransv}) and built with suitable combinations of the vectors ($\bar v^\mu$, $k^\mu$, ${\cal C}^\mu$) and tensors (${{\bf e}}_{\mu\nu}$, $\bar g_{\mu\nu}$) as it is formed by metric fluctuations. Hence, it is necessarily of the form

\be \label{decotmunu}
t_{\mu\nu}\,\equiv\,
\,\left\{
\tau^{(A)}\,{{\A}}_{\mu \nu}+\tau^{(B)} \left(k_\mu {\cal C}_\nu +k_\nu {\cal C}_\mu \right)+
\tau^{(C)} \left[k_\mu \bar v_\nu+ k_\nu \bar v_\mu-\bar g_{\mu\nu} \left(k_\rho \bar v^\rho\right)
\right] +\tau^{(D)} \,k_\mu k_\nu
\right\}\,,
\ee
since the four contributions proportional to the parameters $\tau^{(i)}$ are the only tensors with the desired properties. The last
contribution proportional to $\tau^{(D)}$ will be associated with the self-induced GW energy momentum tensor at second order, a quantity that we shall
not meet any more in this work, hence we set from now on $\tau^{(D)}=0$.
In Eq.~\eqref{decotmunu}, $\tau^{(A)}$, $\tau^{(B)}$, $\tau^{(C)}$  are scalar functions of the space-time coordinates, that depend on the specific theory under consideration,  on the smooth profiles of the quantities $\bar \varphi$ and $\bar g_{\mu\nu}$, and (possibly) on the trace of the polarization tensor and on the GW momentum $k^\mu$. In  Section \ref{sec-ST}, when discussing explicit examples, we shall find realizations for each of  the three contributions to Eq.~\eqref{decotmunu}.

Using the explicit decomposition Eq.~\eqref{decotmunu}, we can re-express the evolution equations for amplitude and polarization:

 \bea
        \bar \nabla_{\rho}\left( k^\rho \,{\cal A}^2\right)&=& \,{\cal A}^2 \tau^{(A)} - \tau^{(C)}\,\left( k \cdot \bar v\right) \A^{\mu\nu} \bar g_{\mu\nu} \label{evams}
        \\
      2\,k_\rho\,\bar \nabla^\rho\,{{\bf e}}_{\mu \nu}&=&\, \tau^{(B)} \left(k_\mu {\cal C}_\nu +k_\nu {\cal C}_\mu \right)+ \tau^{(C)} \left[k_\mu \bar v_\nu+ k_\nu \bar v_\mu-\left(\bar g_{\mu\nu} - {\bf e}_{\mu\nu}{\bf e}\right)\left(k_\rho \bar v^\rho\right) \right]  \label{evpols}
     \eea
      with  the trace $${{\bf e}}\,=\,\bar g^{\mu\nu}\,{{\bf e}}_{\mu\nu}\,.$$  In terms of this EMT decomposition,
the evolution Eq.~\eqref{eqfCmu} for the vector ${\cal C}_\mu\,=\, \bar v^\nu\,{{\cal A}}_{\mu \nu}$ reads
\be \label{evCnu2}
2 k_\rho \bar \nabla^\rho {\cal C}_\mu =  \A_{\mu\nu} \left[ \,2 \,k^\rho\, \bar \nabla_\rho \bar v^\mu\, - \bar v^\mu \, \bar \nabla_\rho k^\rho \right]  +  {\cal C}_\mu \left[ \tau^{(A)} +  \tau^{(B)} \,(v\cdot k )\,\right] + k_\mu \left[  \tau^{(B)} (\bar v \cdot {\cal C}) + \bar v^2 \tau^{(C)} \right]
\ee


We then learn that the evolution equations for GW amplitude and polarization  depend on different contributions to the EMT.  The equation of evolution for the amplitude ${\cal A}$ can be
further decomposed in a tensor and scalar part, as we are going to discuss in what next.

\subsection{Separating the spin-2 and spin-0 degrees of freedom}\label{sec_sep20}

So far, we made use of a polarization tensor $\A_{\mu\nu}$ that satisfy the gauge conditions Eq.~\eqref{harmonic+unitary}.
As scalar and tensor modes propagate at the same speed we assigned to both waves the same phase and,  in the geometric optics limit, we showed that the associated wave vector is a null vector Eq.~\eqref{nullco}. This means that they are {\it null waves}, therefore they can be studied in the framework of null-tetrads as shown in the classic work of \cite{Eardley:1974nw} (see also the textbook \cite{Will:2018bme}).

In particular, in this formalism, there are {\it six independent polarization modes} which are dynamical or not according to the theory taken into account. Such modes correspond to two spin-2 modes, two scalar modes and two vector modes.
In Appendix \ref{app_A} we show how to build these independent polarizations and that in the theory Eq.~\eqref{LagTot} only three of them are dynamical.  This is compatible with the fact that in the theory we are considering vectors are not dynamical and the scalar field is massless.
Moreover, since $t_{\mu\nu}$ contains the rapid fluctuations of the fields as well, it will change after the removal of non-dynamical modes.  In Appendix \ref{app_A} we give a thorough explanation of how to treat these spurious modes.  In particular we show that they decouple from the physical ones and how to build the first order stress energy tensor in terms only of the three physical polarizations. In this Section, thus, we consider only the spin-2 and spin-0 transverse modes and we call $t'_{\mu\nu}$ the first order stress energy tensor after the removal of the spurious modes.

The three propagating modes present in the theory we are considering are the  spin-2  transverse and traceless modes, represented by the polarization tensors $({{\bf e}}^{+}_{\mu\nu},\,{{\bf e}}^{\times}_{\mu\nu})$ , and the spin-0 transverse scalar mode described by ${{\bf e}}^{S}_{\mu\nu}$.
Such polarization tensors are all normalized to $1$, transverse as they have to satisfy Eq.~\eqref{polttra} and orthogonal between each other. Their traces are
\bea
\bar g^{\mu\nu} \,{{\bf e}}^{+}_{\mu\nu}&=&0\,,
\\
\bar g^{\mu\nu} \,{{\bf e}}^{\times}_{\mu\nu}&=&0\,,
\\
\bar g^{\mu\nu} \,{{\bf e}}^{S}_{\mu\nu}&=& \sqrt{2}\,.
\eea
We also assume that the two spin-2 modes have the same amplitude, $\A^T$, as it should be in a non-parity violating theory. The more general case is considered in Appendix \ref{app_B1}. Hence, we decompose the tensor  ${\cal A}_{\mu\nu}$  on the three polarizations as
\be\label{decAa}
{\cal A}_{\mu\nu}\,=\,{\cal A}\,{{\bf e}}_{\mu\nu}\,=\,{{\cal A}^T}\,\left( {{\bf e}}_{\mu\nu}^{+}+{{\bf e}}_{\mu\nu}^{\times} \right)+{\cal A}^S\,{{\bf e}}_{\mu\nu}^{S}
\ee
where
\be
{\cal A}\,=\,\sqrt{
2\,\left({\cal A}^T\right)^2+\left({\cal A}^S\right)^2}\,.
\ee
The trace of Eq.~\eqref{decAa} gives
\be
\label{tracescalar}
{{\bf e}}\,\coloneqq \bar g^{\mu\nu}{{\bf e}}_{\mu\nu} \,= \,\sqrt{2}\,{\cal A}^S/{\cal A}\,,
\ee
which states that {\it the trace of the polarization is proportional to the amplitude of the spin-0 mode}, which we have hidden in the GW's polarization content because of our gauge choice.
The first order stress-energy tensor without the gauge modes reads
\be
t'_{\mu\nu}=\tau^{(A)}  \A_{\mu\nu} - \sqrt{2} \tau^{(C)}\left(k_\rho \bar v^\rho\right)  {\bf e}^S_{\mu\nu} \,
\ee
where $\tau^{(A)} $ and $\tau^{(C)}$ are the ones defined in Eq.~\eqref{decotmunu}. This expression is derived in Appendix  \ref{app_B1}. By plugging Eq.~\eqref{decAa} into Eq.~\eqref{eomG1A} and considering $t'_{\mu\nu}$ as first-order stress energy tensor, we get
\be \label{eomG1Ba}
2 k_\rho\,\bar \nabla^\rho\left(
{\cal A}^T\,{{\bf e}}_{\mu\nu}^{+}+{\cal A}^T\,{{\bf e}}_{\mu\nu}^{\times}+{\cal A}^S\,{{\bf e}}_{\mu\nu}^{S}
\right) +\left(
{\cal A}^T\,{{\bf e}}_{\mu\nu}^{+}+{\cal A}^T\,{{\bf e}}_{\mu\nu}^{\times}+{\cal A}^S\,{{\bf e}}_{\mu\nu}^{S}
\right) \,\,\left(\bar \nabla^\rho k_\rho\right)\,=\, t'_{\mu\nu}\,.
\ee
In Appendix \ref{app_B1}, starting from Eq.~\eqref{eomG1Ba}, we show how to separate the evolution of the scalar and tensor amplitudes, namely
\bea
2 k^\rho \bar \nabla_\rho {\cal A}^S + {\cal A}^S \bar \nabla_\rho k^\rho &=& t'^{\mu\nu} {{\bf e}}_{\mu\nu}^{S}  = \tau^{(A)} \A^S - \sqrt{2} (\bar v^\rho k_{\rho})\tau^{(C)} \,,  \label{eqAS2}\\
2 k^\rho \bar \nabla_\rho {\cal A}^T + {\cal A}^T \bar \nabla_\rho k^\rho &=& \frac{ t'^{\mu\nu}}{2} \left({{\bf e}}_{\mu\nu}^{+} + {{\bf e}}_{\mu\nu}^{\times} \right) = \tau^{(A)} {\cal A}^T \label{eqAT2}
\eea
So we obtain {\it two equations for the amplitudes of the spin-0 and spin-2 modes generically coupled via the evolution of $ t'_{\mu\nu}$}. We notice that  the equation of $\A^T$ admits as solution $\A^T = 0$ regardless of the form of the first order stress-energy tensor. On the contrary, the equation for $\A^S$ has a source term proportional to $\tau^{(C)}$. Thus, $\A^S = 0$ is a solution only if $\tau^{(C)}$ is equal to zero or proportional to $\A^S$ itself.  We will see that for the Lagrangian~\eqref{LagTot} braiding can play the role of source, while conformal couplings generates $\tau^{(C)} \propto \A^S$. In the case $\A^S \neq 0$ the previous two equations are equivalent to
\bea
\bar \nabla_{\rho} \,\left( k^\rho\,\left({\cal A}^S \right)^2\right)&=&  \tau^{(A)} \left(\A^S\right)^2 - \sqrt{2} (\bar v^\rho k_{\rho})\A^S \tau^{(C)}\,,\label{seceqS} \\
 \bar \nabla_{\rho}\left( k^\rho \,\left({\cal A}^T \right)^2\right)&=&\tau^{(A)} \left({\cal A}^T \right)^2 \,.  \label{evetpa}
\eea
\noindent{Hence} the EMT contributions to the evolution of the GW amplitude   is in principle  different for spin-2 and spin-0 modes. In both cases, {\it we can have  non-conservation of the spin-2 and spin-0 currents $ \left( {\cal A}^T
 k^\mu \right)$ and $ \left( {\cal A}^S
 k^\mu \right)$.}

The solutions of  Eq.~\eqref{evetpa} and Eq.~\eqref{seceqS}
  depend on the initial conditions. For example, one can expect
that the initial amplitude of scalar components ${\cal A}^S$ to be smaller than the tensor one ${\cal A}^T$,
thanks to screening mechanisms that reduce the size of  scalar excitation  around the source. For an explicit example where
this happens, see \cite{Dar:2018dra}. This is
an interesting topic  to explore, that goes beyond the scope of this work.

After the removal of the spurious modes, as shown in Appendix \ref{app_B1}, Eq.~\eqref{res2A} for the polarization tensor ${\bf e}_{\mu\nu}$ becomes
\be
k_\rho\,\bar \nabla^\rho\,{{\bf e}}_{\mu \nu}\,=\,\frac{1}{2\,{\cal A}}\,\left[
t'_{\mu\nu}-{{\bf e}}_{\mu\nu} \left( t'_{\rho\sigma} {{\bf e}}^{\rho \sigma} \right)\right] = \,\frac{\left(k_\rho \bar v^\rho\right) \tau^{(C)}}{\sqrt{2}\A} \left[\frac{\A^S}{\A^2} \A_{\mu\nu} - {\bf e}^S_{\mu\nu}\right]\,,\label{evPolNolongitudinal}
\ee
therefore ${\bf e}_{\mu\nu}$ is parallel transported or not according to whether the RHS of Eq.~\eqref{evPolNolongitudinal} vanishes or not. We will show that in the theory described by~\eqref{LagTot}, the RHS of this equation is different from zero, so the total polarization is not parallel transported.

\section{An explicit scalar-tensor realization}\label{sec-ST}

The geometrical optics framework developed in the previous Sections provides us with a general and flexible
formalism that allows us to consistently treat the evolution of high-frequency GWs, and to acquire a transparent
physical understanding of the results. Our formulas depend on the EMT tensor $t_{\mu\nu}$ which is controlled by
 the first derivatives of the high-frequency fields. To investigate our formalism in a concrete setting, we focus
in this Section on a {specific case of the}
scalar-tensor theory called {\it kinetic gravity braiding} \cite{Deffayet:2010qz}, {supplemented with}
  a non-minimal coupling with  the Ricci scalar. Its Lagrangian density can be expressed as
\bea \label{LagTot}
{\cal L}^{\text{(tot)}}&=&{\cal L}^{(F)}+{\cal L}^{(G)}+{\cal L}^{(K)}\,,
\eea
with
\bea
{\cal L}^{(F)}&=&F(\varphi)\,R \label{Flag}\,,
\\
{\cal L}^{(G)}&=&G(\varphi,\,X) \,\Box \varphi \label{Glag}\,,
\\
{\cal L}^{(K)}&=&K(\varphi, X)\label{Klag}\,,
\eea
where $X\,=\,-(\partial_\mu \varphi \partial^\mu \varphi)/2$. This is the most general subset of the Horndeski
 theory  that leads to GW propagation at the speed of light, consistent with observational findings associated
 with the GW170817 event \cite{Abbott_2017}. {In order to apply the formalism
 we developed in the previous Sections, we   focus on a set-up where we assume that the rapidly moving scalar
 fluctuations move with the speed of light. We further discuss this condition in  Appendix \ref{app-scal-vel}. }

  The Lagrangian ${\cal L}^{(F)}$
  is related with the classic Brans-Dicke theory \cite{Brans:1961sx}, and is common to many scalar-tensor theories that allow for a kinetic mixing between
  scalar and tensor fields in the Jordan frame. (Demixing can be obtained
  through a conformal transformation, at the price of introducing non-minimal couplings
  between scalar and matter  degrees of freedom, that we prefer to avoid.)  The Lagrangian ${\cal L}^{(G)}$
   is a generalized cubic Galileon  \cite{Nicolis:2008in}. The contribution ${\cal L}^{(K)}$ corresponds instead
   to K-essence  \cite{ArmendarizPicon:2000ah}.
   It would be interesting to explore further generalizations
 to DHOST theories \cite{Langlois:2015cwa,Langlois:2018dxi,Crisostomi:2016czh,Achour:2016rkg,BenAchour:2016fzp,deRham:2016wji,Langlois:2017mxy} with the same property, or to consider scalar-tensor scenarios aimed to avoid
 constraints from graviton decay into dark energy \cite{Creminelli:2018xsv,Creminelli:2019nok,Creminelli:2019kjy}. We plan to study such generalizations in
 future works.

  The energy-momentum tensor $T_{\mu\nu}$ associated\footnote{The total action for the system can  contain  additional smooth matter fields   that can influence cosmological evolution.  We do not consider them here, since we  focus on GWs  in the geometrical optics limit.} with the Lagrangian density Eq.~\eqref{LagTot} can be expressed as \cite{Kobayashi:2011nu}
\be\label{TmunuTOT}
T_{\mu\nu}^{\text{(tot)}}\,=\,T_{\mu\nu}^{(F)}+T_{\mu\nu}^{(G)}+T_{\mu\nu}^{(K)}\,,
\ee
with
\bea
T_{\mu\nu}^{(F)}&=& \frac{F_\varphi}{F} \left( \nabla_\mu \nabla_\nu \varphi -  g_{\mu\nu}  \Box \varphi \right) + \frac{F_{\varphi\varphi}}{F} \left( \nabla_\mu \varphi \nabla_\nu \varphi + 2 X g_{\mu\nu} \right)\,, \label{TmunuF}
\\
T_{\mu\nu}^{(G)}&=&\frac{1}{2 F} \left( G_X\,\Box \varphi\,\nabla_\mu \varphi \nabla_\nu \varphi + \nabla_\mu G \nabla_\nu \varphi + \nabla_\nu G \nabla_\mu \varphi - g_{\mu\nu}\nabla_\lambda G \nabla^\lambda \varphi \right)\,, \label{TmunuG}\\
T_{\mu\nu}^{(K)}&=&\frac {1}{2 F} \left( K_X\,\nabla_\mu \varphi \nabla_\nu \varphi+\,K\,g_{\mu\nu}\right), \label{TmunuK}
\eea


\smallskip
The dynamics of the system is also controlled by the scalar equations of motion. We have proved that the relevant scalar equations are
satisfied once the gravitational  and the EMT conservation Eq.~\eqref{consEMT} are satisfied, hence we do not
discuss the scalar equation any further.

We notice that ${\cal L}^{(G)}$ and  ${\cal L}^{(F)}$ contain, respectively, derivative scalar
self-couplings and non-minimal couplings of scalar to gravity, which lead to derivative contributions to the EMT  involving metric fields.  Hence, we expect them to give relevant contributions to the
system evolution in the limit of geometrical optics.
In order to extract the high frequency contributions to the linearized EMT tensor, i.e. $t_{\mu\nu}$ as
defined in Eq.~\eqref{defHS}, one has to plug in the decompositions \eqref{MetricSplit} and \eqref{ScalarSplit} into Eq.~\eqref{TmunuTOT}, use the gauge conditions discussed in Section~\ref{summary_gauge} and then use the geometric optics Ansatz. Once this is done, one can organize the  $1/\epsilon$ contribution of the linearized EMT in the form of Eq.~\eqref{decotmunu}, that we re-write here
\be \label{decotmunuA}
t^{\text{(tot)}}_{\mu\nu}\,\equiv\,\left\{
\tau^{(A)}\,{\A}_{\mu \nu}+\tau^{(B)} \left(k_\mu {\cal C}_\nu +k_\nu {\cal C}_\mu \right)+
\tau^{(C)} \left[k_\mu \bar v_\nu+ k_\nu \bar v_\mu-\bar g_{\mu\nu} \left(k_\rho \bar v^\rho\right) \right] \right\}\,.
\ee
For the model \eqref{LagTot} the three functions $\tau^{(A)}$, $\tau^{(B)}$ and $\tau^{(C)}$ read
\bea
\tau^{(A)}\&=&-\frac{\,F_{\varphi}}{ F} \,(k_\rho \bar v^\rho)\,,\\
\tau^{(B)}&=&\frac{\,F_{\varphi}}{ F}\,,\\
\tau^{(C)} &=&-\frac{G_X}{2\,F}\,\left(\bar v_\rho\, {\cal C}^\rho-\frac{\bar v^2}{2}\,{\bar g^{\rho\sigma} \A_{\rho\sigma}}\right)\,-\frac{\,F_{\varphi}}{2F}\,{\bar g^{\rho\sigma} \A_{\rho\sigma}}\,,\label{tauCs}
\eea
Notice that, as anticipated above, only ${\cal L}^{F}$ and ${\cal L}^{G}$  contribute
 to the high-frequency EMT $t_{\mu\nu}$ and that
 \be
 \bar g^{\rho\sigma} \A_{\rho\sigma} = \sqrt{2}\, \A^{S}\,,
 \ee
 because of the trace properties of the polarization tensors.
We can then apply the evolution equations derived in the previous Section to the present instance. We find the following results:
\begin{itemize}
\item[i)]
From Eq.~\eqref{evetpa} an evolution equation for the tensor component of the GW amplitude:
\bea
\bar \nabla_{\rho}\left( k^\rho \,\left({\cal A}^T \right)^2\right)
&=&-\frac{\,F_{\varphi}}{ F}\,(k_\rho \bar v^\rho)\,\left({\cal A}^T \right)^2 \label{evetpa2}\,.
\eea
\item[ii)]
From Eq.~\eqref{seceqS} an evolution equation for the scalar component of the GW amplitude:
\bea
\bar \nabla_{\rho}\left( k^\rho \,\left({\cal A}^S \right)^2\right)
&=& \frac{G_X}{2\,F} (k_\rho \bar v^\rho) {\cal A}^S \,\left(\sqrt{2}\bar v_\rho\, {\cal C}^\rho-\bar v^2\,\A^{S}\right) \label{evescam2}
\eea
\item[iii)]
From Eq.~\eqref{evPolNolongitudinal} an evolution equation for the polarization tensor:
\bea
2\, k_\rho\,\bar \nabla^\rho\,{\bf e}_{\mu \nu}&=& -\left[\frac{G_X}{2\,F}\,\left(\bar v_\rho\, {\cal C}^\rho-\frac{\bar v^2}{\sqrt{2}}\,\A^S\right)\,+\frac{\,F_{\varphi}}{\sqrt{2} F}\,\A^S \right] \frac{\left(k_\rho \bar v^\rho\right)}{\sqrt{2}\A} \left[\frac{\A^S}{\A^2} \A_{\mu\nu} - {\bf e}^S_{\mu\nu}\right] \nonumber \\ &&\label{pol_tot}
\eea
\item[iv)] By contracting Eq.~\eqref{eomG1Ba} with $\bar v^\nu$, an evolution equation for the vector ${\cal C}_\mu$:
\bea
2 k^\rho \bar{\nabla}_\rho \,{\cal C}_\mu
&=& \A_{\mu\nu} \left[ \,2 \,k^\rho\, \bar \nabla_\rho \bar v^\mu\, - \bar v^\mu \, \bar \nabla_\rho k^\rho \right]  - \frac{F_\varphi}{F}(\bar v^\rho k_\rho) \left[ {\cal C}_\mu - \frac{\A^S}{\sqrt{2}} \bar v^\nu {\bf e}^S_{\mu\nu}\right] + \nonumber \\
&& + \frac{G_X (\bar v^\rho k_\rho)}{2 F} \left[\sqrt{2} \bar v^\rho {\cal C}_\rho - \bar v^2 \A^S \right]\bar v^\nu {\bf e}^S_{\mu\nu}
\eea
\end{itemize}
{The four equations above are the most general result obtained in this work. They describe the evolution of both the GW scalar and tensor modes in the considered theory of gravity}. They show  that  the GW propagation depends on the assumed scalar-tensor theory (through the dependence on the functions $F_\varphi$, $G_X$), and on the background quantities $\bar g_{\mu\nu}$
and $\bar v^\mu\,=\,\bar \nabla^\mu\,\bar \varphi$. {Moreover, from Eq.~\eqref{evescam2} we note that the evolution of the scalar amplitude $\mathcal{A}^S$ cannot be decoupled from that of the tensor amplitude $\mathcal{A}^T$, due to the presence of the total amplitude $\mathcal{A}$ in Eq.~\eqref{evescam2} (${\cal C}_\mu = \bar v^\nu \A_{\mu\nu}$). However, Eq.~\eqref{evetpa2} shows that the tensor amplitude $\mathcal{A}^T$ does not depend on $\mathcal{A}^S$. In addition, Eq.~\eqref{pol_tot} implies that the tensor and scalar modes of polarization are in general coupled one to each other.}

\smallskip

The evolution equation for the tensor amplitude, Eq.~\eqref{evetpa2} is proportional to $F_\varphi$. This is a quantity that in a cosmological
setting is associated with the time-dependence of the effective Planck mass. In theories as the
ones we consider here -- with second order equations of motion, and GW  unit speed --
 the failure of conservation of the current $({\cal A}^T)^2\,k^\mu$
is associated with the rate of change of the effective Planck mass, a result well known in the literature
 (see e.g. the discussion in the recent \cite{Belgacem:2019pkk} and references therein).

\smallskip

The evolution equation for the scalar part of the amplitude, Eq.~\eqref{evescam2} is more complex, and depends both on $F$ and $G_X$. As previously stated,  in the absence of braiding $\A^S = 0$ is a solution of its evolution equation~\eqref{eqAS2}. This implies that, if the scalar mode is not produced by a source, then it is absent. This could  be the case, for instance, if screening at the source is very effective.

In Eq.~\eqref{pol_tot}, the RHS quantifies the failure of the GW polarization tensor of being parallel transported along the GW geodesics. We can see that, only in the specific sub-case in which $G_X =  0 $ and $\A^S = 0$, the polarization tensor is parallel transported. In fact, only when these two conditions are met $\tau^{(C)} = 0$.

\section{GW propagation on a cosmological  space-time}\label{c}

One feature of the geometrical optics formalism we developed in the previous Sections is that it can be applied to
GWs travelling over an arbitrary space-time, as long as the scale of variation of the background geometry is well larger
than the GW wavelength. In this Section, we specialize to GWs travelling over a perturbed FRW universe, and apply the
formulas derived in the previous Section for a specific scalar-tensor model.
 Using  the {\it Cosmic Rulers} formalism~\cite{Schmidt:2012ne} (see the discussion in the Introduction)
we will show how to derive an expression for the GW luminosity distance  from the
analysis of the amplitude of the tensor modes, whose dynamics is governed by Eq.~\eqref{evetpa2}.


We report here the main equations we will use in this Section. As we wish to focus only on tensor modes, these are
\bea
\bar \nabla_{\rho}\left( k^\rho \,\left({\cal A}^T \right)^2\right)&=&-\frac{\,F_{\varphi}}{ F}\,(k_\rho \bar v^\rho)\,\left({\cal A}^T \right)^2 \,, \qquad \qquad  \label{evetpa3} \\
2\, k_\rho\,\bar \nabla^\rho\,{\bf e}_{\mu \nu}&=& -\left[\frac{G_X}{2\,F}\,\left(\bar v_\rho\, {\cal C}^\rho-\frac{\bar v^2}{\sqrt{2}}\,\A^S\right)\,+\frac{\,F_{\varphi}}{\sqrt{2} F}\,\A^S \right] \frac{\left(k_\rho \bar v^\rho\right)}{\sqrt{2}\A} \left[\frac{\A^S}{\A^2} \A_{\mu\nu} - {\bf e}^S_{\mu\nu}\right]\,,\quad\label{nopolt1} \\
2 k^\rho \bar{\nabla}_\rho \,{\cal C}_\mu &=&  \A_{\mu\nu} \left[ \,2 \,k^\rho\, \bar \nabla_\rho \bar v^\mu\, - \bar v^\mu \, \bar \nabla_\rho k^\rho \right]  - \frac{F_\varphi}{F}(\bar v^\rho k_\rho) \left[ {\cal C}_\mu - \frac{\A^S}{\sqrt{2}} \bar v^\nu {\bf e}^S_{\mu\nu}\right] + \nonumber \\
&& \qquad\qquad\qquad+ \frac{G_X (\bar v^\rho k_\rho)}{2 F} \left[\sqrt{2} \bar v^\rho {\cal C}_\rho - \bar v^2 \A^S \right]\bar v^\nu {\bf e}^S_{\mu\nu}\,. \label{concmuf}
\eea
We wish to stress that, in general, the GW polarization is not parallely transported in modified gravity. However, in order to keep the discussion simpler, when studying the polarization tensor in this Section we will choose the simplified scenario $G_X = 0$ and assume that the scalar wave is not produced at the source in order to set also $\A^S = 0$.
Eq.~\eqref{nopolt1} and Eq.~\eqref{concmuf} inform us that for the case of an {\it unperturbed} FRW space-time, we are allowed to make the choice ${\cal C}_\mu=0$.
Indeed, in a homogeneous FRW space-time, the scalar background configuration depends only time only, and $\bar v^\mu\,\propto\,\delta^\mu_{\,\,0}$. If  ${\cal C}^\mu=0$, then the  polarization tensors satisfy the condition ${{\bf e}}_{\mu0}\,=\,0$, and they have non-vanishing spatial components only. Since ${{\bf e}}_{ij}$ is transverse to the GW direction, we can without loss of generality choose a frame in which the GW propagates along the $z$-direction, and ${{\bf e}}_{ij}$ has non-vanishing components in the $(x,\,y)$-directions only. A simple calculation shows that, in this case,   the first term in the RHS of Eq.~\eqref{concmuf} vanishes. This implies that   the choice ${\cal C}_\mu=0$ satisfies this equation, and also implies that the RHS of Eq.~\eqref{nopolt1} vanishes.
  Hence, when focusing on an homogeneous,
   unperturbed
FRW space-time, if only the GW tensor components propagate and ${\cal C}^\mu=0$, the geometrical optics equations prescribe that the polarization tensor is parallel propagated along geodesics as the RHS of Eq.~\eqref{nopolt1} is zero in this case.
For the case of a {\it perturbed} space-time, instead, the situation is more complex, since all quantities depend both on time and space, and
the previous arguments do not hold.

We  use as background (i.e. slowly varying) space-time quantities
\begin{align}
    & ds^2_{\bar{g}} \,=\,\bar g_{\mu\nu} d x^\mu d x^\nu\,=\, a^2(\eta) [ -(1 + 2\phi) d\eta^2 + (1-2\psi) \delta_{ij} dx^i dx^j] \,, \label{MetricPoisson} \\
    & \bar{\varphi} = \varphi_0 (\eta) + \delta \varphi(x^\mu)\,. \label{scalarfieldFRW}
\end{align}
All fluctuations appearing in the previous Eqs.~\eqref{MetricPoisson} and~\eqref{scalarfieldFRW} are assumed to have long wavelengths, much larger than the high frequency GW modes that travel on such space-time.
Cosmological perturbations have the effect to alter the estimate of astrophysical parameters derived for the binary system. For example, the inferred luminosity distance of the source of the GWs will differ from the actual one because of the presence of inhomogeneities. It is important to clarify whether there are degeneracies
between these effects and modified gravity.
The propagation of short-wavelength GWs  over a metric which includes long-wavelength scalar fluctuations   has been also investigated in
\cite{Copeland:2018yuh}, with the specific purpose of analyzing some consequences of a
  novel proposal  for  evading  GW170817 constraints on scalar-tensor scenarios.

\subsection{Cosmic rulers for gravitational waves}
In order to study how cosmological perturbations affect the propagation of gravitational waves we follow \cite{Bertacca:2017vod}, but in the more general scalar-tensor theory presented in the previous Section.
In  \cite{Bertacca:2017vod} the authors use the so called \textit{Cosmic Rulers} formalism~\cite{Schmidt:2012ne} to compute the correction to the luminosity distance of the spiraling binary.
This formalism has been first formulated for electromagnetic radiation \cite{Jeong_2012}, \cite{Schmidt:2012ne} but it can be immediately extended to gravitational radiation in the geometrical optics limit.
In the Cosmic Rulers formalism, the observer frame, called the \textit{Redshift-GW frame} (RGW), is used as reference system.
Such frame is different from the {\it real-frame} (or equivalently called the {\it physical-space}) which is the frame where we computed Eqs.~\eqref{evetpa3}, ~\eqref{nopolt1} and ~\eqref{concmuf}.

Since we will use as background metric Eq.~\eqref{MetricPoisson}, it is convenient to perform a conformal transformation,
\begin{align}\label{ConformalTransform}
    \begin{cases}
    &\, \Bar{g}_{\mu\nu} \rightarrow \hat{g}_{\mu\nu} = \Bar{g}_{\mu\nu} / a^2\,, \qquad
    \: \Bar{g}^{\mu\nu} \rightarrow \hat{g}^{\mu\nu} = \Bar{g}^{\mu\nu} a^2\,, \\
   & \, k^\mu \rightarrow \hat{k}^\mu = k^\mu  a^2\,, \qquad\quad \hat{k}_\mu = k_\mu\, = -\bar \nabla_\mu \theta \,,\\
   \end{cases}
\end{align}
in this way $\hat{g}_{\mu\nu} = \eta_{\mu\nu} +   \delta  \hat g_{\mu\nu}$.
Under the transformation~\eqref{ConformalTransform} the connection coefficients transform as
\begin{align}
    \Gamma^\mu_{\,\nu\rho} &= \frac{1}{2 a^2}\hat{g}^{\mu\alpha} \bigg( (a^2\hat{g}_{\nu\alpha})_{,\rho}+(a^2\hat{g}_{\rho\alpha})_{,\nu}-(a^2\hat{g}_{\nu\rho})_{,\alpha}\bigg) = \nonumber\\
    &= \hat{\Gamma}^\mu_{\,\nu\rho} + \frac{1}{2} \bigg( \delta^\mu_\nu \, \partial_\rho \ln a^2  + \delta^\mu_\rho \, \partial_\nu \ln a^2  - \delta_{\nu\rho} \, \hat{g}^{\mu\alpha}\partial_\alpha \ln a^2  \bigg)\,,
\end{align}
where $\hat{\Gamma}^\mu_{\,\nu\rho}$ is the connection symbol associated to $\hat{g}_{\mu\nu}$.

\smallskip
In this Section we use the notation in which barred quantities belong to the RGW-frame, while unbarred quantities belong to the physical space. As we performed the conformal transformation~\eqref{ConformalTransform} and decomposed the scalar field as~\eqref{scalarfieldFRW} there is no confusion with the objects defined in the previous Sections.

We define $x^\mu(\chi)$ as the comoving coordinate in the \textit{real frame}, where $\chi$ is the comoving distance to the observer. Using the notation just introduced, $\bar x^\mu(\bar \chi)$ is the coordinate in the  RGW-frame.
In the RGW-frame we use coordinates that flatten our past gravitational wave-cone, therefore the geodesic of the emitted gravitational waves in the RGW-frame is given by
\be
\bar{x}^\mu (\bar \chi) = (\bar{\eta}, \bar{\mathbf{x}}) = (\eta_0 - \bar{\chi}\,, \, \bar{\chi} \hat{\mathbf{n}})\,,
\ee
where $\eta_0$ is the conformal time at observation, $\bar{\chi}(z)$ is the comoving distance to the observed redshift in the observer frame and $\hat{\mathbf{n}}$ is the observed direction of arrival in the sky, i.e. $\hat{n}^i = \bar{x}^i / \bar{\chi} = \delta^{ij} (\,\partial \bar{\chi} / \partial \bar{x}^j\,)$.
The observed coordinate will be different from the one in the real frame and the total derivative along the part gravitational wave-cone is
\begin{equation}
    \frac{\text d}{\text d \bar{\chi}} = - \frac{\partial }{\partial \bar{\eta}} + \hat{n}^i \frac{\partial}{\partial \bar{x}^i}\,.
\end{equation}
It is convenient to define parallel and perpendicular projectors operator with respect the observed line-of-sight direction.
As shown in ~\cite{Schmidt:2012ne, Bertacca:2017vod}, for any spatial vector  $B_i$ and tensor $A_{ij}$, in the RGW-space we have:
\begin{enumerate}
    \item $A_{||} = \hat{n}^i \hat{n}^j A_{ij}$
    \item $B^i_{\perp} = \mathcal{P}^{ij}B_{j} = B^i - \hat{n}^i B_{||}$
    \item $\bar{\partial}_{||} = \hat{n}^i \bar{\partial}_i$
    \item $\bar{\partial}^2_{||} = \bar{\partial}_{||}\,\bar{\partial}_{||}$
    \item $\bar{\partial}_{\perp\,i} = \mathcal{P}^j_i = \bar{\partial}_i -\hat{n}_i \bar{\partial}_{||}$
    \item $\displaystyle{\bar{\partial_i}\hat{n}^j} = \bar{\partial}_i \displaystyle{\left(\frac{\bar{x}^j}{\bar{\chi}}\right)} = \frac{1}{\bar{\chi}} \left(\delta^j_i - \frac{\bar{x}^j}{\bar{\chi}}\frac{\partial \bar{\chi}}{\partial \bar{x}^i}\right) = \frac{1}{\bar{\chi}} \left(\delta^j_i - \hat{n}^j\hat{n}_i\right) = \frac{1}{\bar{\chi}}\mathcal{P}^j_i$
    \item $\displaystyle{\frac{\text d }{\text d \bar{\chi}} \partial^i_\perp = \bar{\partial}^i_\perp \frac{\text d }{\text d \bar{\chi}} -\frac{1}{\bar{\chi}}\partial^i_\perp}$
    \item $\displaystyle{\frac{\partial B^i}{\partial \bar{x}^j} =\bar{\partial}_{j\perp}B^i_\perp +\hat{n}_j \bar{\partial}_{||} B^i_\perp +\frac{1}{\bar{\chi}}\mathcal{P}^j_i B_{||} + \hat{n}^i [\bar{\partial}_{j\perp} + \hat{n}_j \bar{\partial}_{||}]B_{||}} $
\end{enumerate}
where we have used $\bar{\partial}_i = \partial/ \partial \bar{x}^i$ and $\mathcal{P}^i_j = \delta^i_j - \hat{n}^i \hat{n}_j$. Note also that partial derivatives in the RGW-frame do not commute as shown in~\cite{Schmidt:2012ne}.

\smallskip
We define $\bar{k}^\mu$ as the null geodesic vector in the redshift frame at zeroth order,
\begin{equation}\label{kappabarra}
    \bar{k}^\mu = \frac{\text d \bar{x}^\mu}{\text d \bar{\chi}} = (-1, \,\hat{\mathbf{n}})\,,
\end{equation}
which satisfies $\text d \bar{k}^\mu / \text d \bar{\chi}=0$.

\smallskip
Our aim is to compute GW observables in the RGW-space. In the previous section we derived the evolution equations of the quantities related to the GWs in the real frame, therefore we need to build a map to relate the two frames. We do so relying on the fact that the difference between the real-frame and the observer-frame is due to the presence of the linear perturbations describing the LSS present in the Universe.  In fact, if the Universe was homogeneous and isotropic then, after the conformal transformation~\eqref{ConformalTransform}, the RGW-space would coincides with the real-frame.
We define $\delta \chi$ to be the difference of the affine parameters $\chi$ and $\bar \chi$, so that
\be
\chi = \bar \chi + \delta \chi\,,
\ee
in this way the relation between the coordinates of the geodesic in the real-frame and the RGW-frame is given by
\begin{align}
x^\mu(\chi) = x^\mu (\bar \chi + \delta \chi) &= x^\mu (\bar \chi ) +  \frac{\text d \bar x^\mu}{\text d \bar \chi} \delta \chi  =  \nonumber\\
&= \bar x^\mu (\bar \chi ) +  \delta x^\mu (\bar \chi ) +  \frac{\text d \bar x^\mu}{\text d \bar \chi} \delta \chi =  \bar x^\mu (\bar \chi ) + \delta x^\mu (\bar{\chi}) + \bar{k}^\mu  \delta \chi
\end{align}
where we used Eq.~\eqref{kappabarra} in the last step and introduced $\delta x^\mu(\bar \chi)$ which is the perturbation to the graviton's geodesics at fixed affine parameter, i.e. $ \delta x^\mu (\bar \chi ) \coloneqq x^\mu(\bar \chi) - \bar x^\mu(\bar \chi) $.
We define the total correction to the GW's geodesic as
\begin{equation}\label{Deltax}
    \Delta x^\mu (\bar{\chi}) \coloneqq \delta x^\mu (\bar{\chi}) + \bar{k}^\mu  \delta \chi\,.
\end{equation}
so that the mapping between $x^\mu(\chi)$ and $\bar x^\mu(\bar \chi)$ is given by
\begin{align} \label{Map}
   x^\mu (\chi) = \bar{x}^\mu (\bar{\chi}) + \Delta x^\mu (\bar{\chi})
\end{align}
as also discussed in~\cite{Schmidt:2012ne}.
The correction $\Delta x^\mu (\bar{\chi})$ has two contributions: one at fixed observed-comoving distance, $\delta x^\mu (\bar{\chi})$, and the other one proportional to $\delta \chi$ in the direction of $\bar{k}^\mu$. The first contribution accounts for the perturbation of the GW's geodesics at $\bar \chi$ due to the presence of linear structure in the Universe, while the second one for the difference between the affine parameters of the geodesics in the two frames. In practice, using $\bar \chi$ instead of $\chi$ makes us "misread" the value of the affine parameter along the GW's geodesics at which we compute a certain quantity. This is actually true for all of the quantities characterizing the propagating GW (amplitude, phase and polarization) and the dark energy field, and not only the geodesics it follows. Hence, from now on, the symbol $\Delta$ stands for the sum of both these two kinds of contribution.

The wave-vector is defined as the vector tangent to  the GW geodesic in the real-space $x^\mu(\chi)$, i.e. $ \hat{k}^\mu  = \text d x^\mu(\chi) /  \text d \chi $. After some manipulations it is possible to see that it is given by
\begin{align}
    \hat{k}^\mu &= \frac{\text d x^\mu(\chi)}{\text d \chi} = \frac{\text d \bar{\chi}}{\text d \chi} \frac{\text d x^\mu({\chi}) }{\text d \bar{\chi}} = \left(1 - \frac{\text d  \delta \bar{\chi}}{\text d \chi} \right)  \frac{\text d }{\text d \bar{\chi}} (\bar{x}^\mu(\bar{\chi}) + \Delta x^\mu(\bar{\chi})) = \bar{k}^\mu + \frac{\text d  \delta x^\mu}{\text d \bar{\chi}}(\bar \chi) \,,
\end{align}
at linear order. We notice that the correction to $ \hat{k}^\mu $ is only of the first kind, i.e.  $\Delta \hat k^\mu ( \bar{\chi})= \delta  \hat k^\mu( \bar{\chi})$, and it does not show any contribution  $\propto \bar k^\mu \delta \chi$, as it was instead the case for $ \Delta x^\mu (\bar{\chi})$.
Therefore, we can find $\delta k^\mu$ through
\begin{equation}
    \hat{k}^\mu = \bar{k}^\mu + \delta k^\mu = \bar{k}^\mu + \frac{\text d  \delta x^\mu}{\text d \bar{\chi}}\,,
\end{equation}
which leads to
\begin{align}
   \delta k^\mu=  \frac{\text d  \delta x^\mu(\bar{\chi})}{\text d \bar{\chi}}  \coloneqq (\delta \nu(\bar \chi),\, \delta \mathbf{n} (\bar \chi))  \,, \label{firstorderk}
\end{align}
where  $\delta \nu$  represents the correction to the frequency of the wave, and $\delta n^i$ the one to the direction of arrival.
The wave-vector in the real-space is given by
\begin{equation}
    \hat{k}^\mu(\bar{\chi}) = (-1 + \delta \nu(\bar \chi),\, \hat{n}^i + \delta n^i(\bar \chi))\,.
\end{equation}
We point out also that the correction $\delta k^\mu$ satisfies the linearized geodesics equation and,  in the next Section, we will use this fact to find $\delta \nu$ and  $\delta n^i$ as a function of the gravitational potentials in Poisson's gauge.
We can obtain $\delta x^\mu$ upon integration of Eq.~\eqref{firstorderk}
\be\label{deltaxdeltak}
\delta x^0(\bar{\chi}) = \int^{\bar{\chi}}_0 \text d \Tilde{\chi}\,\delta \nu(\Tilde{\chi}) \qquad \mbox{and} \qquad \delta x^i(\bar{\chi}) = \int^{\bar{\chi}}_0 \text d \Tilde{\chi}\,\delta {n}^i(\Tilde{\chi})\,,
\ee
with the boundary condition $ \delta x^\mu_o \coloneqq \delta x^\mu(\bar \chi = 0) = 0$ set at the observer position $\bar \chi = 0$. The real-space scale factor is given by
\be \label{deltascale}
\frac{a(x^0(\chi))}{\bar a} = \frac{a(\bar{x}^0 + \Delta x^0)}{\bar a }= 1 + \Delta \ln a  = 1 + \mathcal{H} \Delta x^0  \,,
\ee
where $\bar a = a(\bar x^0)$\footnote{In principle we should include the perturbation of the scale factor at observation $\delta a_o= a_o- 1$. This extra term is an additive constant that we neglect for simplicity. For further details see, e.g.,  \cite{Bertacca:2019fnt}).}.

We restrict ourselves to the \textit{local wave zone} where the GW's wavelength is small with respect to the comoving distance from the observer. In this portion of the space-time we introduce the frame of reference of a comoving observer, described by the tetrad $e^\mu_{{\hat a}}$, which we extend until the source position.
 The time-like vector of the orthonormal basis can be chosen using the four-velocity $u^\mu$ of the observer
\begin{equation}
    u_\mu = e_{\hat{0}\mu} = a E_{\hat{0} \mu} \qquad \mbox{and} \qquad  u^\mu = e^\mu_{\hat{0}}=  a^{-1} E^\mu_{\hat{0}} \,,
\end{equation}
where $E^{\hat{a}}_\mu$ is the tetrad in the comoving frame.
The other components are constructed using the relations
\begin{equation}
   \Hat{g}_{\mu\nu} = \eta_{\Hat{a}\Hat{b}}E^{\hat{a}}_\mu E^{\hat{b}}_\nu\,, \quad \eta^{\Hat{a}\Hat{b}} = \Hat{g}_{\mu\nu} E_{\hat{a}}^\mu E_{\hat{b}}^\nu\,, \quad \Hat{g}^{\mu\nu} E^{\hat{a}}_\mu = E^{\hat{a} \nu}\,, \quad \eta_{\Hat{a}\Hat{b}}E^{\hat{a}}_\mu = E_{\hat{b}\mu} \,.
\end{equation}
At linear order in Poisson's gauge $u_\mu = - a (1+\phi, - v_i)$, so that the comoving tetrad is given by
\begin{equation}\label{Tetrad}
E_{\hat{0}\mu}= \bigg(
\begin{array}{cc}
-1-\phi \\
 v_i \\
\end{array}
\bigg)\,,\quad
E_{\hat{a}0} = - v_{\Hat{a}}\quad \mbox{and} \quad
E_{\hat{a}i} = \delta_{\hat{a}i}(1-\psi) \,.
\end{equation}
We can use this tetrad to compute the correction of the scale factor in Eq.~\eqref{deltascale}. Calling "e" and "o"  the emitted and observed positions, respectively,  and denoting the GW frequency as $f$, the scale factor is related to the observed redshift as
\begin{equation}\label{1+z}
    1 + z = \frac{f_{e}}{f_{o}} = \frac{(e_{\hat{0} \mu} k^\mu) |_{e}}{(e_{\hat{0} \mu} k^\mu)_{o}}= \frac{a_0}{a(\chi_{e})}\frac{(E_{\hat{0} \mu} \hat{k}^\mu) |_{e}}{(E_{\hat{0} \mu} \hat{k}^\mu)_{o}} \,.
    \end{equation}
Setting $a_o = \bar{a}_o = 1$ and $(E_{\hat{0} \mu} k^\mu)|_o = 1$, the previous equation becomes
\be
 1 + z = \frac{(E_{\hat{0} \mu} \hat{k}^\mu) |_{e}}{a(\chi_{e})} =  \frac{(E_{\hat{0} \mu} \hat{k}^\mu)^{(0)}|_{e} + (E_{\hat{0} \mu} \hat{k}^\mu)^{(1)}|_{e}}{\bar{a}(1 + \Delta \ln a)} = (1+z)\,\frac{1+(E_{\hat{0} \mu} \hat{k}^\mu)^{(1)}}{1 + \Delta \ln a}\,,
 \ee
where we used  $\bar{a} = 1/ (1+z)$. Hence
\be
\Delta \ln a = (E_{\hat{0} \mu} \hat{k}^\mu)^{(1)}|_{e} = -E_{\hat{0}0}^{(1)} + \hat{n}^i E_{\hat{0}i}^{(1)} -\delta \nu\,.
\ee

\subsection{Gravitational waves in the {observer} frame}
In this Section we follow the same prescription developed in \cite{Bertacca:2017vod}, and we extend their results by including modified gravity effects.
  In particular, we compute the effects of large-scale structures on GW's waveforms (e.g. the phase and the amplitude) of the GW in MG accounting for lensing, Sachs-Wolfe, integrated Sachs-Wolfe, time delay and volume effects.

\subsubsection{The GW  phase}
We showed that in the theory of gravity considered, the wave-vector is null and tangent to a null geodesic of the background space-time. After the conformal transformation defined in Eq.~\eqref{ConformalTransform}, the wave-vector in the conformal space satisfies
\be \label{EqPhase2}
\hat k^\mu \hat k_\mu =  0\,,
\ee
and using $\hat k_\mu = k_\mu = - \bar \nabla_\mu \theta$ we find
\be \label{phase2}
0 = \hat k^\mu \bar \nabla_\mu \theta = \frac{\text d \theta}{\text d \chi} \,.
\ee
We wish to expand the equation above at linear order to find the effects of the structure present in the Universe on the phase of GWs. In order to do so we keep in mind that, when expanding to first order, every quantity gets two types of contribution: one at {\it fixed} comoving distance $\bar \chi$, denoted with $\delta$ and one that accounts for the due correction of the affine parameter, therefore proportional to $\delta \chi$.
Accordingly, if $\bar \theta(\bar \chi )$ is the phase of the wave at zero order, i.e. the one in the RGW-frame, we define $ \delta  \theta(\bar x(\bar \chi)) = \theta(\bar x(\bar \chi)) - \bar \theta\bar x(\bar \chi) )$
so that the total correction to the phase is
\begin{align}
 \theta(x ) &=\theta(\Bar{x}+\Delta x) =   \theta(\Bar{x}(\bar \chi)) + \Delta x^\mu \Bar{\partial}_\mu  \Bar{\theta}(\bar{x}(\bar{\chi}))  = \Bar{\theta}(\Bar{x}(\bar \chi))  + \delta \theta (\Bar{x}(\bar \chi))  + \Delta x^\mu\Bar{\partial}_\mu  \Bar{\theta}(\bar{x}(\bar{\chi})) =  \nonumber \\
 & =\Bar{\theta}(\Bar{x}(\bar \chi))  + \delta \theta (\Bar{x}(\bar \chi)) - \Delta x^\mu \Bar{k}_\mu = \Bar{\theta}(\Bar{x}(\bar \chi)) + \Delta \theta (\Bar{x}(\bar \chi))
\label{DeltaTheta}
\end{align}
where we used that at zero order $ \bar k_\mu = - \bar \partial_\mu \bar \theta$. We defined $\Delta \theta (\Bar{x}(\bar \chi))  \coloneqq   \delta \theta (\Bar{x}(\bar \chi))  - \Delta x^\mu \Bar{k}_\mu$ as the sum of the two contributions to the correction of the phase.

Using this prescription and Eqs.~\eqref{Map} we can expand at linear order Eq.~\eqref{phase2} as
\begin{align*}
    0 =  \frac{\text d \theta (x^\mu (\chi))}{\text d \chi}  &= \frac{\text d}{\text d \chi} \theta(\Bar{x}^\mu+\Delta x^\mu) =  \bigg(1- \frac{\text d \delta \Bar{\chi}}{\text d \chi}\bigg)\frac{\text d}{\text d \Bar{\chi}} \bigg( \Bar{\theta}(\Bar{x}) + \Delta \theta (\Bar{x}) \bigg) = \\
    &=\frac{\text d \Bar{\theta}}{\text d \Bar{\chi}} + \frac{\text d \delta\theta}{\text d \Bar{\chi}} + \delta k^\mu \Bar{\partial}_\mu \Bar{\theta} + \delta x^\mu \frac{\text d }{\text d \Bar{\chi}}\Bar{\partial}_\mu \Bar{\theta} = \frac{\text d \Bar{\theta}}{\text d \Bar{\chi}} + \frac{\text d \delta\theta}{\text d \Bar{\chi}} - \delta k^\mu \Bar k_\mu \,,
\end{align*}
where in the last step we used that $- \Bar{\partial}_\mu \Bar{\theta} = \bar{k}_\mu$  and $d \bar{k}^\mu / d \bar{\chi}=0$. We can extract the zero and first order contributions from the equation above as they have to be satisfied separately,
\begin{align}
    \frac{\text d \Bar{\theta}}{\text d \Bar{\chi}} = 0 \qquad \mbox{and} \qquad \frac{\text d \delta\theta}{\text d \Bar{\chi}} = \delta k^\mu \bar{k}_\mu\,.
\end{align}
The last equation allows us to get $\delta \theta$ from $\delta k_\mu$ upon integration
\begin{equation}
    \delta \theta(\bar{\chi}) = \delta \theta_o + \int^{\bar{\chi}}_0 d \chi\,(\delta \nu + \delta n_{||}) \,,
\end{equation}
where $\delta \theta_o$ is the value of $\delta \theta$ at the observer position who is located at $\bar{\chi}=0$.
The full correction to the phase, $\Delta \theta(\bar{x}(\bar{\chi}))$, is then given by
\begin{align}
    \Delta \theta(\bar{x}(\bar{\chi})) & = \delta \theta(\bar{x}(\bar{\chi})) - \Delta x^\mu \Bar{k}_\mu = \delta \theta(\bar{x}(\bar{\chi})) - \Delta x^0 \bar{k}_0 - \Delta x^i \bar{k}_i =\delta \theta -(\Delta x^0 + \Delta x_{||}) \,.
\end{align}
since $\bar{k}_0 = \eta_{\mu0} \bar{k}^\mu = + 1$.
We can define $T$ as
\begin{equation}\label{T}
     T = - (\Delta x^0 + \Delta x_{||}) = - (\delta x^0 + \delta x_{||})\,,
\end{equation}
therefore
\begin{equation}\label{DeltaPhase}
    \Delta \theta(\bar{\chi}) = \delta \theta(\bar{\chi}) + T  = \delta \theta |_o  + \int^{\bar{\chi}}_0 d \chi\,(\delta \nu + \delta n_{||}) + T = \delta \theta |_o\,,
\end{equation}
\noindent{where} we used Eq.~\eqref{deltaxdeltak} and Eq.~\eqref{T} in the last equality. This result is  in agreement with \cite{Bertacca:2017vod}.

\subsubsection{The GW  amplitude}
In the previous section we derived the evolution equation of the amplitude of the tensor part of the GW in the most general Horndeski theory satisfying the constraint $c_T = 1$ and in which the scalar wave propagates at the same speed of the tensor modes.
The amplitude $\A^T$ of waves is inversely proportional to the luminosity distance of the spiraling binary, and thus we are interested in deriving the expression of the observed amplitude. To this aim, it is useful to rewrite Eq.~\eqref{evetpa2} as
 \begin{equation}\label{EqAmplitude1}
    k^\rho \bar{\nabla}_\rho \ln(\mathcal{A}^T) = - \frac{1}{2}\bigg(\Bar{\nabla}^\rho k_\rho\, + k^\rho \bar{\nabla}_\rho \ln F[\varphi] \, \bigg)\,,
\end{equation}
where we have used the definition of $\bar{v}_\mu $ in terms of the slowly-varying configuration of the extra scalar field.
After the transformation Eq.~\eqref{ConformalTransform}, the latter becomes
\begin{equation}\label{EqAmplitudeConformal}
    \hat{k}^\rho \hat{\nabla} _\rho\ln (\mathcal{A}^T\,a) = - \frac{1}{2} \bigg( \hat{\nabla}_\rho \hat{k}^\rho + \hat{k}^\rho \hat{\nabla} _\rho\ln F \bigg) \, ,
\end{equation}
where $\hat{\nabla}_\rho$ is the covariant derivative respect $\hat{g}_{\mu\nu}$.
At the zero order Eq.~\eqref{EqAmplitudeConformal} is
\begin{align}
     \Bar{k}^\rho \Bar{\partial} _\rho\ln &(\Bar{\mathcal{A}}^T\Bar{a}) =\frac{\text d}{\text d \bar{\chi}} \ln (\Bar{\mathcal{A}}^T\Bar{a})  = - \frac{1}{2} \bigg( \Bar{\partial}_\rho \Bar{k}^\rho + \Bar{k}^\rho \Bar{\partial} _\rho\ln F_0 \bigg) = \nonumber \\
     &= - \frac{1}{2} \bigg( [\Bar{\partial}_0 \Bar{k}^0 + \Bar{\partial}_i \Bar{k}^i]+ \Bar{k}^\rho \Bar{\partial}_\rho \ln F_0\bigg) = - \frac{1}{2} \bigg( \Bar{\partial}_i \hat{n}^i+ \frac{\text d}{\text d\Bar{\chi}} \ln F_0\bigg) =- \frac{1}{2} \bigg( \frac{\mathcal{P}^i_{\,i}}{\Bar{\chi}} + \frac{\text d}{\text d \Bar{\chi}} \ln F_0\bigg)= \nonumber \\
     &= -  \frac{\text d}{\text d \Bar{\chi}} \bigg( \ln \Bar{\chi} + \ln \sqrt{F_0} \bigg) \nonumber \,,
\end{align}
where we called $F_0 = F[\varphi_0]$. We can rewrite the latter as
\begin{equation}\label{AmplitudeZero}
     \frac{\text d}{\text d \bar{\chi}} \ln (\Bar{\mathcal{A}}^T\Bar{a} \sqrt{F_0} \Bar{\chi}) = 0 \qquad \longrightarrow \qquad \Bar{\mathcal{A}}^T(\bar{x}^0,\bar{\chi}) = \frac{\mathcal{Q}}{\Bar{a}(\bar{x}^0)\Bar{\chi}\sqrt{F_0}} \, \,.
\end{equation}
where $\mathcal{Q}$ is an integration constant determined by the local-wave zone solution and constant along the null geodesic. The GR equivalent of Eq.~\eqref{AmplitudeZero} is\footnote{To be precise, the integration constant $\mathcal{Q}$ could be different in modified gravity with respect to GR.}
\begin{equation}
    (\Bar{\mathcal{A}}^T)^{\tiny{GR}}(\bar{x}^0,\bar{\chi}) = \frac{\mathcal{Q}}{\Bar{a}(\bar{x}^0)\Bar{\chi}}\,,
\end{equation}
where there is a factor $1/\sqrt{F_0}$ of difference with respect to the result in modified gravity.

As in the case of the phase, we write the GW amplitude as a sum of the value it would have on the unperturbed background and a total linear correction
\begin{align} \label{ExpansionAmplitude}
 \ln \A^T(\Bar{x}^\mu + \Delta x^\mu) &= \ln \bar \A^T(\bar x(\bar \chi)) + \Delta \ln A^T (\Bar{x}(\bar \chi)) = \nonumber \\
   & =\ln \Bar{\A^T}(\Bar{x}(\bar \chi))  + \delta \ln  \A^T (\Bar{x}(\bar \chi))  + \Delta x^\mu \Bar{\partial}_\mu  \ln \Bar{\A^T}(\bar{x}(\bar{\chi}))
\end{align}
where $ \delta \ln  \A^T (\Bar{x}(\bar \chi))  =   \ln  \A^T (\Bar{x}(\bar \chi))  -   \ln \bar \A^T (\Bar{x}(\bar \chi)) $ and plug this expansion into Eq.~\eqref{EqAmplitudeConformal}. The linearization of  Eq.~\eqref{EqAmplitudeConformal} is made of four different contributions, three of them do not depend explicitly on the extra scalar field and can be found also in the case of GR \cite{Bertacca:2017vod}. These are
\begin{align*}
    &1)\: \frac{\text d}{\text d \chi} \ln (\mathcal{A}^T(\Bar{x}^\mu + \Delta x^\mu))  = \bigg(1-\frac{\text d \delta\chi}{\text d \bar{\chi}} \bigg)\bigg[\frac{\text d}{\text d \Bar{\chi}}\ln \Bar{\mathcal{A}}^T + \frac{\text d}{\text d \Bar{\chi}}\delta \ln \mathcal{A}^T +\frac{\text d}{\text d \Bar{\chi}}( \Delta x^\mu \Bar{\partial}_\mu \ln\Bar{\mathcal{A}}^T)  \bigg] = \\
   & \qquad\qquad\qquad\qquad= \frac{\text d \ln \Bar{\mathcal{A}}^T}{\text d \Bar{\chi}} + \frac{\text d \delta\ln \Bar{\mathcal{A}}^T}{\text d \Bar{\chi}} + \delta\chi  \frac{\text d^2 \ln \Bar{\mathcal{A}}^T}{\text d \Bar{\chi}^2} + \delta k^\mu  \Bar{\partial}_\mu \ln\Bar{\mathcal{A}}^T + \delta x^\mu\frac{\text d \Bar{\partial}_\mu \ln\Bar{\mathcal{A}}^T}{\text d \Bar{\chi}}\,,  \\\\
   & 2)\:\frac{\text d}{\text d \chi} \ln a(\bar{x}^0 + \Delta x^0) =   \bigg(1-\frac{\text d \delta\chi}{\text d \bar{\chi}} \bigg)\frac{\text d}{\text d \Bar{\chi}}\ln [\Bar{a}(1+\mathcal{H}\Delta x^0)] = \\
    &\qquad\qquad\qquad\qquad = -\mathcal{H} \bigg(1-\frac{\text d \delta\chi}{\text d \Bar{\chi}}\bigg) + \frac{\text d}{\text d \Bar{\chi}}(\mathcal{H}\Delta x^0) = -\mathcal{H}[1-\delta k^0] -\mathcal{H}'[ \delta x^0 - \delta \chi]\,,
\end{align*}
\begin{align*}
   3)&\:-\frac{1}{2}\hat{\nabla}_\rho \hat{k}^\rho  =  -\frac{1}{2} \bigg[ \frac{\partial \Bar{x}^\mu}{\partial x^\rho}\Bar{\partial}_\mu (\Bar{k}^\rho + \delta k^\rho) + \delta \hat{\Gamma}^\rho_{\,\rho\mu} \Bar{k}^\mu \bigg] = \nonumber\\  \nonumber\\
    & =\frac{-1}{2} \bigg[\delta \hat{\Gamma}^\rho_{\,\rho\mu} \Bar{k}^\mu + \frac{2}{\Bar{\chi}}(1+ \delta k_{||}) + \frac{\text d \delta k_{||}}{\text d \Bar{\chi}} + \Bar{\partial}_0 (\delta k^0 + \delta k_{||})  \Bar{\partial}_{i\bot}\delta k^i_{\bot}-\frac{2}{\Bar{\chi}^2}(\delta\chi + \delta x_{||})-\frac{1}{\Bar{\chi}}\Bar{\partial}_{i\bot}\delta x^i_{\bot}  \bigg]\,,
\end{align*}
where $\delta \hat{\Gamma}^\rho_{\,\rho\mu}$ is the linear order connection coefficient associated to $\hat{g}_{\mu\nu}$.
The new part with respect to GR is
\begin{align*}
   4)&\:-\frac{1}{2} \frac{\text d}{\text d \chi} \ln F (\varphi_0 + \Delta \varphi)= -\frac12 \bigg(1-\frac{\text d \delta\chi}{\text d \Bar{\chi}} \bigg) \frac{\text d}{\text d \Bar{\chi}} \ln \bigg[F_0\bigg(1 + \frac{F_{\varphi 0}}{F_0} \Delta \varphi\bigg)\bigg] = \\ \\
    & = -\frac{1}{2} \bigg[\bigg(\frac{F_{\varphi0}}{F_0} + \Delta\varphi\bigg(\frac{F_{\varphi\varphi0}}{F_0}- \frac{F_{\varphi0}^2}{F_0^2}\bigg)\bigg)\frac{\text d \varphi_0}{\text d \Bar{\chi}}+\frac{F_{\varphi0}}{F_0}\bigg(\frac{\text d \delta\varphi}{d \Bar{\chi}}+ \delta\chi \frac{\text d^2 \varphi_0}{d \Bar{\chi}^2} + \delta k^0 \varphi'_0 + \delta x^0 \frac{\text d \varphi'_0}{d \Bar{\chi}}\bigg)\bigg]\,, \nonumber \\
\end{align*}
were we used the short notations $ (\text d F / \text d \varphi) |_{\varphi_0} = F_{\varphi0}$, $ (\text d^2 F / \text d \varphi^2) |_{\varphi_0} = F_{\varphi\varphi0}$, $\varphi'_0 = \bar{\partial}_0 \varphi_0$ and $\Delta \varphi = \delta \varphi + \Delta x^\mu \bar{\partial}_\mu \varphi_0$.
Combining together the four contributions yields
\begin{align}\label{AmplitudeFirstOrder}
    &\frac{\text d }{\text d \Bar{\chi}} \delta\ln (\mathcal{A}^T)(\bar{\chi}) = -\frac{1}{2} \bigg[\delta \hat{\Gamma}^\rho_{\,\rho\mu} \Bar{k}^\mu + \frac{\text d \delta k_{||}}{\text d \Bar{\chi}} + \Bar{\partial}_0 (\delta k^0 + \delta k_{||}) -2 \frac{\text d \kappa}{\text d \Bar{\chi}}+ \frac{\text d }{d \Bar{\chi}}\left( \frac{F_{\varphi0}}{F_0} \delta\varphi\right)\bigg] \,, \nonumber \\
\end{align}
where $\kappa$ is the weak lensing convergence term defined as
\begin{equation*}
    \kappa = - \frac{1}{2}\Bar{\partial}_{i\bot}\Delta x^i_{\bot} \, .
\end{equation*}
Eq.~\eqref{AmplitudeFirstOrder} can be integrated in order to obtain $\delta \ln \mathcal{A}^T(\bar{\chi})$. The full correction to $\A^T$, defined through Eq.~\eqref{ExpansionAmplitude}, then reads
\begin{align} \label{DeltaAmplitude}
    \Delta \ln \mathcal{A}^T &= \delta \ln \mathcal{A}^T(\bar{\chi})+ \Delta x^0 \Bar{\partial}_0\ln \Bar{\mathcal{A}}^T+ \Delta x^i \Bar{\partial}_i\ln \Bar{\mathcal{A}}^T = \nonumber\\
     &= \delta \ln \mathcal{A}^T(\bar{\chi}) - \Delta \ln a \left(\,1 + \frac{F_{\varphi0} \varphi_0'}{2 F_0\mathcal{H}} - \frac{1}{\Bar{\chi}\mathcal{H}}\right) + \frac{T}{\Bar{\chi}} \,,
\end{align}
\noindent{where} $T$ is given in Eq.~\eqref{T}.  Besides contributions depending on cosmological fluctuations,
we find an {\it explicit additional term due to modified gravity}.

\subsection{Perturbations in the Poisson gauge}
The expressions obtained so far never actually used the explicit form of the metric Eq.~\eqref{MetricPoisson}, therefore they are valid for cosmological perturbations in any gauge.
However, by using the Poisson gauge we can rewrite Eq.~\eqref{AmplitudeFirstOrder}
 in a  more physically transparent and understandable way.
As far as the phase is concerned, its corrections will not change in this particular gauge, hence we focus on the GW amplitude only.

The corrections $\delta k^\mu$ are related to the two scalar gravitational potentials $\phi$ and $\psi$ via the geodesic equation satisfied by the wave-vector. At first order it is
\begin{align}
    \frac{\text d}{\text d \Bar{\chi}} \delta k^\mu(\Bar{\chi}) + \delta \hat{\Gamma}^\mu_{\alpha\beta}\Bar{k}^\alpha (\Bar{\chi}) \Bar{k}^\beta (\Bar{\chi}) = 0\,,
\end{align}
where $\delta \hat{\Gamma}^\mu_{\alpha\beta}$ are given by
\begin{equation}
    \begin{matrix}
    \delta \hat{\Gamma}^0_{00} = \phi'\,, & \delta \hat{\Gamma}^i_{00} = \bar{\partial}^i \phi\,, & \delta \hat{\Gamma}^0_{i0}= \bar{\partial}_i\phi\,, \\
    \delta \hat{\Gamma}^0_{ij} = -\delta_{ij}\psi'\,, & \delta \hat{\Gamma}^i_{j0} = - \delta^i_j \psi'\,, & \delta \hat{\Gamma}^i_{jk} = -\delta^i_k \bar{\partial}_j \psi-\delta^i_j \bar{\partial}_k \psi +\delta_{jk} \bar{\partial}^i \psi\,.
    \end{matrix}
\end{equation}
We can then derive  the set of equations
\begin{equation}\label{EqGeodFirstOrder}
     \frac{\text d }{\text d \Bar{\chi}} (\delta \nu- 2\phi) =  \phi' + \psi' \,,\qquad\qquad
     \frac{\text d }{\text d \Bar{\chi}} (\delta n^i- 2 \hat n^i \psi) = -\Bar{\partial}^i (\phi + \psi)\,.
\end{equation}
Moreover, in the observer frame
\begin{align}
     E_{\Hat{0}\mu} \hat{k}^\mu = -(1+\phi)(-1+\delta \nu )+v_i(\hat n^i + \delta n^i) = 1 - \delta \nu + \phi + v_{||}\,,
\end{align}
therefore the condition $(E_{\hat{0}\mu}\,\hat{k}^\mu)|_{o} = 1$ sets
\begin{equation}\label{InitialConditionFrequency}
     \delta \nu_{o} = \phi_{o} + v_{|| o}\,,
\end{equation}
which we need as initial condition in order to integrate Eqs.~\eqref{EqGeodFirstOrder}.
The initial condition for the spatial part of the wave vector is
\begin{equation}\label{InitialConditionSpatial}
\delta n^{\Hat{a}}_{o} = -v^{\Hat{a}}_{o} + \hat n^{\Hat{a}}\psi_{o} \quad \rightarrow \quad \delta n^i_{o} = -v^i_{o} + \hat n^i\psi_{o}
\,.
\end{equation}
Using Eqs.~\eqref{InitialConditionFrequency}--\eqref{InitialConditionSpatial}, Eq.~\eqref{EqGeodFirstOrder} can be integrated
\begin{align}
    &\delta k^0 =  \delta \nu = 2\phi - (\phi_{o}-v_{|| o}) + \int^{\Bar{\chi}}_0 d \Tilde{\chi}\, (\phi' + \psi')\,,\\
    &\delta k^i = \delta n^i = - v^i_{o} - \hat n^i \psi_{o} + 2 \hat n^i\psi -\int^{\Bar{\chi}}_0 d \Tilde{\chi}\, \Bar{\partial}^i (\phi + \psi) = \hat{n}^i \delta n_{||} + \delta n^i_{\perp}\,,
\end{align}
and
\begin{equation}
    \delta n_{||} = \phi_o - v_{|| o} - \phi +\psi +2I\,, \qquad \delta n^i_\perp = - v^i_{\perp o} + 2 S^i_\perp\,,
\end{equation}
where we defined
\be
I = -\frac{1}{2}\int^{\Bar{\chi}}_0 d \Tilde{\chi}\,(\phi' + \psi')\,, \qquad S^i_\perp = -\frac{1}{2} \int^{\bar{\chi}}_0 d \Tilde{\chi} \bar{\partial}^i_\perp (\phi+ \psi)\,.
\ee
We recognize in $I$ the integrated  Sachs-Wolfe effect. Furthermore we have
\begin{equation}
    \Bar{\partial}_0\,(\delta k^0 + \delta k_{||}) =  \phi' + \psi'\,,
    \qquad \delta \Gamma^\rho_{\,\rho\mu} \Bar{k}^\mu = -\phi' + 3\psi' + \hat n^i \bar{\partial}_i \phi - 3  \hat n^i \bar{\partial}_i \psi = \frac{\text d \phi}{\text d \bar{\chi}} -3 \frac{\text d \psi}{\text d \bar{\chi}}\,.
   \end{equation}
With this expressions we can find $\delta \ln \mathcal{A}^T$ in the  Poisson gauge by integrating Eq.~\eqref{AmplitudeFirstOrder}
\begin{align}
    \delta\ln \mathcal{A}^T = \delta\ln \mathcal{A}_{o} + \kappa + \psi - \psi_{o} -\frac12 \bigg[ \bigg(\frac{F_{\varphi0}}{F_{0}}\delta \varphi \bigg )(\Bar{\chi}) - \left(\frac{F_{\varphi0}}{F_{0}}  \delta \varphi \right)_{o}\bigg]\,, \nonumber \\
\end{align}
where we used $\kappa_{o} = I_{o} = 0$ as it should be as these two are integrated effects. 
In order to obtain $\Delta \ln \mathcal{A}^T$ as in Eq.~\eqref{DeltaAmplitude}, we need also $\Delta \ln a$ and $T$, respectively they are
\begin{align}
    & \Delta \ln a = \mathcal{H} \Delta x^0 = E^{(1)}_{\hat{0} \mu}\bar{k}^\mu + E_{\hat{0} \mu}^{(0)}\delta k^\mu = (\phi_{o} - v_{|| o}) - (\phi - v_{||}) +2I\,,  \label{hDeltaxzero}\\
    &T = - \int^{\bar{\chi}}_0 d \Tilde{\chi}(\phi + \psi)\,.
\end{align}
In this way, the total correction to $\A^T$ is\footnote{We have also used that by construction $\delta\ln \mathcal{A}^T_{o} = \psi_{o} +\frac12 \left(\frac{F_{\varphi0}}{F_{0}}  \delta \varphi \right)_{o}$}
\begin{multline} \label{DeltaAmplitudePoisson}
    \Delta \ln \mathcal{A}^T(\bar{\chi}) =  \kappa + \psi -\frac12 \left(\frac{F_{\varphi0}}{F_{0}}  \delta \varphi \right)(\bar{\chi})-\frac{1}{\bar{\chi}}\int^{\bar{\chi}}_0 d\Tilde{\chi}\,(\phi + \psi)\\
    -\bigg(\phi_{o} - v_{|| o} -\phi + v_{||} +2I\bigg)\left(\,1 + \frac{F_{\varphi0} \varphi_0'}{2 F_{0}\mathcal{H}} - \frac{1}{\Bar{\chi}\mathcal{H}}\right)\,.\\
\end{multline}

\noindent{The} weak-lensing convergence term was $\kappa$ can be also written in the more familiar way
\begin{align}
    \kappa &= - \frac12\bar{\partial}_{i\,\perp} \Delta x^i_\perp =\frac12 \int^{\bar{\chi}}_0 d\Tilde{\chi} \,\frac{(\bar{\chi}-\Tilde{\chi})}{\Tilde{\chi} \bar{\chi}} \triangle _{\Omega}(\phi+\psi)\,, \nonumber\\&
\end{align}
where $\triangle_{\Omega}= \Tilde{\chi}^2 \bar{\nabla}^2_\perp$, using
\begin{align}
    \Delta x^i_\perp &=  \delta x^i_\perp = \int^{\bar{\chi}}_0 d\chi \, \delta n^i_\perp = -v^i_{\perp o} \bar{\chi} - \int^{\bar{\chi}}_0 d\Tilde{\chi} \,(\bar{\chi}-\Tilde{\chi})\Tilde{\partial}^i_\perp (\phi+\psi)\,.
\end{align}

\subsection{The luminosity distance}
We can use the above results to write the full form of the gravitational wave in the geometrical optics limit
\begin{equation}
    \bar{h}^{TT}_{\mu\nu} = \mathcal{A}^T e^{TT}_{\mu\nu}\,e^{i \theta / \epsilon} = \Bar{\mathcal{A}}^T(\bar{x}^0,\bar{\chi})\bigg(1+\Delta \ln \mathcal{A}^T\bigg) e^{TT}_{\mu\nu}\,e^{i (\Bar{\theta}(\bar{\chi}) + \Delta \theta) / \epsilon} \,.
\end{equation}
The amplitude at emission is given by
\begin{align}
    \mathcal{A}^T(\bar{\eta}_e, \Bar{x}^i_e) &= \Bar{\mathcal{A}}^T\bigg(1+\Delta \ln \mathcal{A}^T\bigg) = \frac{\mathcal{Q}\,(1+z)^2}{\Bar{D}^{GR}_L\sqrt{F_0}}\bigg(1+\Delta \ln (\mathcal{A}^T)\bigg)\, ,
\end{align}
where we used Eq.~\eqref{AmplitudeZero} and $\Bar{a}(\eta_e) = (1+z)^{-1}$. $\bar{D}^{GR}_L = (1+z)\bar{\chi}$ is the observed average luminosity distance taken over all the sources with the same observed redshift in GR.
We can define the following quantities
\begin{align}
    &\bar{D}^{gw}_L = \Bar{D}^{GR}_L\sqrt{F_0}\,, \label{DlGWBarra}\\
    & D^{gw}_L = \frac{\Bar{D}^{GR}_L\sqrt{F_0}}{(1+\Delta \ln \mathcal{A}^T)} = \bar{D}^{gw}_L (1-\Delta \ln \mathcal{A}^T)\,,
\end{align}
therefore, the relative correction to the gravitational luminosity distance is
\begin{equation}\label{DeltaDLGW}
    \frac{\Delta D^{gw}_L}{\Bar{D}^{gw}_L} = \frac{D^{gw}_L - \Bar{D}^{gw}_L}{\Bar{D}^{gw}_L} = - \Delta \ln \mathcal{A}^T(\bar{\chi})\,.
\end{equation}
The gravitational wave at the detector is also red-shifted.
We can use Eq.~\eqref{DeltaDLGW} and Eq.~\eqref{DeltaAmplitudePoisson} to write
\begin{align}
     \frac{\Delta D^{gw}_L}{\bar{D}^{gw}_L} &= \bigg(1-\frac{1}{\mathcal{H}\bar{\chi}} +\frac{F_{\varphi0} \varphi_0'}{2 F_0\mathcal{H}} \bigg) v_{||} - \frac12 \int^{\bar{\chi}}_0 d\Tilde{\chi} \,\frac{(\bar{\chi}-\Tilde{\chi})}{\Tilde{\chi} \bar{\chi}} \triangle _{\Omega}\, (\phi+\psi)  \nonumber\\
    &+\phi \bigg(\frac{1}{\Bar{\chi}\mathcal{H}} -\frac{F_{\varphi0} \varphi_0'}{2 F_0\mathcal{H}}\bigg) - \left(\,1 + \frac{F_{\varphi0} \varphi_0'}{2 F_0\mathcal{H}} - \frac{1}{\Bar{\chi}\mathcal{H}}\right) \int^{\bar{\chi}}_0 d\Tilde{\chi}\,(\phi' + \psi') \nonumber\\
    &- \bigg(\phi+\psi- \frac{F_{\varphi0}}{2F_0}\delta \varphi(\bar{\chi}) \bigg) +\frac{1}{\bar{\chi}}\int^{\bar{\chi}}_0 d\Tilde{\chi}\,(\phi + \psi)\,,\label{DeltaDL}
\end{align}
\noindent{where} we dropped the unobservable constant contribution evaluated at the observer position.
If we choose $F = const$, 
 we recover the correct limit to GR, which can be found in \cite{Bertacca:2017vod}; on the other hand, Eq.~\eqref{DeltaDL} shows how modified gravity alters the GW luminosity distance.

In fact, in Eq.~\eqref{DeltaDL} it is possible to distinguish six different contributions:
\begin{itemize}
    \item a peculiar velocity contribution, explicitly modified by dynamical dark energy;
    \item a weak-lensing contribution, not explicitly altered;
    \item a Sachs-Wolfe effect, explicitly modified;
    \item an integrated Sachs-Wolfe effect, explicitly modified;
    \item volume effects, explicitly modified;
    \item a contribution from Shapiro time delay, not explicitly altered dynamical dark energy.
\end{itemize}

\noindent{We} stress that, on top of the explicit modifications in Eq.~\eqref{DeltaDL}, due to the presence of the extra degree of freedom, also the dynamics of $\phi$ and $\psi$ is also potentially  different with respect to GR.

\subsection{The polarization tensor}
We turn now to the evolution equation of the polarization tensor ${\bf e}_{\mu\nu}$.
In order to avoid cumbersome computations we will focus on the subclass in which $G_X = 0$, which is compatible with the gauge conditions given by Eqs.~\eqref{polttra}--\eqref{citra}. \ Only in this sub-case we can invoke screening mechanisms to fix ${\cal A}^S = 0 $ so that Eq.~\eqref{nopolt1} becomes
\be
k^\rho \bar{\nabla}_\rho {\bf e}_{\mu\nu} = 0
\label{nopolt1_simpler}
\ee
After the conformal transformation, Eq.~\eqref{ConformalTransform}, Eq.~\eqref{nopolt1_simpler} becomes
\begin{equation}
    \hat{k}^\rho \hat{\nabla}_\rho ({\bf e}_{\mu\nu} a^{-2}) = 0 \label{EqPolarizationConformal}
\end{equation}
In the transformed spacetime the polarization ${\bf e}_{\mu\nu}$ has to satisfy the constraint
\be \label{gaugeVeCConformal}
\hat{k}^\mu {\bf e}_{\mu\nu} = 0\,,
\ee
coming from the harmonic gauge.
 As we did for the phase and the amplitude of the GW, we expand the polarization tensor as
 \begin{align}
  {{\bf e}}_{\mu\nu}(x^\mu(\chi)) = {{\bf e}}_{\mu\nu}(\Bar{x}^\mu+ \Delta x^\mu) &=  \bar{{\bf e}}_{\mu\nu}(\bar{x}^\mu (\bar \chi)) + \Delta {\bf e}_{\mu\nu} (\Bar{x}^\mu(\bar \chi)) = \nonumber \\
   & = \bar{{\bf e}}_{\mu\nu}(\Bar{x}^\mu(\bar \chi))  + \delta {{\bf e}}_{\mu\nu}(\Bar{x}^\mu(\bar \chi))  + \Delta x^\rho \bar \partial_\rho   \bar{{\bf e}}_{\mu\nu}(\bar{x}^\mu (\bar{\chi}))
\label{DeltaE}
 \end{align}
where $\bar{{\bf e}}_{\mu\nu}$ is the value that ${\bf e}_{\mu\nu}$ would have on a unperturbed space-time, while $\Delta {\bf e}_{\mu\nu} =  \delta {{\bf e}}_{\mu\nu}  +\Delta x^\rho \bar \partial_\rho   \bar{{\bf e}}_{\mu\nu}$ is the linear correction due to the presene of inhomogeneities. As in the previous cases, we defined $\delta {{\bf e}}_{\mu\nu}(\Bar{x}(\bar \chi)) \coloneqq {{\bf e}}_{\mu\nu}(\Bar{x}(\bar \chi)) - \bar {{\bf e}}_{\mu\nu}(\Bar{x}(\bar \chi))$.
The expansion of Eq.~\eqref{EqPolarizationConformal} at linear order is
\begin{align}
    \quad\hat{k}^\rho\hat{\nabla}_\rho ({\bf e}_{\mu\nu} a^{-2})&= \frac{\text d}{\text d {\chi}} ({\bf e}_{\mu\nu} a^{-2}) = \left(1 -\frac{\text d \delta \chi}{\text d \chi}\right) \frac{\text d}{\text d \Bar{\chi}}\left((\Bar{{\bf e}}_{\mu\nu} + \Delta {\bf e}_{\mu\nu}) \Bar{a}^{-2}(1-2 \Delta \ln a) \right) = \nonumber\\ \nonumber \\
    & = \frac{\text d(\Bar{{\bf e}}_{\mu\nu}\Bar{a}^{-2})}{\text d \Bar{\chi}} + \frac{\text d(\delta {\bf e}_{\mu\nu}\Bar{a}^{-2})}{\text d \Bar{\chi}} + \frac{\text d(\delta x^\rho \bar \partial_\rho (\Bar{ {\bf e}}_{\mu\nu}\Bar{a}^{-2}))}{\text d \Bar{\chi}} + \delta \chi \frac{\text d^2(\Bar{{\bf e}}_{\mu\nu}\Bar{a}^{-2})}{\text d \Bar{\chi}^2}\,,
\label{DeltaEmunu}
\end{align}
where we have considered that in $\hat \nabla_\rho$ the connection coefficients are of linear order in the metric perturbation as $\hat \nabla_\rho$ is the covariant derivative associated to the metric $\hat g_{\mu\nu} = \eta_{\mu\nu} + \delta \hat g_{\mu\nu}$. Therefore in Eq.~\eqref{DeltaEmunu}, $\Delta x^\rho \hat{\nabla}_\rho (\Bar{ {\bf e}}_{\mu\nu}\Bar{a}^{-2}) = \Delta x^\rho \bar \partial_\rho (\Bar{ {\bf e}}_{\mu\nu}\Bar{a}^{-2})$ at the order we are restricting ourselves to, as $\Delta x^\rho$ is already a first order quantity.
Consequently, the  zeroth order of Eq.~\eqref{EqPolarizationConformal} is
\be
\frac{\text d(\Bar{{\bf e}}_{\mu\nu}\Bar{a}^{-2})}{\text d \Bar{\chi}} = 0 \quad \rightarrow \quad \Bar{{\bf e}}_{\mu\nu} =  \Bar{a}^{2} {\cal Q}_{\mu\nu} \label{ZeroOrderPolGzero}
\ee
where ${\cal Q}_{\mu\nu}$ is a constant, transverse tensor.
The linear order of Eq.~\eqref{EqPolarizationConformal}'s expansion is
\begin{align}
\frac{\text d(\delta {\bf e}_{\mu\nu}\Bar{a}^{-2})}{\text d \Bar{\chi}} &+ \frac{\text d(\delta x^\rho \bar \partial_\rho (\Bar{ {\bf e}}_{\mu\nu}\Bar{a}^{-2}))}{\text d \Bar{\chi}} + \delta \chi \frac{\text d^2(\Bar{{\bf e}}_{\mu\nu}\Bar{a}^{-2})}{\text d \Bar{\chi}^2} = 0
\end{align}
The latter can be simplified using the result of Eq.~\eqref{ZeroOrderPolGzero}
\begin{align}
\frac{\text d(\delta {\bf e}_{\mu\nu}\Bar{a}^{-2})}{\text d \Bar{\chi}} = 0 \,.\label{FinalPolFirstOrder}
\end{align}
This equation allows us to obtain $\delta {\bf e}_{\mu\nu}\Bar{a}^{-2}$ by integration.
Then, the full correction to the GW polarization, i.e. $\Delta {\bf e}_{\mu\nu}(\bar{\chi})$ is given by
\begin{equation}
    \Delta {\bf e}_{\mu\nu} = \delta {\bf e}_{\mu\nu} + 2 {\cal H} \Delta x^0\,\bar{\bf e}_{\mu\nu}\,.
\end{equation}
where we used Eq.~\eqref{ZeroOrderPolGzero} and that Christoffel symbols are null at zero order. We point out that $ {\cal H} \Delta x^0$ contains, as given in Eq.~\eqref{hDeltaxzero}, effects such as the ISW and doppler.
We plan to further study the behavior of the polarization tensor in the case of braiding, where it is not parallel-transported in future works.

\section{Conclusions}

The detection of gravitational waves (GWs) propagating over cosmological
distances will offer new opportunities for probing cosmological structures in
our Universe, as well as for testing possible departures from GR.
In this work we established tools for studying GW
propagation in scalar-tensor theories of gravity, adopting a geometrical ansatz, and focusing
on high frequency spin-2 and spin-0 modes travelling at the light speed on slowly-varying backgrounds, gauging away all the spurious modes.
Our approach can apply to scenarios with non-minimal couplings between scalar and tensor
degrees of freedom: we took particular care in discussing and motivating our gauge choices-- {see Eqs.~\eqref{gu1a}--\eqref{gd1a}} -- and
to develop general arguments that do not rely on the particular choice of the scalar-tensor theory.

We determined the general structure of the evolution equations for the GW amplitude and polarization
 tensor.
We found that  they can be different with respect to  GR -- see in particular Eqs.~\eqref{res1A}--\eqref{res2A}, {and Eqs.~\eqref{evetpa}--\eqref{seceqS}}.  In theories
 which preserve the graviton number,
   the equation for the amplitude of GWs  can differ from GR if the effective Planck scale varies
   over the GW geodesics. Interestingly, we also find that the GW polarization tensor can fail to be
    parallel  transported along the GW geodesics,
and discussed
  physical interpretations of our results.
We applied our general formulas to a representative
 example of scalar-tensor scenarios {considering, for the first time in the literature, scalar waves, SWs, and tensor waves, GWs, both travelling with unit speed}.
In this context, Eqs.~\eqref{evetpa2}--\eqref{pol_tot} represent the most general results of this work.
 They show  that  the GW propagation depends both on background quantities and the assumed scalar-tensor theory, that the amplitude $\mathcal{A}^T$ of the tensor mode does not depend on its scalar counterpart $\mathcal{A}^S$,
and,
  that the tensor and scalar modes of polarization are generally coupled to each other. See also \cite{Dalang et al. 2020a} for a recent paper focussing on the propagation of tensor modes.

An important advantage of our approach is that it can be applied to study the propagation of GWs
 on perturbed cosmological space-times, provided that the wavelength of cosmological fluctuations is well larger
 than the GW wavelength. In fact, under the simplified assumption that $\mathcal{A}^S$ is suppressed at the GW emission and absence of braiding, we derived the general evolution equations controlling GW propagation
 in such a set-up, distinguishing and identifying the distinct contributions associated with cosmological
 fluctuations, and with scalar-tensor effects.

More precisely, {at first order in perturbation theory}, we determined
%
%
%
%
several corrections to standard formulas for the GW luminosity distance --  {Eq.~\eqref{DeltaDL}}.
We found that the contribution to the GW luminosity distance from the peculiar velocity, Sachs-Wolfe, integrated Sachs-Wolfe and volume effects {are explicitly modified in the presence of scalar-tensor interactions. On the other hand, weak-lensing and Shapiro time delay are only implicitly altered, via the modified dynamics of $\phi$ and $\psi$, due to the presence of the extra degree of freedom.

Finally, under the assumption $G_X=0$ and $\A^S = 0$, we obtained the evolution of the transverse polarization tensor, Eq.~\eqref{FinalPolFirstOrder}, in the linear regime.
We argued that only in this specific sub-case the polarization is parallel transported along the geodesics of the GW. We plan to investigate the dynamics of the polarization in presence of braiding, where it is not parallel transported, in a future work.

\smallskip
Starting from the general results presented here,
much work is left for the future. It would be interesting to apply our method to more general scalar-tensor
theories than the representative examples discussed here, or to extend our approach
to  modified gravity scenarios with multiple fields. It would be interesting to study in more detail the physical
consequences of our findings for GWs propagating over cosmological distances, and analyze in detail
degeneracies between modified gravity and effects of non-linearities as well as understanding the effect of the the self-induced GW energy momentum tensor in theories alternative to GR.  We hope to return soon to these
topics.

\subsection*{Acknowledgments}


It is a pleasure to  thank  Nicola Bartolo, Giulia Cusin, Charles Dalang, Pierre Fleury, Alexander Ganz, Luigi Guzzo, Lucas Lombriser, Angelo Ricciardone and Nicol\'{a}s Yunes for useful discussions. C.C. warmly thanks Andrew~J.~S. Hamilton for very useful clarifications.
  The work of G.T. is partially supported by STFC grant ST/P00055X/1. D.B. and S.M. acknowledge partial financial support by ASI Grant No.
2016-24-H.0.

\bigskip
\begin{appendix}
\section{Conditions to ensure a unit  speed for scalar fluctuations}
\label{app-scal-vel}


In this Appendix we briefly  discuss some constraints  that the compatibility
condition in Eq.~\eqref{CompatibilityCondition1} imposes
on the scalar-tensor systems and the background solutions under consideration.
For definiteness, we consider
the scalar-tensor theory
\bea \label{LagTot1}
{\cal L}^{tot}&=&F(\varphi)\,R + G(\varphi,\,X) \,\Box \varphi + K(\varphi, X)\,,
\eea
and derive a constraint on $K$ and $G$ that will guarantee the compatibility condition to be satisfied.
In order to do se we linearize to first order in $\varphi_r$ and $\hat h_{\mu\nu}$ the scalar field equation, $\partial {\cal L}^{tot} / \partial \varphi = 0$
\begin{multline}
    (\bar{F}_\varphi \bar{g}^{\mu\nu} - \bar{G}_X \bar{\varphi}^\mu\bar{\varphi}^\nu) \left[ R^{(1)}_{\mu\nu}\right]_{\epsilon^{-2}} + \\ + \frac{(\varphi_{\mu\nu})^{(1)}_{\epsilon^{-1}}}{\bar{F}} \bigg[ \:\bar{g}^{\mu\nu}(3 \bar{F}^2_\varphi+ 2 \bar{X} \bar{F}_\varphi \bar{G}_X  + \bar{X}^2\bar{G}^2_X +\bar{F} A )  - \frac{\bar{F} \bar{G}_{XX}}{2} \left(\bar{X}^\mu \bar{\varphi}^\nu + \bar{X}^\nu \bar{\varphi}^\mu \right)- \\- \bar{\varphi}^\mu \bar{\varphi}^\nu (  2\bar{F}_\varphi \bar{G}_X +\bar{X} \bar{G}^2_X + \bar{F} B)  -2\bar{F}\bar{G}_X \bar{\varphi}^{\mu\nu} \bigg] + {\cal O}(\epsilon^0)=0 \\
\end{multline}
with $A =(\bar \varphi^\mu \bar\nabla_\mu \bar G_X + \bar K_X + 2\bar G_\varphi+ 2\bar G_X \bar \Box \bar \varphi)$ , $B =( \bar K_{XX} + 2 \bar G_{X\varphi} + \bar G_{XX} \bar \Box \bar \varphi)$ and $\varphi_\nu = \bar{\nabla}_\nu \varphi$, $\varphi_{\mu\nu} = \bar{\nabla}_\nu \bar{\nabla}_\mu \varphi$ .
At leading order, ${\cal O} (\epsilon^{-2})$, the equation above gives
\be
\left[ R^{(1)}_{\mu\nu}\right]_{1 / \epsilon^{2}} = 0\,,
\ee
which is consistent with the gravitational field equations. At next-to-leading order, ${\cal O} (\epsilon^{-1})$,
\begin{multline}
    \frac{(\varphi_{\mu\nu})^{(1)}_{\epsilon^{-1}}}{\bar{F}} \bigg[ \:\bar{g}^{\mu\nu}(3 \bar{F}^2_\varphi+ 2 \bar{X} \bar{F}_\varphi \bar{G}_X  + \bar{X}^2\bar{G}^2_X +\bar{F} A ) - \frac{\bar{F} \bar{G}_{XX}}{2} \left(\bar{X}^\mu \bar{\varphi}^\nu + \bar{X}^\nu \bar{\varphi}^\mu \right)- \\- \bar{\varphi}^\mu \bar{\varphi}^\nu (  2\bar{F}_\varphi \bar{G}_X +\bar{X} \bar{G}^2_X + \bar{F} B)  -2\bar{F}\bar{G}_X \bar{\varphi}^{\mu\nu} \bigg] =0\,,
\end{multline}
which, in the case $\bar K_{XX}=0$ and $G_X=0$ reduces to
\be \label{EQvarphiNoBraiding}
 \frac{(3 \bar{F}^2_\varphi + \bar{F} \bar{K}_X+ 2\bar{F} \bar{G}_\varphi)}{\bar{F}}  (\Box \varphi)^{(1)}_{1 / \epsilon^1}  = 0 \,.
\ee
We can safely assume $\bar{F} \bar{K}_X \neq 0$, otherwise neither $g_{\mu\nu}$ nor $\varphi$ have a kinetic term in the Lagrangian to begin with, so that
\be
(\Box \varphi)^{(1)}_{1 / \epsilon^1}  = 0\,,
\ee
so that the compatibility of the gauge choices is guaranteed by the equation of motion.
This result can be generalized to a group of theories larger than those satisfying $G_X =0$ by choosing $G(\varphi, X)$ such that
\be
   \frac{\bar{F} \bar{G}_{XX}}{2} \left(\bar{X}^\mu \bar{\varphi}^\nu + \bar{X}^\nu \bar{\varphi}^\mu \right) + \bar{\varphi}^\mu \bar{\varphi}^\nu (  2\bar{F}_\varphi \bar{G}_X +\bar{X} \bar{G}^2_X + \bar{F} B)  + 2\bar{F}\bar{G}_X \bar{\varphi}^{\mu\nu} = 0\,.\\
\ee
{It is important to point out that this condition is quite restrictive and not only sets constraints
on the structure of the functions $F$, $G$, and $B$, but also on the slowly-varying background profile of $\bar \varphi$. Depending on the details of the system,
one would need to investigate whether this condition can be satisfied
and is compatible with the scalar equations of   motion, which can also include additional contributions from couplings
of the slowly-varying scalar with additional degrees of freedom. A complete
analysis of this issue goes beyond the scope of this work.}


\section{Defining spin-2 and spin-0 polarization tensors using tetrads}\label{app_A}
This Appendix aims to explain the method we use in the main text for disentangling the evolution equations for the amplitude of tensor and scalar modes.

We  introduce null tetrads  as in \cite{Dolan:2018ydp,Dolan:2018nzc}
\be
z_a^\mu\,=\,\left( k^\mu,\,n^\mu,\,m^\mu,\,\bar m^\mu\right)\,;
\ee
 $k^\mu$ and $n^\mu$ are
real,
$m^\mu$ complex, and $\bar m^\mu$ its complex conjugate. These vectors  are null with respect to background metric $\bar g_{\mu\nu}$, that is (raising/lowering with metric $\bar g_{\mu\nu}$):
\be
0\,=\,k_\mu\,k^\mu\,=\,m_\mu\,m^\mu\,=\,{\bar m}_\mu\,{\bar m}^\mu\,=\,n_\mu\,n^\mu
\ee
The only other non-vanishing products are
\be
\bar g_{\mu\nu}\,m^\mu\,\bar m^\nu\,=\,1 \qquad \mbox{and}\qquad \bar g_{\mu\nu}\,k^\mu\,n^\nu\,=\,-1
\ee
The vector $k^\mu$ is identified with the GW direction.
We can decompose a rank 2 symmetric tensor on the null-tetrad basis as
\begin{align}
    A_{\mu\nu} = \alpha_{AB}\,\Theta^{AB}_{\mu\nu}\label{NullDec}
\end{align}
with
\be
\Theta^{AB}_{\mu\nu} = \frac12 (A_\mu B_\nu + A_\nu B_\mu )\qquad\mbox{and}\qquad A_\mu,B_\mu = ( k_\mu, n_\mu, m_\mu, \bar{m}_\mu)\,,
\ee
which shows that $ \A_{\mu\nu} $ has ten independent component as it should be for a rank 2 symmetric tensor.  Using the orthogonality relations between the elements of $z_a^\mu$, it is possible to show that the objects  $\Theta^{AB}_{\mu\nu} $ satisfy
\bea\label{thetaprop}
&& \Theta^{mm}_{\mu\nu} \Theta_{\bar{m}\bar{m}}^{\mu\nu} = \Theta^{\bar{m}\bar{m}}_{\mu\nu} \Theta_{mm}^{\mu\nu}=  \Theta^{nn}_{\mu\nu} \Theta_{kk}^{\mu\nu} =  1\,, \quad   \Theta^{nm}_{\mu\nu} \Theta_{k \bar m}^{\mu\nu}= \Theta^{n \bar{m}}_{\mu\nu} \Theta_{ k m}^{\mu\nu} = - \Theta^{m\bar{m}}_{\mu\nu} \Theta_{m\bar{m}}^{\mu\nu} = - \frac12 \,,  \nonumber \\
\eea
while the other products vanish.
Since we are interested in the geometric-optics limits to study the propagation of GWs, we apply the null-tetrad formalism to $\A_{\mu\nu}$  which is defined as $\hat h_{\mu\nu} = \A_{\mu\nu} e^{i \theta / \epsilon} $ and satisfy the gauge conditions given in Eqs.~\eqref{polttra}, \eqref{gaugeCb} and \eqref{citra}.
In this case, the tensor $\A_{\mu\nu}$ represents a null-wave so the ten independent components have a clear physical meaning as they are related to the Newmann-Penrose scalars (see \cite{Eardley:1974nw}).
In particular
\bea
 \Psi_2 = \frac{1}{12} \ddot{\A}_{kk} &\:\rightarrow\:& \mbox{Longitudinal scalar mode s = 0}\,, \in \mathbb{R}\\
 \Phi_{22} = \frac12 \ddot{\A} _{m\bar{m}}&\:\rightarrow\: &\mbox{Transverse scalar mode  s= 0}\,, \in \mathbb{R}\\
 \Psi_3 = \frac14 \ddot{\A}_{km}&\:\rightarrow\:&\mbox{Vector, mixed scalar/longitudinal, s}  = \pm 1 \,, \in \mathbb{C} \rightarrow \exists \bar{\Psi}_3\\
 \Psi_4 = \frac12 \ddot{\A}_{mm}&\:\rightarrow\:&\mbox{Tensor, transverse and traceless , s} \pm 2\,, \in \mathbb{C} \rightarrow \exists \bar{\Psi}_4
\eea
while $\Theta^{nn}_{\mu\nu}\,\,, \Theta^{nk}_{\mu\nu}\,,\, \Theta^{nm}_{\mu\nu}$ and $\Theta^{n \bar m}_{\mu\nu}$ are not associated to any physical observable. In fact it is possible to show that the condition $k^\mu \A_{\mu\nu} = 0$ implies $\alpha_{nn} = \alpha_{nk} =\alpha_{nm} =\alpha_{n \bar m} =0  $ because the only non vanishing scalar product is $k^\mu n_\mu = -1$.
So a null transverse wave actually has only six independent components
\begin{align}
    \A_{\mu\nu} &=\alpha_{kk}\,\Theta^{kk}_{\mu\nu}+\alpha_{mm}\,\Theta^{mm}_{\mu\nu}+\alpha_{\bar{m}\bar{m}}\,\Theta^{\bar{m}\bar{m}}_{\mu\nu}
     +\alpha_{km}\,\Theta^{km}_{\mu\nu}+\alpha_{k\bar{m}}\,\Theta^{k\bar{m}}_{\mu\nu}+ \alpha_{m\bar{m}}\,\Theta^{m\bar{m}}_{\mu\nu} =  \nonumber\\
     &= \alpha_{CD}\,\Theta^{CD}_{\mu\nu}\,, \label{NullDecompositionTT}
\end{align}
with $C_\mu,D_\mu = ( k_\mu, m_\mu, \bar{m}_\mu)$.
In the main text we used ${\bf e}^+_{\mu\nu}$, ${\bf e}^\times_{\mu\nu}$ and  ${\bf e}^S_{\mu\nu}$ which are defined as
\be \label{polarizationtotetrad}
{\bf e}^+_{\mu\nu} = \frac{1}{\sqrt{2}} \left( \Theta^{mm}_{\mu\nu} + \Theta^{\bar m \bar m}_{\mu\nu}\right) \,,\qquad {\bf e}^\times_{\mu\nu}= \frac{i}{\sqrt{2}} \left( \Theta^{\bar m \bar m}_{\mu\nu} - \Theta^{mm}_{\mu\nu}\right) \,, \qquad {\bf e}^S_{\mu\nu}=\sqrt{2}  \Theta^{m \bar m}_{\mu\nu}\,.
\ee
and they are all normalized to $1$ and they are orthogonal to each other.
The amplitudes $\A^+$, $\A^\times$  and $\A^S$ are related to $\alpha_{CD}$ as
\be \label{amplitudestoalpha}
\A^+ =\frac{1}{\sqrt{2}} \left( \alpha_{mm} + \alpha_{\bar m \bar m} \right)  \,, \qquad \A^\times =\frac{i}{\sqrt{2}} \left( \alpha_{mm} - \alpha_{\bar m \bar m} \right)  \,, \qquad  \mbox{and} \qquad \A^S = \frac{\alpha_{m \bar m}}{\sqrt{2}}\,.
\ee

\subsection{Gauge conditions and null tetrads}\label{app_gauge}
We can use the decomposition of $\A_{\mu\nu}$ on the null-tetrad basis to treat the vector $\Psi_3$ and  scalar-longitudinal mode $\Psi_2$, namely $\alpha_{km}, \alpha_{k \bar m}$ and $\alpha_{kk}$.
In fact, the null tetrad basis allows us to nicely distinguish the non-physical modes that we would like to gauge away.
In Section \ref{sec-gau} of the main text we claimed that we are entitle to make a last gauge transformation using the vector field $\sigma_\mu$, which has to satisfy $\bar \Box \sigma_\mu = \bar v^\mu \sigma_\mu = 0$, to fix 3 components of the metric perturbation. We use this last gauge freedom to choose a particular form of the vector $\C_\mu  =  \bar v^\mu \A_{\mu\nu}$ such that  $\alpha_{km}, \alpha_{k \bar m}, \alpha_{kk}$ are fixed.
From Eq.~\eqref{NullDecompositionTT}  it follows
\bea
\C_\mu &=& \bar v^\mu \A_{\mu\nu} = \bar v^\nu\, \left(\alpha_{CD}\,\Theta^{CD}_{\mu\nu}\right) = \label{CmuFromA} \\
&=& \frac12 \bigg[ \:k_\mu \bigg(2 \alpha_{kk} (\bar v\cdot k) + \alpha_{km} (\bar v \cdot m) + \alpha_{k\bar{m}} (\bar v \cdot \bar{m}) \bigg) + \nonumber \\
&&\quad + m_\mu \bigg(2 \alpha_{mm} (\bar v\cdot m) + \alpha_{km} (\bar v \cdot k) + \alpha_{m\bar{m}} (\bar v \cdot \bar{m}) \bigg)  + \nonumber \\
&&\quad + \bar{m}_\mu \bigg(2 \alpha_{\bar{m}\bar{m}} (\bar v\cdot \bar{m}) + \alpha_{k\bar{m}} (\bar v \cdot k) + \alpha_{m\bar{m}} (\bar v \cdot m) \bigg)\bigg] \,. \label{CmuFromA2}
\eea
We can decompose $\C_\mu$ on the null tetrad basis as
\be
\C_\mu = \C^k k_\mu  + \C^m m_\mu + \C^{\bar{m}} \bar{m}_\mu\,,\label{NullDecompositionT}
\ee
where we already used $\C^n = 0$ because $k^\mu \C_\mu = 0$ as in Eq.~\eqref{citra}.
Now we compare Eq.~\eqref{NullDecompositionT} with Eq.~\eqref{CmuFromA2}  and, using the fact that $(k_\mu, m_\mu, \bar m_\mu)$ constitute a basis we find the system of equations
\bea
2\,\C^k &=& 2 \alpha_{kk} (\bar v\cdot k) + \alpha_{km} (\bar v \cdot m) + \alpha_{k\bar{m}} (\bar v \cdot \bar{m}) \\
2\, \C^m &=& 2 \alpha_{mm} (\bar v\cdot m) + \alpha_{km} (\bar v \cdot k) + \alpha_{m\bar{m}} (\bar v \cdot \bar{m})\\
2 \,\C^{\bar{m}} &=& 2 \alpha_{\bar{m}\bar{m}} (\bar v\cdot \bar{m}) + \alpha_{k\bar{m}} \bar (v \cdot k) + \alpha_{m\bar{m}} (\bar v \cdot m)
\eea
which can be inverted for $\alpha_{kk} ,  \alpha_{mm}$ and $\alpha_{\bar{m}\bar{m}}$ in the case of $(v\cdot k) \neq 0$
  \bea
        \alpha_{km} &=& \frac{1}{(\bar v \cdot k)} \bigg[2\, \C^m - 2 \alpha_{mm} (\bar v\cdot m) -  \alpha_{m\bar{m}} (\bar v \cdot \bar{m})\bigg] \label{alphakm}\\
        \alpha_{k\bar{m}} &=& \frac{1}{(\bar v \cdot k)} \bigg[2 \,\C^{\bar{m}} - 2 \alpha_{\bar{m}\bar{m}} (\bar v\cdot \bar{m}) - \alpha_{m\bar{m}} (\bar v \cdot m)\bigg]\label{alphakmbar}\\
        \alpha_{kk} &=& \frac{1}{(\bar v \cdot k)^2} \bigg[\,\C^k (\bar v \cdot k) - \,\C^m (\bar v \cdot m) -\,\C^{\bar{m}} (\bar v \cdot \bar{m})+ \nonumber\\
        &&\quad\qquad+ \alpha_{mm} (\bar v\cdot m)^2 + \alpha_{\bar{m}\bar{m}} (\bar v\cdot \bar{m})^2 +  \alpha_{m\bar{m}} (\bar v \cdot m) (\bar v \cdot \bar{m})\bigg] \nonumber\\ \label{alphakk}
     \eea
so that  $\alpha_{kk}, \,\alpha_{km}, \,\alpha_{k\bar{m}}$ are entirely written in terms of $\C^k,\, \C^m, \,\C^{\bar{m}}$ and the other free coefficients $\alpha_{mm},\, \alpha_{\bar{m}\bar{m}}, \,\alpha_{m\bar{m}}$.
The assumption $(v\cdot k) \neq 0$ is well justified because
\be
k^\mu \bar v_\mu = k^\mu \bar \nabla_\mu \bar{\varphi} = \frac{\text d \bar{\varphi}}{\text d \lambda} \,,
\ee
and we expect the smooth - slowly varying profile of the scalar field to change along the geodesics.

Using the gauge freedom associated to $\sigma_\mu$ one can fix the three independent components of $\C^\mu$ on the null tetrad basis, namely $\C^k, \C^m$ and $ \C^{\bar{m}}$. The transformation law of $\C^\mu$ under $x^\mu \rightarrow x^\mu + \sigma^\mu$  is given in the main text.  Doing so means that $(\alpha_{kk}, \alpha_{km}, \alpha_{k\bar{m}})$ can be entirely written in terms of $(\C^k,\, \C^m,\,\C^{\bar{m}})$ and $(\alpha_{mm}, \alpha_{\bar m\bar m}, \alpha_{m\bar{m}})$ through Eqs.~\eqref{alphakm} -~\eqref{alphakk}.
Note that we can choose $(\C^k,\, \C^m,\,\C^{\bar{m}})$ such that $\alpha_{kk} = \alpha_{km} = \alpha_{k\bar{m}} = 0$ or  not. In the first case one removes directly the vector and scalar-longitudinal modes form $\A_{\mu\nu}$ but also has to be careful to remove the same modes in the first-order stress energy tensor, $t_{\mu\nu}$. This can be achieved using the equations of motion  Eqs.~\eqref{EQmm1} -~\eqref{EQnn1}. In the second case,  $(\alpha_{kk},\, \alpha_{km},\,\alpha_{k\bar{m}} \neq 0)$, then it is still possible to decouple the dynamics of the physical modes, i.e.  $(\alpha_{mm}, \alpha_{\bar m\bar m}, \alpha_{m\bar{m}})$ from the three gauge ones $(\alpha_{kk}, \alpha_{km} ,\alpha_{k\bar{m}})$.
We show this second procedure in Appendix \ref{app_B1}.

\subsection{Decoupling of amplitude's evolution equations}\label{app_B1}
In the main text we computed the equations of motions for the amplitude and polarization tensor of a GW containing only the spin-2 and transverse spin-0 mode. In particular, we claimed that the amplitude of these modes satisfy Eq.~\eqref{eqAS2} and Eq.~\eqref{eqAT2} while, the total polarization tensor satisfies Eq.~\eqref{evPolNolongitudinal}.
Here we show how to separate the evolution equations of the physical modes from the spurious ones starting from the general decomposition of  $\A_{\mu\nu}$ over the null tetrad basis, namely Eq.~\eqref{NullDecompositionTT}.
In this section we use primes to denote quantities after the removal of the gauge modes  ($\A'_{\mu\nu}$, $\A'$, ${\bf e}'_{\mu\nu}$,..) to make a distinction with the previous part of this Appendix. In the main text from section \ref{sec_sep20} on, we always use $\A'_{\mu\nu}$ and ${\bf e }'_{\mu\nu}$ even though primes are omitted.

\smallskip
We start by plugging the decomposition of  $\A_{\mu\nu}$~\eqref{NullDecompositionTT} in Eq.~\eqref{eomG1A}
\be
\Theta^{CD}_{\mu\nu} \B_{CD} + 2 \alpha_{CD}\,k^\rho \,\bar{\nabla}_\rho \Theta^{CD}_{\mu\nu} = t_{\mu\nu} \label{EOMgeomoptDecomposed}
\ee
where
\be \label{BCD}
\B_{CD} = 2 k^\rho \bar{\nabla}_\rho\,\alpha_{CD} + \alpha_{CD} \bar{\nabla}_\rho k^\rho
\ee
In order obtain separate equations for the different modes, we contract Eq.~\eqref{EOMgeomoptDecomposed} with $\Theta_{AB}^{\mu\nu}$ for different $A,B = \{n , k, n, \bar m \}$. Using the properties~\eqref{thetaprop}, we find
\bea
 0 &=& \Theta_{kk}^{\mu\nu} t_{\mu\nu} \label{EQkk} \\
 0 &=& \Theta_{km}^{\mu\nu} t_{\mu\nu} \label{EQkm} \\
 0 &=& \Theta_{km}^{\mu\nu} t_{\mu\nu} \label{EQkm} \\
 0 &=& \Theta_{k\bar{m}}^{\mu\nu} t_{\mu\nu}\label{EQkbarm} \\
 \B_{\bar{m}\bar{m}} + 4 \alpha_{\bar{m}\bar{m}} k^\rho m^\mu \bar{\nabla}_\rho \bar{m}_\mu &=& \Theta_{mm}^{\mu\nu} t_{\mu\nu}  \label{EQmm}\\
 \B_{mm} + 4 \alpha_{mm} k^\rho \bar{m}^\mu \bar{\nabla}_\rho m_\mu &=&\Theta_{\bar{m}\bar{m}}^{\mu\nu} t_{\mu\nu} \label{EQbarmbarm} \\
  \B_{m\bar{m}} &=& 2\,\Theta_{m\bar{m}}^{\mu\nu} t_{\mu\nu} \label{EQmbarm}\\
     - \frac12 \B_{k \bar m} -\alpha_{k \bar m} k^\rho m^\mu \bar{\nabla}_\rho \bar m_\mu + \alpha_{m \bar m}  k^\rho n^\mu \bar{\nabla}_\rho m_\mu +2 \alpha_{\bar m \bar m} k^\rho n^\mu \bar{\nabla}_\rho \bar m_\mu  &=& \Theta_{n m}^{\mu\nu} t_{\mu\nu} \label{EQmn} \\
                    - \frac12 \B_{k m} -\alpha_{k  m} k^\rho \bar m^\mu \bar{\nabla}_\rho m_\mu + \alpha_{m \bar m}  k^\rho n^\mu \bar{\nabla}_\rho \bar m_\mu +2 \alpha_{ m  m} k^\rho n^\mu \bar{\nabla}_\rho  m_\mu  &=& \Theta_{n\bar{m}}^{\mu\nu} t_{\mu\nu}  \label{EQnbarm}       \\
                  \B_{k k} -  2 \alpha_{km} k^\rho n^\mu \bar{\nabla}_\rho m_\mu - 2 \alpha_{k \bar m} k^\rho n^\mu \bar{\nabla}_\rho \bar m_\mu &=& \Theta_{nn}^{\mu\nu} t_{\mu\nu} \label{EQnn}
\eea
We can decompose the stress energy tensor on the null tetrad basis as well
\be
t_{\mu\nu} = \T_{CD} \Theta^{CD}_{\mu\nu}
\ee
with $C,D \in (k, m, \bar{m})$ because $t_{\mu\nu}$ is such that $k^\mu t_{\mu\nu} =0$ so that its component on the basis $z^\mu_a$ proportional to $n^\mu$ have to vanish.
We can relate the coefficients $ \T_{CD}$ to suitable contractions of $t_{\mu\nu}$ with the elements $\Theta^{CD}_{\mu\nu}$
\begin{align}
&\Theta_{m\bar{m}}^{\mu\nu} t_{\mu\nu} = \frac12 \T_{m\bar{m}} \qquad  &\Theta_{\bar{m}\bar{m}}^{\mu\nu} t_{\mu\nu} = \T_{mm} \qquad &\Theta_{mm}^{\mu\nu} t_{\mu\nu} =\T_{\bar{m}\bar{m}} \,,\label{componentstprime}\\
&\Theta_{m n}^{\mu\nu} t_{\mu\nu} = -\frac12 \T_{k\bar{m}} \qquad &\Theta_{\bar m n}^{\mu\nu} t_{\mu\nu} = -\frac12 \T_{k m} \qquad &\Theta_{nn}^{\mu\nu} t_{\mu\nu} = \T_{kk} \,,
\end{align}
while the other components are null. Since $t_{\mu\nu}$ is transverse the following relations hold
\be
\Theta_{kk}^{\mu\nu} t_{\mu\nu} = 0 \qquad \Theta_{km}^{\mu\nu} t_{\mu\nu} = 0 \qquad  \Theta_{k\bar{m}}^{\mu\nu} t_{\mu\nu} = 0\,,
\ee
so that we see that Eqs.~\eqref{EQkk} -~\eqref{EQkbarm} are compatible with the condition $k^\mu t_{\mu\nu} =0$.
Eqs.~\eqref{EQmm} -~\eqref{EQnn} in order become
\begin{align}
 &\B_{m m} + 4 \alpha_{mm} k^\rho \bar{m}^\mu \bar{\nabla}_\rho m_\mu = \T_{mm} \label{EQmm1}\\
 &\B_{\bar m\bar m} - 4 \alpha_{\bar{m}\bar{m}} k^\rho \bar{m}^\mu \bar{\nabla}_\rho m_\mu = \T_{\bar{m}\bar{m}}\label{EQbarmbarm1}\\
 &\B_{m \bar m} = \T_{m\bar{m}}\label{EQmbarm1} \\
 &\B_{k \bar m} + 2\alpha_{k \bar m} k^\rho m^\mu \bar{\nabla}_\rho \bar m_\mu -2 \alpha_{m \bar m}  k^\rho n^\mu \bar{\nabla}_\rho m_\mu -4 \alpha_{\bar m \bar m} k^\rho n^\mu \bar{\nabla}_\rho \bar m_\mu  =  \T_{k\bar{m}}  \label{EQmn1} \\
 &\B_{k m} +2 \alpha_{k  m} k^\rho \bar m^\mu \bar{\nabla}_\rho m_\mu -2 \alpha_{m \bar m}  k^\rho n^\mu \bar{\nabla}_\rho \bar m_\mu -4 \alpha_{ m  m} k^\rho n^\mu \bar{\nabla}_\rho  m_\mu  = \T_{k m} \label{EQnbarm1}\\
 &\B_{k k} -  2 \alpha_{km} k^\rho n^\mu \bar{\nabla}_\rho m_\mu - 2 \alpha_{k \bar m} k^\rho n^\mu \bar{\nabla}_\rho \bar m_\mu = \T_{kk} \label{EQnn1}
\end{align}

\smallskip
We are interested in the evolution equation of $\A_{\mu\nu}$ after the removal of the modes $kk$, $km$ and $k \bar m$, i.e. $\A'_{\mu\nu}$ .
The decomposition of $\A'_{\mu\nu}$ on the basis $z^\mu_a$ is given by
\be
 \A'_{\mu\nu} = \alpha_{mm}\,\Theta^{mm}_{\mu\nu}+\alpha_{\bar{m}\bar{m}}\,\Theta^{\bar{m}\bar{m}}_{\mu\nu}+\alpha_{m\bar{m}}\,\Theta^{m\bar{m}}_{\mu\nu} = \alpha_{EF}\,\Theta^{EF}_{\mu\nu}
 \ee
 with $E,F = \{m, \bar m\}$.
We can  find the equation of motion of $\A'_{\mu\nu}$ in two ways: either exploiting the gauge freedom associated to $\sigma_\mu$ as described in \ref{app_gauge}  on $\A_{\mu\nu}$ or by moving the terms proportional to $\Theta^{kk}_{\mu\nu}, \Theta^{km}_{\mu\nu}$ and $\Theta^{k \bar m}_{\mu\nu}$ in Eq.~\eqref{EOMgeomoptDecomposed} from the LHS to its RHS and use the equations just derived. Both of the procedures lead to the same result and here we show the second methodology.
We also choose a parallel transported null-tetrad along the rays to make the computations less involved, so that  Eqs.~\eqref{EQmm1} -~\eqref{EQnn1} reduce to
\be\label{EQnulltetradparallel}
\B_{m m} = \T_{m m}, \quad \B_{\bar m\bar m} = \T_{\bar m\bar m},\quad \B_{m \bar m} = \T_{m \bar m},\quad \B_{k \bar m}= \T_{k \bar m},\quad \B_{k m} = \T_{k m}, \quad \B_{k k}  = \T_{k k} \,.
\ee
In the case of parallel transported null-tetrad Eq~\eqref{EOMgeomoptDecomposed} is given by
\bea \label{EOMgeomoptDecomposedParallel}
\Theta^{CD}_{\mu\nu} \B_{CD} &=& t_{\mu\nu} = \T_{CD} \Theta^{CD}_{\mu\nu}\,,
\eea
where $C,D = \{k, m, \bar m\}$.
Now we move from the LHS to the RHS the terms proportional to $\Theta^{kk}_{\mu\nu}, \Theta^{km}_{\mu\nu}$ and $\Theta^{k \bar m}_{\mu\nu}$  in Eq~\eqref{EOMgeomoptDecomposedParallel}
\begin{align}\label{thetabef}
\Theta^{EF}_{\mu\nu}\, \B_{EF} = \Theta^{mm}_{\mu\nu} \B_{mm} +& \Theta^{\bar m \bar m}_{\mu\nu} \B_{\bar m \bar m} + \Theta^{m \bar m}_{\mu\nu} \B_{m \bar m} =\nonumber \\
&=  \T_{CD} \Theta^{CD}_{\mu\nu} - \Theta^{kk}_{\mu\nu} \B_{kk} + \Theta^{k m}_{\mu\nu} \B_{k \bar m} + \Theta^{k \bar m}_{\mu\nu} \B_{k \bar m} = \nonumber \\
&= \Theta^{mm}_{\mu\nu} \T_{mm} + \Theta^{\bar m \bar m}_{\mu\nu} \T_{\bar m \bar m} + \Theta^{m \bar m}_{\mu\nu} \T_{m \bar m} + \Theta^{k m}_{\mu\nu} \left(\T_{k m} - \B_{k m} \right) + \nonumber  \\
  & \quad+ \Theta^{ k\bar m }_{\mu\nu} \left(\T_{k \bar m} - \B_{k \bar m} \right)+ \Theta^{kk}_{\mu\nu} \left(\T_{k k} - \B_{k k} \right)\, = \nonumber\\
   & =\Theta^{mm}_{\mu\nu} \T_{mm} + \Theta^{\bar m \bar m}_{\mu\nu} \T_{\bar m \bar m} + \Theta^{m \bar m}_{\mu\nu} \T_{m \bar m}\,,
\end{align}
where in the last step we used Eqs.~\eqref{EQnulltetradparallel} and $E,F = \{m, \bar m\}$.
We define the first-order stress energy tensor deprived of the gauge modes
\be\label{deftprime}
t'_{\mu\nu} \coloneqq \Theta^{mm}_{\mu\nu} \T_{mm} + \Theta^{\bar m \bar m}_{\mu\nu} \T_{\bar m \bar m} + \Theta^{m \bar m}_{\mu\nu} \T_{m \bar m}\,.
\ee
It is possible to show that this is the same form of $t'_{\mu\nu}$ one would find by imposing the gauge fixing described in \ref{app_gauge} to eliminate $\alpha_{kk}, \alpha_{km}$ and $\alpha_{k \bar m}$.
Finally, we find the equation of motion of $ \A'_{\mu\nu}$  by realizing that
\begin{align}\label{aprime}
\Theta^{EF}_{\mu\nu}\,\B_{EF}  &= \Theta^{EF}_{\mu\nu}\, \left( 2 k^\rho \bar{\nabla}_\rho\,\alpha_{EF} + \alpha_{EF} \bar{\nabla}_\rho k^\rho  \right) = 2 k^\rho \bar{\nabla}_\rho\,\left( \alpha_{EF}  \Theta^{EF}_{\mu\nu}\,\right)+ \alpha_{EF} \Theta^{EF}_{\mu\nu}\, \bar{\nabla}_\rho k^\rho \nonumber \\
&=  2 k^\rho \bar  \nabla_\rho \,\A'_{\mu\nu} +  \A'_{\mu\nu} \bar \nabla_\rho k^\rho\,,
\end{align}
where we used the definition of $\B_{CD}$ given in Eq.~\eqref{BCD} and the fact that the tetrad is parallel transported so that $k^\rho \bar \nabla_\rho \Theta^{EF}_{\mu\nu} = 0 $.
Combining  Eq.~\eqref{aprime} and Eq.~\eqref{thetabef} yields
\begin{align}\label{eqAprime}
  2 k^\rho \bar  \nabla_\rho \,\A'_{\mu\nu} +  \A'_{\mu\nu} \bar \nabla_\rho k^\rho = t'_{\mu\nu}
\end{align}
which is the desired result.
The equation of motion of ${\bf e }'_{\mu\nu} $ can then be found by contracting Eq.~\eqref{eqAprime} with ${\bf e }'^{\mu\nu}$
\bea\label{eqAprimecontract}
 {\bf e }'^{\mu\nu} [2 k^\rho \bar  \nabla_\rho \,\A'_{\mu\nu} +  \A'_{\mu\nu} \bar \nabla_\rho k^\rho ]&=& t'_{\mu\nu} {\bf e }'^{\mu\nu} \nonumber\\
2 k^\rho \bar  \nabla_\rho \,\A' +  \A' \bar \nabla_\rho k^\rho &=& t'_{\mu\nu} {\bf e }'^{\mu\nu}
\eea
where we used ${\bf e }'^{\mu\nu} {\bf e }'_{\mu\nu} = 1$ so that $ {\bf e }'^{\mu\nu}  \A'_{\mu\nu} = \A' $.
Subtracting Eq.~\eqref{eqAprime} and Eq.~\eqref{eqAprimecontract} and using  $ \A'_{\mu\nu} = \A' {\bf e }'_{\mu\nu} $ yields
\be\label{PolAppendix}
 k^\rho \bar \nabla_\rho {\bf e }'_{\mu\nu} = \frac{1}{2 \A'} \left[t'_{\mu\nu} - {\bf e }'_{\mu\nu}\left(t'_{\rho\sigma} {\bf e }'^{\rho\sigma}\right) \right]\,,
\ee
which is Eq.~\eqref{evPolNolongitudinal}.
In the case of a stress-energy tensor given by
\be
t_{\mu\nu}\,\equiv\,
\left\{
\tau^{(A)}\,\A_{\mu \nu}+\tau^{(B)} \left(k_\mu {\cal C}_\nu +k_\nu {\cal C}_\mu \right)+
\tau^{(C)} \left[k_\mu \bar v_\nu+ k_\nu \bar v_\mu-\bar g_{\mu\nu} \left(k_\rho \bar v^\rho\right) \right] \right\}\,,
\ee
we can use Eq.~\eqref{componentstprime} to find the components of $t'_{\mu\nu}$ in the null tetrad basis explicitly. These are given by
\bea
\T_{m\bar{m}} &=& \tau^{(A)} \alpha_{m\bar{m}} - 2 \tau^{(C)}\left(k_\rho \bar v^\rho\right)\,, \\
\T_{mm}&=&  \tau^{(A)} \alpha_{m{m}}\,,\\
\T_{\bar{m}\bar{m}} &=&  \tau^{(A)} \alpha_{\bar{m}\bar{m}}\,,
\eea
so that $t'_{\mu\nu} $, defined as in Eq.~\eqref{deftprime}, takes the form
\bea \label{tprimetau}
t'_{\mu\nu} &=& \tau^{(A)} \alpha_{m{m}} \Theta^{mm}_{\mu\nu} +   \tau^{(A)} \alpha_{\bar{m}\bar{m}} \Theta^{\bar m \bar m}_{\mu\nu}  + \left(\tau^{(A)} \alpha_{m\bar{m}} - 2 \tau^{(C)}\left(k_\rho \bar v^\rho\right) \right) \Theta^{m \bar m}_{\mu\nu}  \nonumber\\
&=& \tau^{(A)}  \A'_{\mu\nu} - 2 \tau^{(C)}\left(k_\rho \bar v^\rho\right)  \Theta^{m \bar m}_{\mu\nu} =\tau^{(A)}  \A'_{\mu\nu} - \sqrt{2} \tau^{(C)}\left(k_\rho \bar v^\rho\right)  {\bf e}^S_{\mu\nu}  \,,
\eea
where, in the last step, we used ${\bf e}^S_{\mu\nu} = \sqrt{2} \Theta^{m \bar m}_{\mu\nu} $ as claimend in Eq.~\eqref{polarizationtotetrad}.
Using these expressions Eqs.~\eqref{EQmm1}, \eqref{EQbarmbarm1} and \eqref{EQmbarm1}  become
\bea
2 k^\rho \bar{\nabla}_\rho \alpha_{mm} +  \alpha_{mm} \bar{\nabla}_\rho  k^\rho &=& \tau^{(A)} \alpha_{m{m}}\,, \label{EQmm2}\\
2 k^\rho \bar{\nabla}_\rho \alpha_{\bar m\bar m} +  \alpha_{\bar m\bar m} \bar{\nabla}_\rho  k^\rho &=& \tau^{(A)} \alpha_{\bar m\bar m}\,, \label{EQbarmbarm2}\\
2 k^\rho \bar{\nabla}_\rho\,\alpha_{m\bar{m}} + \alpha_{m\bar{m}} \bar{\nabla}_\rho k^\rho &=& \tau^{(A)} \alpha_{m\bar{m}} - 2 \tau^{(C)}\left(k_\rho \bar v^\rho\right)\,. \label{EQmbarm2}
\eea
Using Eqs.~\eqref{amplitudestoalpha} and \eqref{polarizationtotetrad} into the sum and the difference of Eq.~\eqref{EQmm2} and Eq.~\eqref{EQbarmbarm2}, and setting $\A^+ = \A^\times = \A^T$, then one may recover the evolution equation of $\A^T$ present in the main text, i.e. Eq.~\eqref{eqAT2}.
Eq.~\eqref{EQmbarm2} is in turn equivalent to  Eq.~\eqref{eqAS2} after substituting $\alpha_{m \bar m} = \sqrt{2} \A^S$.

Now we turn to the evolution equation of the polarization tensor after the removal of the modes $kk$, $km$ and $k \bar m$, Eq.~\eqref{PolAppendix}.
Note that Eqs.~\eqref{amplitudestoalpha} and \eqref{polarizationtotetrad} allow us to decompose $\A'_{\mu\nu}$ both on the basis $\Theta^{EF}_{\mu\nu}$ or $({\bf e}^+_{\mu\nu}, {\bf e}^\times_{\mu\nu}, {\bf e}^S_{\mu\nu})$ equivalently, i.e.
\be
\A'_{\mu\nu} = \alpha_{mm}\,\Theta^{mm}_{\mu\nu}+\alpha_{\bar{m}\bar{m}}\,\Theta^{\bar{m}\bar{m}}_{\mu\nu}+\alpha_{m\bar{m}}\,\Theta^{m\bar{m}}_{\mu\nu}  = \A^+ {\bf e}^+_{\mu\nu} + \A^\times {\bf e}^\times_{\mu\nu}+ \A^S {\bf e}^S_{\mu\nu}\,.
\ee
 For $t'_{\mu\nu}$ of the form in Eq.~\eqref{tprimetau} we have
\begin{align}
{\bf e }'^{\mu\nu}t'_{\mu\nu} &=  \tau^{(A)}  \A'  - \sqrt{2} \tau^{(C)}\left(k_\rho \bar v^\rho\right)  {\bf e }'^{\mu\nu}  {\bf e}^S_{\mu\nu}  =  \tau^{(A)}  \A'  - \sqrt{2} \tau^{(C)}\left(k_\rho \bar v^\rho\right) \frac{ \A'^{\mu\nu} {\bf e}^S_{\mu\nu}}{\A'} \nonumber \\
& = \tau^{(A)}  \A'  - \sqrt{2} \tau^{(C)}\left(k_\rho \bar v^\rho\right) \frac{ \A'^{\mu\nu} {\bf e}^S_{\mu\nu}}{\A'} = \tau^{(A)}  \A'  - \sqrt{2} \tau^{(C)}\left(k_\rho \bar v^\rho\right) \frac{ \A^S}{\A'}
\end{align}
so that
\begin{align}
t'_{\mu\nu} - {\bf e }'_{\mu\nu}\left(t'_{\rho\sigma} {\bf e }'^{\rho\sigma}\right) &= \sqrt{2} \tau^{(C)}\left(k_\rho \bar v^\rho\right) \frac{ \A^S}{\A'} {\bf e }'_{\mu\nu}
 - \sqrt{2} \tau^{(C)}\left(k_\rho \bar v^\rho\right)  {\bf e}^S_{\mu\nu} = \nonumber \\
 & = \sqrt{2} \tau^{(C)}\left(k_\rho \bar v^\rho\right) \left[ \frac{ \A^S}{\A'^2} \A'_{\mu\nu} - {\bf e}^S_{\mu\nu} \right]\,,
\end{align}
and
 \bea
 k^\rho \bar \nabla_\rho {\bf e }'_{\mu\nu} &=& \frac{1}{2 \A'} \left[t'_{\mu\nu} - {\bf e }'_{\mu\nu}\left(t'_{\rho\sigma} {\bf e }'^{\rho\sigma}\right) \right] = \frac{\left(k_\rho \bar v^\rho\right) \tau^{(C)}}{\sqrt{2}\A'} \left[\frac{ \A^S}{\A'^2} \A'_{\mu\nu} - {\bf e}^S_{\mu\nu}\right]\label{EQpolAllGauge}
\eea
which is the final step in Eq.~\eqref{evPolNolongitudinal}.
Note that, in general, the LHS of Eq.~\eqref{EQpolAllGauge} is non zero even if the null tetrad is chosen parallel transported along the rays because
\be
  k^\rho \bar \nabla_\rho  {\bf e }'_{\mu\nu} = k^\rho \bar \nabla_\rho \left( \frac{\A'_{\mu\nu}}{\A'} \right) = k^\rho \bar \nabla_\rho \left( \frac{\alpha'_{EF} \Theta^{EF}_{\mu\nu}}{\A'} \right)  = k^\rho \, \Theta^{EF}_{\mu\nu}\, \bar \nabla_\rho \left( \frac{\alpha_{EF} }{\A'} \right)
\ee
where $E, F = \{ m, \bar m\}$. The Eq. above states that  the transport properties of ${\bf e}'_{\mu\nu}$ depend on the evolution equations of the amplitude of the modes present. In general ${\bf e}'_{\mu\nu}$ is not parallel transported whenever the amplitudes of the single components of the wave decay differently between each other. In the theory we considered, the spin-0 mode satisfies an evolution equation ( i.e. Eq.~\eqref{eqAS2}) which is different from the one of the two tensor modes (i.e. Eq.~\eqref{eqAT2}). This fact has as consequence that the total polarization is not parallel transported along the GW's geodesic.
In the case of GR, where the RHS of Eq.~\eqref{EQpolAllGauge} is zero, we see that also the LHS is zero. In fact, in GR the mode $m \bar m$ is not present and the plus and cross polarization modes have the same amplitude so that $\A'_{\mu\nu} = \A^+ {\bf e}^+_{\mu\nu} + \A^\times {\bf e}^\times_{\mu\nu} = \A^T \left({\bf e}^+_{\mu\nu} + {\bf e}^\times_{\mu\nu} \right) $ and $\A' = \sqrt{2} \A^T$ . Thus
\begin{align}
k^\rho \, \Theta^{EF}_{\mu\nu}\, \bar \nabla_\rho \left( \frac{\alpha_{EF} }{\A'} \right) &= k^\rho \,  \left({\bf e}^+_{\mu\nu} + {\bf e}^\times_{\mu\nu} \right) \, \bar \nabla_\rho \left( \frac{\A^T}{ \sqrt{2} \A^T} \right) =   k^\rho  \left({\bf e}^+_{\mu\nu} + {\bf e}^\times_{\mu\nu} \right)\, \bar \nabla_\rho \left( \frac{1}{\sqrt{2}} \right) = 0\,.
\end{align}
%
%
%
%
%
%
%
%


\end{appendix}
\providecommand{\href}[2]{#2}\begingroup\raggedright\endgroup

\end{document}